\documentclass[pre,eqsecnum,preprint,showpacs]{revtex4}
\usepackage{amsmath}
\usepackage{epsfig}
\usepackage{graphicx}

\def\hangin{\begin{list}{}{\setlength{\leftmargin}{1.1in}
\setlength{\itemindent}{-1.1in}}}
\def\endhangin{\end{list}}
\input{tcilatex}

\begin{document}

\title{Nonequilibrium Statistical Mechanics and Hydrodynamics for a Granular
Fluid}
\author{James W. Dufty}
\affiliation{Department of Physics, University of Florida, Gainesville, FL 32611}

\begin{abstract}
Granular fluids consist of collections of activated mesoscopic or
macroscopic particles (e.g., powders or grains) whose flows often appear
similar to those of normal fluids. To explore the qualitative and
quantitative description of these flows an idealized model for such fluids,
a system of smooth inelastic hard spheres, is considered. The single feature
distinguishing granular and normal fluids being explored in this way is the
inelasticity of collisions. The dominant differences observed in real
granular fluids are indeed captured by this feature. Following a brief
introductory description of real granular fluids and motivation for the
idealized model, the elements of nonequilibrium statistical mechanics are
recalled (observables, states, and their dynamics). Peculiarities of the
hard sphere interactions are developed in detail. The exact microscopic
balance equations for the number, energy, and momentum densities are derived
and their averages described as the origin for a possible macroscopic
continuum mechanics description. This formally exact analysis leads to
closed, macroscopic hydrodynamic equations through the notion of a
\textquotedblleft normal" state. This concept is introduced and the
Navier-Stokes constitutive equations are derived, with associated Green-Kubo
expressions for the transport coefficients. A parallel description of
granular gases is described in the context of kinetic theory, and the
Boltzmann limit is identified critically. The construction of the
\textquotedblleft normal\textquotedblright\ solution to the kinetic equation
is outlined, and Navier-Stokes order hydrodynamic equations are re-derived
for a low density granular gas.

These are notes prepared as the basis for six lectures at the Second Warsaw
School on Statistical Physics held in Kazimierz, Poland, June 2007.
\end{abstract}

\date{\today }
\maketitle

\address{Department of Physics, University of Florida, Gainesville, Florida
32611}

\section{ Introduction}

\label{sec1}

Granular fluids are ubiquitous in nature \cite{General,Review}. To
illustrate with the familiar first, go to your kitchen and take out the
mustard seeds, pepper corns, salt, and rice. Put 100 grains of each in small
jars and tumble these jars in various directions. The gravitational field
induces temporary flows of groups of grains in each case. While the flows
appear to be locally convective it is clear that each grain's motion is more
complex due to interactions with other grains. This is qualitatively similar
to macroscopic flows of normal fluids (those composed of atoms or molecules)
in which the coarse grained convection is the collective effect of complex
atomic collisional motion. The objective here is to explore to what extent
the well developed methods to describe the macroscopic dynamics of normal
fluids \cite{Hansen} can be extended to the flow of such compositions of
grains \cite{Haff}. Clearly, there are significant differences in the
physical systems. It will be seen that many properties of interest are
insensitive to some of the most obvious differences, and that others can be
incorporated in realistic models for the system of grains.

Normal fluids incorporate a wide range of different systems. Simple atomic
fluids are well represented by spherically symmetric central point forces
between the atomic constituents, and Navier-Stokes hydrodynamics (defined
precisely below) applies to most macroscopic nonequilibrium states of
interest in this case. The interactions in low molecular weight fluids no
longer have spherical symmetry, but at the macroscopic level this is a
quantitative rather than a qualitative effect (e.g., only the values of
transport coefficients in the Navier-Stokes description change). Mixtures of
these types of fluids also have the same qualitative macroscopic behavior.
On the other hand complex fluids, such as those composed of high molecular
weight (polymers), generally exhibit quite different macroscopic behavior
which can depend on both the system and the class of macroscopic states
considered.

Some of this diversity is evident for systems of grains as well. The mustard
seeds are more mono disperse and smooth than the pepper corns, but both are
roughly spherical. On the other hand rice is asymmetric to differing extents
(e.g., large aspect ratio in China, small in Spain). As with normal fluids,
their macroscopic flows are quite similar. Salt has irregular shape and
furthermore is complicated by being hydrophilic - any moisture in the air
changes dramatically the interaction between grains. This can change its
motion as well, as everyone knows about salt shakers at the seaside. Thus,
the medium in which the grains move can be important. Structurally complex
grains such as collections of filaments will not be considered here. Still,
structurally simple granular systems can behave as complex fluids in many
nonequilibrium states of interest. This is an important difference between
normal fluids and granular systems that provides some of the most difficult
challenges and also some of the most interesting opportunities.

Beyond the kitchen, granular systems include objects of interest to the
phamacutical industry (pills and associated powders generated in their
production), the agricultural industry (storage and transport of edible
grains), geology (rock and snow avalanches), and extra-terrestrial systems
(Saturn's rings, regolith on Mars). Two general classes of states for these
granular systems are distinguished, \textit{compact} and \textit{activated}.
Unshaken, the rice in the jar appears at rest. In fact, each grain has some
kinetic energy due to the temperature of the room. However, on Earth the
gravitational potential energy relative to the bottom of the container for
heights $h$ greater than the dimension of the grain is much larger than this
energy of motion, due to the large mass of typical grains. Thus, they pack
at the bottom of the container. Questions about the possible distribution of
packing configurations constitutes a field of current activity \cite{Halsey}%
, and is of central practical interest for the storage of grains. For
example, granular storage in silos commonly leads to explosions whose
mitigation requires knowledge of the distribution of forces in the packed
grains. Here, attention is limited to the second class of \textit{activated}
grains. Work is done on the system (e.g., shaking) to provide a kinetic
energy greater than that required to overcome the compactification by
gravity. In addition, when the number of grains is large the activation
typically induces an apparent random component to the granular motion due to
frequent collisions among the grains. Systems of activated, collisional
grains constitute the qualitative definition of \textit{granular fluids} for
the discussion here.

Phenomenologically, Navier-Stokes hydrodynamics has been applied to a wide
class of granular flows in practice with qualitative success in many cases.
To what extent can such a description be justified from a more fundamental
basis? If justified, how can the parameters of this description (equation of
state, energy loss function, transport coefficients) be given fundamental
definitions as functions of the state conditions? If justified, what is the
context and limitations of this description? These are the issues that are
addressed here. As with the history of progress on these same questions for
normal fluids, initial attention is focused on an idealized fluid and states
where conceptual problems can be isolated and addressed cleanly. Steps to
describe more realistic granular fluids can then proceed with increased
confidence and guidance. In the next section this idealized granular fluid
is defined and the overview of its exploration is summarized.

\section{Overview of the Presentation}

\label{sec2}

The mustard seeds are hard, spherical, and monodisperse. This system can be
refined for experiments using carefully machined spheres of glass or metal
with empirically determined binary collision properties. Although hard, the
binary collisions are quite complex in detail since each is composed of a
large number of composite molecules. While in contact, the shape of each
grain is distorted and energy is redistributed between the kinetic energy of
their centers of mass and that of their internal degrees of freedom. On
separation, they again move freely but with some energy lost to internal and
rotational degrees of freedom. The "hard" interaction means that the contact
time is short, so effectively the motion is that of free streaming
punctuated by velocity changes on collision with a consequent loss of energy
in each case. This suggests the idealization of a system of hard spheres
(zero collision time) with inelastic collisions. A further simplification is
the neglect of transfer between kinetic and rotational motion. This
idealized fluid then consists of a system of smooth, inelastic hard spheres.
It is similar to the idealized hard sphere fluid for normal atomic systems,
with the sole difference being an energy loss on pair collisions.

In the next section the context of this ideal granular fluid is elaborated
further to clarify the extent to which its properties should represent both
qualitatively and quantitatively many properties of real granular fluids.
The elements of nonequilibrium statistical mechanics are recalled briefly
and applied to this system of hard, inelastic, smooth spheres \cite%
{Brey97,Dufty00,vanN01} . Since the forces are singular, the usual form for
the generator of dynamics for piecewise continuous forces no longer applies
and the necessary changes are described in Appendix A. As a consequence of
these changes the generators in different representations for the time
dependence are different (e.g., that for the Liouville equation and that for
the observables). Three different generators are identified in Appendix A.
The average energy for an isolated system is shown to decrease monotonically
due to the loss of energy on each binary collision. Consequently, there is
no equilibrium Gibbs solution to the Liouville equation. Instead, it
supports a "homogeneous cooling state" (HCS) in which the dynamics of the
energy loss appears only through a scaling of the velocities (hence the
terminology "cooling") and associated normalization factors. Empirically
(i.e., in molecular dynamics simulations) it is found that a wide class of
homogeneous states approach the HCS after a few collisions per particle \cite%
{Lutsko02}. Consequently, the rapid velocity relaxation in normal fluids
leading to the Gibbs distribution has its counterpart in this granular fluid
in the approach on a similar time scale to the HCS distribution. The scaling
form of the HCS suggests a related dimensionless representation for the
Liouville equation in a more general context.

As described above, a main target of this investigation is the possibility
of a macroscopic hydrodynamics. This refers to closed equations for the
hydrodynamic observables: number density, energy density, and momentum
density. As a fundamental starting point, the exact microscopic phase
functions corresponding to these fields are identified and the balance
equations for them are obtained in Appendix B. For a normal fluid, these are
the local microscopic conservation laws. Section \ref{sec4} discusses the
average of these equations, the macroscopic balance equations, which is the
framework in which the corresponding hydrodynamic equations can exist.
Contributions to the fluxes in these equations due to convection are
identified from a local Galilean transformation \cite{McL}. The remaining
contributions are due to "dissipative" exchanges of density, energy, and
momentum for fluid cell at rest. The derivation of explicit expressions for
this remainder, called "constitutive equations", is the central problem for
obtaining hydrodynamic equations. A change of variables from density, energy
density, and momentum density to density, temperature, and flow velocity is
defined and introduced. Finally, the macroscopic state corresponding to the
HCS is identified from these equations.

A more precise definition of hydrodynamics is considered in Section \ref%
{sec5}, where the notion of "normal" states is introduced and motivated. It
is shown how this concept leads to constitutive equations which a
functionals of the hydrodynamic fields, giving in principle a set of five
closed equations for the determination of these fields. The special case of
weakly inhomogeneous states is considered in Section \ref{sec6} as an
example, leading to the phenomenological Navier-Stokes equations for a
granular fluid. Two important differences between normal and granular fluids
are noted, long wavelength instability and the existence of new stationary
states.

The formal construction of a normal solution to the Liouville equation is
considered in Section \ref{sec7} for weakly inhomogeneous states by an
expansion in the local spatial gradients. The leading order reference state
is a local HCS distribution, similar to the local equilibrium Gibbs
distribution for normal fluids. The constitutive equations at this
approximation provide definitions for the hydrostatic pressure and the local
cooling rate in each fluid cell. The formal solution to the Liouville
equation including first order in the gradients provides constitutive
equations characterized by the transport coefficients for a granular fluids.
This extends recent results obtained from a solution to the Liouville
equation for small initial perturbations (linear response) \cite{DBB06,DBB07}%
. Expressions for these transport coefficients, the Green-Kubo formulas for
granular fluids, also are discussed in Section \ref{sec7}. The specific case
of shear viscosity is explored further for illustration. The simplest test
of these formal developments is provided by the dynamics of a single
impurity in a granular fluid in its HCS, corresponding to impurity motion in
an equilibrium fluid. The expected hydrodynamics in this case is simple
diffusion of the particle's probability density. The Green-Kubo expression
for the diffusion coefficient is noted, and a practical short time
approximation for the associated velocity autocorrelation function is
demonstrated \cite{Dufty02}.

A different approach to the determination of constitutive equations is
provided by kinetic theory, a representation in terms of the reduced single
particle distribution function. This approach is described in Section \ref%
{sec8} where the objects of interest are shown to be determined exactly by
the single particle distribution function, and the hierarchy of equations
determining all reduced distribution functions is derived from the Liouville
equation. A formal solution to the hierarchy is obtained systematically at
low density, resulting in a Boltzmann description for a granular gas. The
derivation of hydrodynamics from this kinetic theory is outlined and
compared to the general fluid results.

Finally, some additional perspective on these results is given in the
Discussion Section. The analysis here suggests that hydrodynamics is a
useful concept for granular fluids, but its context and limitations have not
been explored in any detail. Appropriate space and time scales for
hydrodynamics in general and limitations of the Navier-Stokes approximation
in particular are discussed.

As these notes are prepared for a series of lectures the material is focused
on the research of the author and his colleagues. The references quoted are
heavily weighted in that direction as well. Apologies are offered at the
outset for the many excellent contributions to this subject matter not given
explicit recognition.

\section{Idealized Granular Fluid and Statistical Mechanics}

\label{sec3}

\subsection{Hard sphere idealization}

Consider a fluid comprised of mono-disperse, spherical particles with
pairwise additive central interactions (i.e., smooth particles with no
tangential momentum transfer). Also, assume that Newton's third law holds so
that momentum is conserved between colliding pairs. More general cases of
mixtures, non-central, or many-body forces can be incorporated with greater
complexity but no significant conceptual changes. The case of dissipative
"soft" spheres is considered first. The pair force is assumed to be
piecewise continuous for relative distance between particles $r$ and $s$, $%
q_{rs}\equiv \left\vert \mathbf{q}_{r}-\mathbf{q}_{s}\right\vert \leq \sigma 
$ and vanishing for $q_{rs}>\sigma $. As a central force it is directed
along the line of centers $\widehat{\mathbf{q}}_{rs}$ and therefore has the
form 
\begin{equation}
\mathbf{F}\left( \mathbf{q}_{rs},\mathbf{g}_{rs}\right) =\widehat{\mathbf{q}}%
_{rs}\Theta \left( \sigma -q_{rs}\right) f\left( q_{rs},{g}_{rs},\widehat{%
\mathbf{q}}_{rs}\cdot \widehat{\mathbf{g}}_{rs}\right) .  \label{3.1}
\end{equation}%
Here $\mathbf{g}_{rs}=\mathbf{v}_{r}-\mathbf{v}_{s}$ is the velocity of
approach or separation. From spherical symmetry the magnitude of the force
can depend only on the scalars $q_{rs},{g}_{rs},\widehat{\mathbf{q}}%
_{rs}\cdot \widehat{\mathbf{g}}_{rs}$. The functional form of this force is
such as to describe the two physical effects of repulsion and dissipation
during the deformation. For example, it could be the superposition of an
conservative repulsive elastic force plus a drag force proportional to the
normal component of the relative velocity. The amount of deformation is then 
$d\sim \sqrt{e/k}$ where $e$ is the energy per particle and $k$ is the
elastic constant. The conditions of interest here are such that $d/\sigma
<<1 $ and $\tau _{c}/\tau _{m}<<1$, where $\tau _{c}\sim d\sqrt{m/e}$ is the
average contact time and $\tau _{m}\sim n^{-1/3}\sqrt{m/e}$ is the mean free
time between collisions. Consequently, $\left( \tau _{c}/\tau _{m}\right)
\sim \left( n\sigma ^{3}\right) ^{1/3}(d/\sigma )<(d/\sigma )$ since $%
n\sigma ^{3}<1$ for fluid states. These are rough estimates, but they show
that the controllable parameters $e$ and $k$ admit conditions where the
particles behave as hard objects. A force law $f\left( q_{rs}\right) \sim
\left( \sigma /r\right) ^{n}$ describes well the repulsive forces for simple
atomic fluids, with $n>12$, and its properties are known to be accurately
represented by those for hard spheres $n\rightarrow \infty $ \cite{Hansen}.
A similar limit for the conservative part of the deformation potential can
be expected as well.

It is necessary also to retain the fact that there is a total energy loss
during the contact time in this hard sphere limit. This is done by requiring
that the collisions are inelastic. In detail, the total momentum for the
colliding pair is conserved and the relative velocity changes according to%
\begin{equation}
\mathbf{g}_{rs}^{\prime }=\mathbf{g}_{rs}-\left( 1+\alpha \left( \widehat{%
\mathbf{q}}_{rs}\cdot \mathbf{g}_{rs}\right) \right) \left( \widehat{\mathbf{%
q}}_{rs}\cdot \mathbf{g}_{rs}\right) \widehat{\boldsymbol{\sigma }},
\label{3.2}
\end{equation}%
with the corresponding energy change

\begin{equation}
\Delta \left( \frac{1}{2}m\left( v_{r}^{2}+v_{s}^{2}\right) \right) =\frac{1%
}{4}m\left( {g}_{rs}^{\prime 2}-g_{rs}^{2}\right) =-\frac{1}{4}m\left(
1-\alpha ^{2}\left( \widehat{\mathbf{q}}_{rs}\cdot \mathbf{g}_{rs}\right)
\right) \left( \widehat{\mathbf{q}}_{rs}\cdot \mathbf{g}_{rs}\right) ^{2}.
\label{3.3}
\end{equation}%
The scalar parameter $\alpha \left( \widehat{\mathbf{q}}_{rs}\cdot \mathbf{g}%
_{rs}\right) $ is the restitution coefficient and takes on the values $%
0<\alpha \leq 1$, with $\alpha =1$ corresponding to elastic collisions. A
detailed correspondence between inelastic hard spheres and soft dissipative
spheres requires that the restitution coefficient should depend on the
normal component of the relative velocity \cite{Poschel2}. In particular, it
is found that $\alpha \left( \widehat{\mathbf{q}}_{rs}\cdot \mathbf{g}%
_{rs}\right) \rightarrow 1$ as $\widehat{\mathbf{q}}_{rs}\cdot \mathbf{g}%
_{rs}\rightarrow 0$. However, if the conditions studied avoid this limit
(e.g., continual energy input), then it is possible to consider the further
idealization of an average energy loss per collision characterized by a
constant restitution coefficient $\alpha \left( \widehat{\mathbf{q}}%
_{rs}\cdot \mathbf{g}_{rs}\right) \rightarrow \alpha $. This will be the
case studied in all of the following.

There is a mathematical price to be paid for this idealization of smooth,
inelastic, hard spheres. The forces are singular at contact, and therefore
the usual description of the dynamics from Hamilton's equations must be
modified \cite{Ernst,Dom,Resibois,McL}. The modification corresponds to
replacing the effect of the continuous force by a binary scattering operator
that generates the instantaneous momentum change of the pair on contact. The
form of this operator is derived in Appendix A. This complication is common
to the representation by hard spheres of both normal and granular fluids. In
the latter case there is another, perhaps unexpected, effect of the hard
collisions plus dissipation called inelastic collapse \cite{McNamaraYoung} .
To understand this, consider a perfectly elastic hard ball dropped from a
height $h$ on a horizontal table in the gravitational field. Without
interference it bounces indefinitely and it is never in contact with the
table over any finite time interval. In contrast, the inelastic hard sphere
will undergo an infinite number of collisions in a finite time $\tau $ and
come to rest in contact with the table. \ A similar effect occurs in the
inelastic hard sphere granular fluid where groups of particles can cluster
with increasing numbers of collisions among them in a finite time interval.
The issue of inelastic collapse will be revisited at the end of this section.

\subsection{Overview of nonequilibrium statistical mechanics}

For a given physical system its description via statistical mechanics
consists of states (represented by distribution functions over its phase
space), observables (phase space functions), and expectation values
(averages of the observables over the states). Its evolution in time is
given by a mapping of either the states or the observables in phase space.
The system of interest here is a one component fluid of $N$ identical
smooth, inelastic, hard spheres (mass $m$, diameter $\sigma $). The state of
the system at time $t=0$ is completely characterized by the positions and
velocities of all particles, represented by a point in a $6N$ dimensional
phase space $\Gamma _{0}\equiv \left\{ \mathbf{q}_{1}(0),\ldots ,\mathbf{q}%
_{N}(0),\mathbf{v}_{1}(0),\dots ,\mathbf{v}_{N}(0)\right\} $. The dynamics
consists of straight line motion along the direction of the velocity at time 
$t$ (free streaming), until any pair of particles, say $i,j$, is in contact.
The relative velocity $\mathbf{g}_{ij}=\mathbf{v}_{i}-\mathbf{v}_{j}$ of
that pair changes instantaneously according to a given collision rule (\ref%
{3.2}) while their total momentum is unchanged. Subsequently, all particles
continue to stream freely until another pair is at contact, and the
collisional change is repeated for that pair. In this way a unique
trajectory $\Gamma _{t}\equiv \left\{ \mathbf{q}_{1}(t),\ldots ,\mathbf{q}%
_{N}(t),\mathbf{v}_{1}(t),\dots ,\mathbf{v}_{N}(t)\right\} $ is generated in
the phase space for $t>0$, where the configurational degrees of freedom
change continuously while those for the velocities are piecewise constants.

The statistical mechanics for a fluid of inelastic hard spheres has been
described elsewhere \cite{Brey97,Dufty00,vanN01,Kandrup,DBL02}. It is
comprised of the dynamics just described, a macrostate specified in terms of
a probability density $\rho (\Gamma )$, and a set of observables denoted by $%
A(\Gamma )$. The expectation value for an observable at time $t>0$ for a
state $\rho (\Gamma )$ given at $t=0$ is defined by 
\begin{equation}
\langle A(t);0\rangle \equiv \int d\Gamma \rho (\Gamma )A(\Gamma _{t})
\label{3.4}
\end{equation}%
where $A(t)=A(\Gamma _{t})$, and $\Gamma _{t}\equiv \left\{ \mathbf{q}%
_{1}(t),\ldots ,\mathbf{q}_{N}(t),\mathbf{v}_{1}(t),\dots ,\mathbf{v}%
_{N}(t)\right\} $ is the phase point evolved to time $t$ from $\Gamma
=\Gamma _{t=0}$. The dynamics can be represented in terms of a generator $L$
defined by 
\begin{equation}
\langle A(t);0\rangle =\int d\Gamma \,\rho (\Gamma )e^{tL}A(\Gamma ).
\label{3.5}
\end{equation}%
For continuous potentials the generator is easily recognized from Hamilton's
equation as a Poisson bracket operation with the corresponding Hamiltonian.
However, its identification for the discontinuous hard sphere potential is
less direct \cite{Ernst,Dom,Resibois,McL}. There are two components to the
generator, corresponding to the two steps of free streaming and velocity
changes at contact. The first part is the same as for continuous potentials
while the second part replaces the contribution from the singular force by a
\textquotedblright binary collision operator\textquotedblright\ $T(i,j)$ for
each pair $i,j$ \cite{lutsko1} 
\begin{equation}
L=\sum_{i=1}^{N}\mathbf{v}_{i}\cdot \mathbf{\nabla }_{i}+\frac{1}{2}%
\sum_{i=1}^{N}\sum_{j\neq i}^{N}T(i,j).  \label{3.6}
\end{equation}%
The binary collision operator $T(i,j)$ for normal fluids is identified
directly from the Poisson bracket of Hamilton's equations 
\begin{equation}
T(i,j)\rightarrow \theta _{ij}=m^{-1}\mathbf{F}(q_{ij})\cdot \left( 
\boldsymbol{\nabla }_{\mathbf{v}_{i}}-\boldsymbol{\nabla }_{\mathbf{v}%
_{j}}\right) .  \label{3.7}
\end{equation}%
where $\mathbf{F}(q_{ij})$ is a conservative force. For hard spheres, the
position variables are still continuous functions of time but the momenta
are piecewise constant (in the absence of external forces) and
discontinuous. The form for $T(i,j)$ in this case is obtained in Appendix A
with the result 
\begin{equation}
T(i,j)=\Theta (-\mathbf{g}_{ij}\cdot \widehat{\mathbf{q}}_{ij})|\mathbf{g}%
_{ij}\cdot \widehat{\mathbf{q}}_{ij}|\delta (q_{ij}-\sigma )(b_{ij}-1),
\label{3.8}
\end{equation}%
where $\Theta $ is the Heaviside step function, and $b_{ij}$ is a
substitution operator which changes the relative velocity $\mathbf{g}_{ij}$
into its scattered value $\mathbf{g}_{ij}^{\prime }$, given by Eq.\ (\ref%
{3.2}) 
\begin{equation}
b_{ij}A(\mathbf{g}_{ij})=A(\mathbf{g}_{ij}^{\prime }).  \label{3.9}
\end{equation}%
The theta function and delta function in (\ref{3.8}) assure that a collision
takes place, i.e. the pair is at contact and directed toward each other.

An alternative equivalent representation of the dynamics is obtained by
transfering the dynamics from the observable $A(\Gamma )$ to the state $\rho
(\Gamma )$ by the definition 
\begin{equation}
\int d\Gamma \,\rho (\Gamma )e^{tL}A(\Gamma )\equiv \int d\Gamma \,\left(
e^{-t\overline{L}}\rho (\Gamma )\right) A(\Gamma ).  \label{3.10}
\end{equation}%
The representation in terms of a dynamical state is referred to as Liouville
dynamics. The probability density $\rho (\Gamma )$ must vanish for all
configurations of overlapping hard spheres, so the domain of integration on
the left side of (\ref{3.10}) is effectively restricted to non-overlapping
configurations. Thus the generator $L$ is used always in that context.
However, the right side of (\ref{3.10}) no longer has that restriction and
consequently the generator for Liouville dynamics is not the same as that
for observables (as in the case of continuous potentials). Instead, direct
analysis of (\ref{3.10}) leads to the result (see Appendix \ref{appA}) 
\begin{equation}
\overline{L}=\sum_{i=1}^{N}\mathbf{v}_{i}\cdot \mathbf{\nabla }_{i}-\frac{1}{%
2}\sum_{i=1}^{N}\sum_{j\neq i}^{N}\overline{T}(i,j),  \label{3.11}
\end{equation}%
with the new binary collision operator 
\begin{equation}
\overline{T}_{-}(i,j)=\delta (q_{ij}-\sigma )|\mathbf{g}_{ij}\cdot \widehat{%
\mathbf{q}}_{ij}|(\Theta (\mathbf{g}_{ij}\cdot \widehat{\mathbf{q}}%
_{ij})\alpha ^{-2}b_{ij}^{-1}-\Theta (-\mathbf{g}_{ij}\cdot \widehat{\mathbf{%
q}}_{ij})).  \label{3.12}
\end{equation}%
Here $b_{ij}^{-1}$ is the inverse of the operator $b_{ij}$ in (\ref{3.9}).

Next consider time correlation functions for two observables $A$ and $B$%
\begin{equation}
\langle A(t)B;0\rangle \equiv \int d\Gamma \left( e^{tL}A(\Gamma )\right)
\rho (\Gamma )B(\Gamma )=\int d\Gamma A(\Gamma )e^{-t\overline{L}}\left(
\rho (\Gamma )B(\Gamma )\right) .  \label{3.13}
\end{equation}%
A third generator is defined for reversed\ dynamics along the trajectory by%
\begin{equation}
e^{-t\overline{L}}\left( \rho (\Gamma )B(\Gamma )\right) \equiv \left( e^{-t%
\overline{L}}\rho (\Gamma )\right) \left( e^{-tL_{-}}B(\Gamma )\right) .
\label{3.15}
\end{equation}%
This gives%
\begin{eqnarray}
\langle A(t)B;0\rangle &=&\int d\Gamma A(\Gamma )e^{-t\overline{L}}\left(
\rho (\Gamma )B(\Gamma )\right) =\int d\Gamma \rho (\Gamma ,t)A(\Gamma
)e^{-tL_{-}}B(\Gamma )  \notag \\
&=&\langle AB(-t);t\rangle .  \label{3.14}
\end{eqnarray}%
where the reversed dynamics of $B(-t)$ has been defined by%
\begin{equation}
B(-t)\equiv e^{-tL_{-}}B(\Gamma ).  \label{3.14a}
\end{equation}%
The new generator is identified in Appendix \ref{appA} as%
\begin{equation}
L_{-}=\sum_{i=1}^{N}\mathbf{v}_{i}\cdot \mathbf{\nabla }_{\mathbf{q}_{i}}-%
\frac{1}{2}\sum_{i=1}^{N}\sum_{j\neq i}^{N}T_{-}(i,j)  \label{3.16}
\end{equation}%
with%
\begin{equation}
T_{-}(i,j)=\delta (q_{ij}-\sigma )\Theta (\mathbf{\hat{g}}_{ij}\cdot \mathbf{%
\hat{q}}_{ij})|\mathbf{g}_{ij}\cdot \mathbf{\hat{q}}_{ij}|\left(
b_{ij}^{-1}-1\right)  \label{3.18}
\end{equation}

In summary, the problems presented by the singular forces for a fluid of
hard spheres are resolved if Hamilton's equations for observables are
replaced by 
\begin{equation}
\left( \partial _{t}-L\right) A(\Gamma ,t)=0,\hspace{0.3in}\left( \partial
_{t}+L_{-}\right) A(\Gamma ,-t)=0  \label{3.19}
\end{equation}%
and the Liouville equation for probability densities is replaced by 
\begin{equation}
\left( \partial _{t}+\overline{L}\right) \rho (\Gamma ,t)=0,  \label{3.20}
\end{equation}%
for $t\geq 0$, with the respective generators given by (\ref{3.6}), (\ref%
{3.16}), and (\ref{3.11}). The three generators are all different, so some
care must be used to apply them under the correct conditions of their
definitions. Note that the forms of the generators $L$ and $L_{-}$, and
corresponding binary collision operators $T(i,j)$ and $T_{-}(i,j)$, do not
depend on the details of the collision rule defining the operator $b_{ij}$;
the result applies for both elastic and inelastic collisions \cite{lutsko1}.
In contrast, the generator for Liouville dynamics is obtained by a change of
variables that introduces the Jacobian of the transformation between the
variables $\mathbf{g}_{ij}$ and $b_{ij}\mathbf{g}_{ij}$. Hence it depends
explicitly on the collision rule and the restitution coefficient $\alpha $.

For normal fluids the probability of a configuration with any pair of
particles in contact and at rest has vanishing measure. Then the above
description of trajectories as sequences of pair collisions among the
particles is adequate. As noted above, in granular fluids the phenomenon of
inelastic collapse admits the possibility of evolution to a state where
clusters of particles are in contact and at rest. It would appear that the
collision of another particle with that cluster be described would require a
more complex generator for the dynamics than that considered here. Consider
the simplest case of a pair in contact and at rest, with a third particle
incident on one of the two. The generator here would transfer momentum to
one of the pair, as though it were isolated. Then, it would no longer be at
rest with respect to the other member of the pair and a second instantaneous
momentum transfer would occur to the second member. As a result, both
particles originally at contact and in relative rest would experience
relative motion and all three particles would separate. In fact, this is the
correct dynamics for real particles and it avoids the difficulty of
indeterminate dynamics for collisions with clusters. Mathematically,
therefore, there appears to be no difficulty for the dynamics generated here
due to inelastic collapses. In practice, however, this can be difficult to
simulate as the effective time between such collisions can be very small and
require following a large number of collisions.

\subsection{Homogeneous cooling state}

In the absence of external forces there is special solution to the Liouville
equation for normal fluids: the stationary, homogenous (translationally
invariant) equilibrium solution, $\rho _{e}$. For the isolated system
considered here this is a probability density with sharply defined total
energy, total momentum, and number of particles. It is \textquotedblleft
universal\textquotedblright\ in the sense that most other homogeneous
initial preparations rapidly approach this stationary equilibrium solution,
on a time scale of the order of several collisions per particle. This is the
collisional velocity relaxation to Maxwellian distributions for each
velocity degree of freedom. Granular fluids are different in the sense that
the Liouville equation for an isolated system has no stationary solution.
This is due to the loss of energy on each inelastic collision, such that the
total energy decreases monotonically $E(t)<E(0)$. Nevertheless, it appears
there is a universal homogeneous solution $\rho _{h}$ whose time dependence
occurs entirely through a scaling of the velocities for each particle \cite%
{HCSMD}%
\begin{equation}
\rho _{h}\left( \Gamma ;t\right) =(lv_{h}\left( t\right) )^{-Nd}\rho
_{h}^{\ast }\left( \left\{ \frac{\mathbf{q}_{rs}}{l},\frac{\mathbf{v}_{r}-%
\mathbf{U}_{h}}{v_{h}(t)}\right\} \right) .  \label{3.21}
\end{equation}%
Here $\mathbf{q}_{rs}=\mathbf{q}_{r}-\mathbf{q}_{s}$, and $\mathbf{U}_{h}$
is an overall constant velocity of the system. This velocity can be removed
by a Galilean transformation but it is useful to retain it for the
generalization to a corresponding local form defined below. Also, $l$ is an
arbitrary constant characteristic length and $v_{h}(t)$ is a "thermal"
velocity defined in terms of the energy per particle%
\begin{equation}
v_{h}^{2}(t)=\frac{2T_{h}(t)}{m},\hspace{0.3in}T_{h}(t)\equiv \frac{2E_{h}(t)%
}{3N}=\frac{2}{3N}\int d\Gamma \left( \sum_{r}\frac{1}{2}mv_{r}^{2}\right)
\rho _{h}\left( \Gamma ;t\right)  \label{3.22}
\end{equation}%
The second equality defines an associated "temperature" for the system, so
that $v_{h}(t)$ can be interpreted as the average thermal speed. This
temperature definition agrees with that introduced below for one of the
hydrodynamic fields. Then, $\rho _{h}\left( \Gamma ;t\right) =\rho
_{h}\left( \Gamma ;T(t\right) )$ is an example of a "normal" state, also
defined below, whose time dependence occurs entirely through the
hydrodynamic fields.

The specific form for $\rho _{h}^{\ast }$ is determined by the requirement
that $\rho _{h}$ should be a solution to the Liouville equation (\ref{3.20})%
\begin{equation}
\left( \partial _{t}T_{h}\right) \partial _{T_{h}}\rho _{h}+\overline{L}\rho
_{h}=0,  \label{3.23}
\end{equation}%
or%
\begin{equation}
\overline{\mathcal{L}}_{T_{h}}\rho _{h}=0,\hspace{0.3in}\overline{\mathcal{L}%
}_{T}=-\zeta _{h}\left( T\right) T\partial _{T}+\overline{L}  \label{3.23a}
\end{equation}%
The "cooling rate" $\zeta _{h}$ has been introduced by the definition%
\begin{equation}
\zeta _{h}\left( T_{h}(t)\right) \equiv -T_{h}^{-1}(t)\partial _{t}T_{h}(t).
\label{3.26}
\end{equation}%
The scaling property (\ref{3.21}) gives an equivalent alternative form%
\begin{equation}
\overline{\mathcal{L}}\rho _{h}=0,  \label{3.24}
\end{equation}%
where the operator $\overline{\mathcal{L}}$ is defined by%
\begin{equation}
\overline{\mathcal{L}}X=\frac{\zeta _{h}}{2}\sum_{r=1}^{N}\boldsymbol{\nabla 
}_{\mathbf{v}_{r}}\cdot \left[ \left( \mathbf{v}_{r}-\mathbf{U}_{h}\right) X%
\right] +\overline{L}X.  \label{3.25}
\end{equation}%
An explicit expression for the cooling rate in terms of $\rho _{h}$ follows
from differentiation of (\ref{3.22}) with respect to time and using the
Liouville equation to get%
\begin{eqnarray}
\zeta _{h} &=&\frac{1}{T_{h}N}\int d\Gamma \left( \sum_{r}\frac{1}{3}%
mv_{r}^{2}\right) \overline{L}\rho _{h}\left( \Gamma ;t\right)  \notag \\
&=&-\frac{1}{T_{h}N}\int d\Gamma \left( L\sum_{r}\frac{1}{3}%
mv_{r}^{2}\right) \rho _{h}\left( \Gamma ;t\right)  \notag \\
&=&-(N-1)\frac{m}{6T_{h}\left( t\right) }\int d\Gamma \rho _{h}\left( \Gamma
;t\right) T(i,j)\left( v_{i}^{2}+v_{i}^{2}\right) .  \label{3.27}
\end{eqnarray}%
Recalling the explicit form for $T(i,j)$ in (\ref{3.8}), the cooling rate is
seen to be proportional to the energy changes on binary collisions given by (%
\ref{3.3}). The cooling rate then simplifies to%
\begin{equation}
\zeta _{h}=(1-\alpha ^{2})\frac{Nm}{12T_{h}(t)}\int d\Gamma \rho _{h}(\Gamma
,t)(\mathbf{g}_{12}\mathbf{\cdot }\widehat{\mathbf{q}}_{12})^{3}\Theta (%
\mathbf{g}_{12}\cdot \widehat{\mathbf{q}}_{12})\delta (q_{12}-\sigma ).
\label{3.28}
\end{equation}%
Note that since $\rho _{h}\left( \Gamma ;t\right) $ depends on time only
through $T_{h}(t)$ the cooling rate also has this property, as the notation
in (\ref{3.26}) implies.

In summary, the special HCS solution to the Liouville equation $\rho
_{h}\left( \Gamma ;t\right) $ is defined by (\ref{3.24}) together with (\ref%
{3.28}) which is a linear functional of $\rho _{h}\left( \Gamma ;t\right) $.
In this representation, the linear Liouville equation in terms of $\Gamma ,t$
becomes a nonlinear equation parameterized by $T_{h}(t)$. In principle, the
solution is determined in two steps. First, (\ref{3.24}) and (\ref{3.28})
are solved as a function of $T_{h}(t)$. Second $T_{h}(t)$ is determined from
(\ref{3.26}) and substituted into the solution found in the first step. The
time dependence from this second step can be determined quite generally from
the scaling property of his dependence on $T_{h}(t)$ can be made explicit
through the scaling form of (\ref{3.21}) which leads to 
\begin{equation}
\zeta _{h}\left( T_{h}(t)\right) =\frac{v_{h}(t)}{\ell }\zeta _{h}^{\ast },
\label{3.29}
\end{equation}%
where $\zeta _{h}^{\ast }$ is a constant 
\begin{equation}
\zeta _{h}^{\ast }=(1-\alpha ^{2})\frac{N}{6}\int d\Gamma ^{\ast }\rho
_{h}^{\ast }(\Gamma ^{\ast })(\mathbf{g\ast }_{12}\mathbf{\cdot }\widehat{%
\mathbf{q}}_{12})^{3}\Theta (\mathbf{g}_{12}^{\ast }\cdot \widehat{\mathbf{q}%
}_{12})\delta (q_{12}^{\ast }-\frac{\sigma }{\ell }).  \label{3.30}
\end{equation}%
The dimensionless variables for the integration here are given by%
\begin{equation}
\Gamma ^{\ast }\equiv \left\{ \mathbf{q}_{1}^{\ast },\cdot \cdot ,\mathbf{q}%
_{N}^{\ast },\mathbf{v}_{1}^{\ast }\cdot \cdot ,\mathbf{v}_{N}^{\ast
}\right\} ,\hspace{0.3in}\mathbf{q}_{r}^{\ast }=\frac{\mathbf{q}_{r}}{l},%
\hspace{0.3in}\mathbf{v}_{r}^{\ast }=\frac{\mathbf{v}_{r}-\mathbf{U}_{h}}{%
v_{h}(t)}.  \label{3.31}
\end{equation}%
Thus (\ref{3.29}) exposes the explicit temperature dependence of $\zeta
_{h}\left( T_{h}(t)\right) $ and (\ref{3.26}) becomes 
\begin{equation}
\partial _{t}T_{h}^{-1/2}(t)=\frac{\zeta _{h}^{\ast }}{\ell \sqrt{2m}},
\label{3.32}
\end{equation}%
which can be integrated to give \cite{Haff}%
\begin{equation}
T_{h}(t)=T_{h}(0)\left( 1+\frac{v_{h}(0)\zeta _{h}^{\ast }}{2\ell }t\right)
^{-2}\rightarrow \ell ^{2}m\zeta _{h}^{\ast -2}t^{-2}.  \label{3.33}
\end{equation}%
The initial time has been chosen at $t=0$ without loss of generality. The
long time cooling is seen to be algebraic and universal (independent of
initial conditions). Thus, $\rho _{h}\left( \Gamma ;t\right) $ is a
universal function of $T_{h}(t)$ which itself becomes universal at long
times.

\subsection{Dimensionless representations}

These last considerations suggest that the mathematics may be simpler and
the physics better exposed by using a representation in terms of the
dimensionless variables (\ref{3.31}) \cite{DBL02,Kandrup,DBB07}. The
associated dimensionless time is defined through the differential form%
\begin{equation}
ds=\frac{v_{h}(t)}{\ell }dt.  \label{3.34}
\end{equation}%
This can be integrated using (\ref{3.33}) to give%
\begin{equation}
s(t,0)=\frac{2}{\zeta _{h}^{\ast }}\ln \left( \frac{v_{h}(0)\zeta _{h}^{\ast
}}{2\ell }t\right) .  \label{3.35}
\end{equation}%
The parameter $s(t,0)$ is a measure of the average number of collisions per
particle in the interval $\left( 0,t\right) $. In terms of this parameter
the cooling of (\ref{3.33}) becomes exponential%
\begin{equation}
\frac{T_{h}(t)}{T_{h}(0)}=e^{-\zeta _{h}^{\ast }s}.  \label{3.36}
\end{equation}%
The dimensionless form of (\ref{3.24}) is%
\begin{equation}
\overline{\mathcal{L}}^{\ast }\rho _{h}^{\ast }=0,\hspace{0.3in}
\label{3.37}
\end{equation}%
with%
\begin{equation}
\overline{\mathcal{L}}^{\ast }X=\frac{\zeta _{h}^{\ast }}{2}\sum_{r=1}^{N}%
\boldsymbol{\nabla }_{\mathbf{v}_{r}^{\ast }}\cdot \left[ \mathbf{v}%
_{r}^{\ast }X\right] +\overline{L}^{\ast }X.  \label{3.38}
\end{equation}%
This equation is supplemented by the definition of $\zeta _{h}^{\ast }$ in
terms of $\rho _{h}^{\ast }$ in (\ref{3.30}). There is no longer any time
dependence, no dependence on $T_{h}(t)$, only a parameterization of the
solution by the scalar $\zeta _{h}^{\ast }$. Further elaboration on this is
given below.

Now return to the more general solutions to the Liouville equation given by (%
\ref{3.20}). Each solution will be characterized by an initial total energy,
momentum, and particle number. For these global parameters there is an
associated $\rho _{h}$ and $T_{h}(t)$ defined. The equation for general
solutions can be expressed in terms of the same dimensionless variables (\ref%
{3.31}) and (\ref{3.34}) for the associated HCS. The result is \cite{DBL02}%
\begin{equation}
\left( \partial _{s}+\overline{\mathcal{L}}^{\ast }\right) \rho ^{\ast }=0,%
\hspace{0.3in}\rho \left( \Gamma ;t\right) =(lv_{h}\left( t\right)
)^{-Nd}\rho ^{\ast }\left( \Gamma ^{\ast };s\right) .  \label{3.39}
\end{equation}%
The dimensionless generator for the dynamics, $\overline{\mathcal{L}}^{\ast
} $, is the same as that given in (\ref{3.38}). This result is similar in
form to the original representation of the Liouville equation, except that
now the time is measured in terms of the average number of collisions and
the generator for the dynamics has an additional contribution compensating
for collisional cooling as it would occur in the corresponding HCS. Note
that in this representation $\rho _{h}^{\ast }$ is a stationary solution to
the Liouville equation (\ref{3.39}). The differences between normal and
granular fluids have been somewhat mitigated by this dimensionless
representation. For example, notions of "approach to equilibrium" can be
translated into "approach to the HCS", and the universal features of the
equilibrium state can be translated to those of the HCS. MD simulations
suggest that these comparisons are useful in the sense that very different
homogeneous initial conditions approach the HCS on a time scale of several
collisions per particle \cite{Lutsko02}. The corresponding dimensionless
representations for observables corresponding to (\ref{3.19}) are 
\begin{equation}
\left( \partial _{s}-\mathcal{L}^{\ast }\right) A^{\ast }\left( \Gamma
^{\ast };s\right) =0,\hspace{0.3in}\left( \partial _{s}+\mathcal{L}%
_{-}^{\ast }\right) A^{\ast }\left( \Gamma ^{\ast };-s\right) =0,
\label{3.40}
\end{equation}%
\begin{equation}
\mathcal{L}^{\ast }X=-\frac{\zeta _{h}^{\ast }}{2}\sum_{r=1}^{N}\mathbf{v}%
_{r}^{\ast }\cdot \boldsymbol{\nabla }_{\mathbf{v}_{r}^{\ast }}X+L^{\ast }X,%
\hspace{0.3in}\mathcal{L}_{-}^{\ast }X=\frac{\zeta _{h}^{\ast }}{2}%
\sum_{r=1}^{N}\mathbf{v}_{r}^{\ast }\cdot \boldsymbol{\nabla }_{\mathbf{v}%
_{r}^{\ast }}X+L_{-}^{\ast }X  \label{3.41}
\end{equation}

\section{Macroscopic Balance Equations}

\label{sec4}

For a simple one component fluid the relevant macroscopic variables are the
average number density $n$, energy density $e$, and momentum density $%
\mathbf{g}$ 
\begin{equation}
n(\mathbf{r},t)=\left\langle \widehat{n}\left( \mathbf{r}\right)
;t\right\rangle ,\hspace{0.3in}e(\mathbf{r},t)=\left\langle \widehat{e}%
\left( \mathbf{r}\right) ;t\right\rangle ,\hspace{0.3in}\text{\ }\mathbf{g}(%
\mathbf{r},t)=\left\langle \widehat{\mathbf{g}}\left( \mathbf{r}\right)
;t\right\rangle .  \label{4.1}
\end{equation}%
The brackets denotes an average over the ensemble $\rho (\Gamma ,t)$ as
indicated on the right side of (\ref{3.15}) 
\begin{equation}
\left\langle X;t\right\rangle \equiv \int d\Gamma \,\left( e^{-t\overline{L}%
}\rho (\Gamma )\right) X(\Gamma ).  \label{4.2}
\end{equation}%
This set will be referred to as the hydrodynamic fields, as it is their
macroscopic dynamics which is the candidate for a closed description on some
appropriate length and time scale. The specific forms of the phase functions
are given in Appendix \ref{appB} (a caret has been introduced to distinguish
the microscopic and macroscopic fields). Also in that Appendix the exact
microscopic balance equations for the phase functions are derived. Their
averages give the corresponding macroscopic balance equations%
\begin{equation}
\partial _{t}n(\mathbf{r},t)+m^{-1}\nabla _{\mathbf{r}}\cdot \mathbf{g}(%
\mathbf{r,}t)=0  \label{4.3}
\end{equation}%
\begin{equation}
\partial _{t}e(\mathbf{r},t)+\nabla _{\mathbf{r}}\cdot \left\langle \mathbf{s%
}(\mathbf{r});t\right\rangle =\left\langle w(\mathbf{r});t\right\rangle
\label{4.4}
\end{equation}%
\begin{equation}
\partial _{t}g_{\alpha }(\mathbf{r,}t)+\nabla _{\mathbf{r}_{\beta
}}\left\langle h_{\alpha \beta }(\mathbf{r});t\right\rangle =0.  \label{4.5}
\end{equation}%
Here $\left\langle \mathbf{s}(\mathbf{r,}t);0\right\rangle $ is the average
energy flux, $\left\langle w(\mathbf{r,}t);0\right\rangle $ is the average
energy loss function, and $\left\langle h_{\alpha \beta }(\mathbf{r,}%
t);0\right\rangle $ is the average momentum flux. Again, the phase functions 
$\mathbf{s}(\mathbf{r,}t),w(\mathbf{r,}t),$ and $h_{\alpha \beta }(\mathbf{r,%
}t)$ are given explicitly in Appendix \ref{appB} \cite{DBB07}. These exact
equations are the starting point for investigating the possibility of a
hydrodynamic description for a granular fluid.

To further simplify the balance equations it is useful to define the average
flow velocity according to%
\begin{equation}
\mathbf{g}(\mathbf{r,}t)\equiv n(\mathbf{r},t)m\mathbf{U}(\mathbf{r},t).
\label{4.6}
\end{equation}%
Then as shown in Appendix \ref{appC} the purely convective parts of the
energy and fluxes can be extracted by a local Galilean transformation \cite%
{McL} 
\begin{equation}
e(\mathbf{r},t)=e^{\prime }(\mathbf{r},t)\mathbf{+}\frac{1}{2}mn(\mathbf{r}%
,t)U^{2}(\mathbf{r},t)  \label{4.7}
\end{equation}%
\begin{equation}
\left\langle s_{\alpha }(\mathbf{r});t\right\rangle =\left\langle s_{\alpha
}^{\prime }(\mathbf{r});t\right\rangle +U_{\alpha }(\mathbf{r},t)\left(
e^{\prime }(\mathbf{r},t)\mathbf{+}\frac{1}{2}mn(\mathbf{r},t)U^{2}(\mathbf{r%
},t)\right) +\left\langle h_{\alpha \beta }^{\prime }(\mathbf{r}%
);t\right\rangle U_{\beta }(\mathbf{r},t)  \label{4.8}
\end{equation}%
\begin{equation}
\left\langle h_{\alpha \beta }(\mathbf{r});t\right\rangle =\left\langle
h_{\alpha \beta }^{\prime }(\mathbf{r});t\right\rangle +mn(\mathbf{r}%
,t)U_{\alpha }(\mathbf{r},t)U_{\beta }(\mathbf{r},t)  \label{4.9}
\end{equation}%
The phase functions with a prime denote the same function evaluated at $%
\mathbf{v}_{i}\rightarrow \mathbf{v}_{i}^{\prime }=\mathbf{v}_{i}-\mathbf{U}(%
\mathbf{r},t)$. Therefore, $e^{\prime }(\mathbf{r},t)$ is the rest frame
\textquotedblright thermal energy\textquotedblright , $\left\langle \mathbf{s%
}^{\prime }(\mathbf{r});t\right\rangle $\textbf{\ }is the rest frame
\textquotedblright heat flux\textquotedblright , and $\left\langle h_{\alpha
\beta }^{\prime }(\mathbf{r});t\right\rangle $ is the rest frame
\textquotedblright pressure tensor\textquotedblright . This terminology does
not imply any thermodynamic implications, however, and it is useful to
retain the names for comparison with normal fluid forms. In this spirit, the
following notation is introduced%
\begin{equation}
e^{\prime }(\mathbf{r},t)\equiv \frac{3}{2}n(\mathbf{r},t)T(\mathbf{r},t),%
\hspace{0.3in}\zeta (\mathbf{r},t)\equiv -\frac{2}{3n(\mathbf{r},t)T(\mathbf{%
r},t)}\left\langle w(\mathbf{r});t\right\rangle ,  \label{4.10}
\end{equation}%
\begin{equation}
\left\langle \mathbf{s}^{\prime }(\mathbf{r});t\right\rangle \equiv \mathbf{q%
}(\mathbf{r},t),\hspace{0.3in}\left\langle h_{\alpha \beta }^{\prime }(%
\mathbf{r});t\right\rangle \equiv P_{\alpha \beta }(\mathbf{r,}t)
\label{4.11}
\end{equation}%
Equation (\ref{4.10}) defines the \textquotedblright granular
temperature\textquotedblright . As a definition, $T(\mathbf{r},t)$ is
meaningful for any state of the system as a measure of the local kinetic
energy $e^{\prime }(\mathbf{r},t)$ - it simply constitutes a change of
variables from the pair $\left( n,e^{\prime },\mathbf{U}\right) $ to $\left(
n,T,\mathbf{U}\right) $ as the hydrodynamic fields of interest. However, its
relationship to any given measuring device (\textquotedblright
thermometer\textquotedblright ) must be considered with care. Similarly, (%
\ref{4.11}) gives a microscopic definition for the heat flux $\mathbf{q}(%
\mathbf{r},t)$ and pressure tensor $P_{\alpha \beta }(\mathbf{r,}t)$. The
interpretation of $\zeta (\mathbf{r},t)$ as a "cooling rate" appears in (\ref%
{4.13}) directly below.

In terms of these new variables the macroscopic balance equations become 
\begin{equation}
D_{t}n(\mathbf{r},t)+n(\mathbf{r,}t)\nabla _{\mathbf{r}}\cdot \mathbf{U}(%
\mathbf{r},t)=0  \label{4.12}
\end{equation}%
\begin{equation}
\left( D_{t}+\zeta (\mathbf{r},t)\right) T(\mathbf{r},t)+\frac{2}{3n(\mathbf{%
r},t)}\left( P_{\alpha \beta }(\mathbf{r},t)\partial _{\alpha }U_{\beta }(%
\mathbf{r},t)+\nabla \cdot \mathbf{q}(\mathbf{r},t)\right) =0,  \label{4.13}
\end{equation}%
\begin{equation}
D_{t}U_{\alpha }(\mathbf{r},t)+(mn(\mathbf{r},t))^{-1}\partial _{\beta
}P_{\alpha \beta }(\mathbf{r},t)=0,  \label{4.14}
\end{equation}%
where $D_{t}=\partial _{t}+\mathbf{U}\cdot \nabla $ is the material
derivative. These macroscopic balance equations are still exact. They have
the same form as for a normal fluid, with the only change being the presence
of the cooling rate. These are not a closed set of equations for $n,$ $T,$
and $\mathbf{U}$ since the heat flux, pressure tensor, and cooling rate have
not been explicitly represented. Clearly, however, at this point these
equations apply equally well to both granular and normal fluids under the
most general fluid state conditions.

\section{"Normal" States, Constitutive Relations, and Hydrodynamics}

\label{sec5}

The most general notion of a hydrodynamic description is a closed set of
equations for the hydrodynamic fields, $\left\{ y_{\alpha }\right\}
\Leftrightarrow \left\{ n,T,\mathbf{U}\right\} .$ This follows from the
exact macroscopic balance equations if the cooling rate, heat flux, and
pressure tensor can be represented as functionals of these fields%
\begin{equation}
\zeta (\mathbf{r},t)\rightarrow \zeta (\mathbf{r},t\mid \left\{ y_{\alpha
}\right\} ),\hspace{0.3in}\mathbf{q}(\mathbf{r},t)\rightarrow \mathbf{q}(%
\mathbf{r},t\mid \left\{ y_{\alpha }\right\} ),\hspace{0.3in}P_{\alpha \beta
}(\mathbf{r},t)\rightarrow P_{\alpha \beta }(\mathbf{r},t\mid \left\{
y_{\alpha }\right\} )  \label{5.1}
\end{equation}%
These are known as constitutive relations. A comment on notation is
appropriate at this point: $f\left( \mathbf{r},t,\left\{ y_{\alpha }\left( 
\mathbf{r},t\right) \right\} \right) $ denotes a \emph{function} of $\mathbf{%
r},t$ and the fields $\left\{ y_{\alpha }\left( \mathbf{r},t\right) \right\} 
$ at the point $\mathbf{r}$, while $f\left( \mathbf{r},t\mid \left\{
y_{\alpha }\right\} \right) $ denotes a function of $\mathbf{r},t$ and a 
\emph{functional} of $\left\{ y_{\alpha }\right\} $ at all space points.
With such relations the macroscopic balance equations (\ref{4.12}) - (\ref%
{4.14}) become hydrodynamic equations, i.e.%
\begin{equation}
\partial _{t}y_{\alpha }\left( \mathbf{r},t\right) =N_{\alpha }(\mathbf{r}%
,t\mid \left\{ y_{\alpha }\right\} ).  \label{5.2}
\end{equation}%
The conditions for the existence of constitutive equations then constitute
the context for a hydrodynamic description. Certainly, it cannot be expected
in general that the fluxes can be characterized entirely by the hydrodynamic
fields for all length and time scales. On the other hand important examples
exist, such as the pressure tensor for an equilibrium fluid in an arbitrary
external potential (density functional theory).

The connection of this question to the statistical mechanical basis for
hydrodynamics follows from the fact that the cooling rate and fluxes are
linear functionals of the solution to the Liouville equation, i.e. averages
of the form (\ref{4.2}). Therefore, a sufficient condition for constitutive
equations is for the distribution function to be characterized by the
hydrodynamic fields. A class of \textquotedblright normal\textquotedblright\
distributions is defined by the functional forms%
\begin{equation}
\rho _{n}\left( \Gamma ,t\right) =\rho _{n}\left( \Gamma \mid \left\{
y_{\alpha }\right\} \right)  \label{5.4}
\end{equation}%
This means that all time dependence and all the breaking of translational
invariance occurs only through the hydrodynamic fields. A familiar example
of a normal distribution for real fluids is the local canonical distribution%
\begin{equation}
\rho _{e\ell }\left( \Gamma \mid \left\{ y_{\alpha }\right\} \right) =\exp
\left\{ q-\int d\mathbf{r}T^{-1}\left( \mathbf{r},t\right) \left( \widehat{e}%
^{\prime }\left( \mathbf{r}\right) -\mu \left( \mathbf{r},t\right) \widehat{n%
}(\mathbf{r})\right) \right\}  \label{5.4a}
\end{equation}%
where $q$ is a normalization constant, and $\mu \left( \mathbf{r},t\right) $
is the chemical potential (as a specified function of the density and
temperature for the hydrodynamic fields chosen here). If a normal solution
to the Liouville can be found then the constitutive equations follow
immediately%
\begin{equation}
\zeta (\mathbf{r},t\mid \left\{ y_{\alpha }\right\} )=\frac{2}{3n(\mathbf{r}%
,t)T(\mathbf{r},t)}\int d\Gamma \rho _{n}\left( \Gamma \mid \left\{
y_{\alpha }\right\} \right) w(\mathbf{r})  \label{5.6}
\end{equation}%
\begin{equation}
\mathbf{q}(\mathbf{r},t\mid \left\{ y_{\alpha }\right\} )=\int d\Gamma \rho
_{n}\left( \Gamma \mid \left\{ y_{\alpha }\right\} \right) \mathbf{s}%
^{\prime }(\mathbf{r})  \label{5.7}
\end{equation}%
\begin{equation}
P_{\alpha \beta }(\mathbf{r},t\mid \left\{ y_{\alpha }\right\} )=\int
d\Gamma \rho _{n}\left( \Gamma \mid \left\{ y_{\alpha }\right\} \right)
h_{\alpha \beta }^{\prime }(\mathbf{r})  \label{5.8}
\end{equation}

The origin of hydrodynamics has now been \textquotedblright
reduced\textquotedblright\ to finding conditions for the existence of a
normal solution to the Liouville equation. Its time derivative in the
Liouville equation can be expressed in terms of the hydrodynamic equations (%
\ref{5.2})%
\begin{equation*}
\partial _{t}\rho _{n}=\int d\mathbf{r}\frac{\delta \rho _{n}}{\delta
y_{\alpha }\left( \mathbf{r},t\right) }\partial _{t}y_{\alpha }\left( 
\mathbf{r},t\right) =\int d\mathbf{r}\frac{\delta \rho _{n}}{\delta
y_{\alpha }\left( \mathbf{r},t\right) }N_{\alpha }(\mathbf{r},t\mid \left\{
y_{\alpha }\right\} ).
\end{equation*}%
Substitution of (\ref{5.4}) into the Liouville equation gives the form for
normal solutions%
\begin{equation}
\int d\mathbf{r}\frac{\delta \rho _{n}}{\delta y_{\alpha }\left( \mathbf{r}%
,t\right) }N_{\alpha }(\mathbf{r},t\mid \left\{ y_{\alpha }\right\} )+%
\overline{L}\rho _{n}=0.  \label{5.9}
\end{equation}%
This intimate connection of constructing a hydrodynamic description and
finding a normal solution to the Liouville equation is in fact a single
self-consistent problem. For specified fields, (\ref{5.9}) is an equation
for the $\Gamma $ dependence of the normal phase space density as a function
of the fields. This dependence then allows determination of the normal forms
in (\ref{5.6}) - (\ref{5.8}). Finally, solution of the hydrodynamic
equations (\ref{5.2}), with suitable initial and boundary conditions,
provides the explicit forms for the fields, and completes the normal
solution. The existence and determination of this solution is the central
problem for establishing a hydrodynamic description for both normal and
granular fluids.

The concept of a normal solution and its use in the macroscopic balance
equations makes no special reference to the possible inelasticity of
collisions. Nor does the concept refer to states near homogeneity or the
requirement of representation as local partial differential equations. As
will be seen below, the familiar Navier-Stokes equations represent a special
case of this more general idea. For normal fluids, the simple form of the
Navier-Stokes equations applies for a wide range of structurally simple
fluids, with rheology as a counter example for more complex fluids. As noted
in the Introduction, even structurally simple granular fluids can exhibit
behavior like complex real fluids for which the Navier-Stokes representation
fails \cite{Santos04}. However, the generality of the discussion here shows
that the failure of a Navier-Stokes approximation should not be taken as the
absence of a more complex hydrodynamic description.

To clarify the conditions under which a normal solution could be expected,
consider first a normal fluid with elastic collisions in an initial
non-equilibrium state with specified hydrodynamic fields $\left\{ y_{\alpha
}\left( \mathbf{r},t=0\right) \right\} $, whose values vary smoothly across
the system. In each small cell the phase space density $\rho (\Gamma ,t)$
approaches a local Gibbs distribution characterized by the hydrodynamic
fields at its central point $\mathbf{r}$, such as is given by (\ref{5.4a}).
However, this is not a solution to the Liouville equation due to the
differences in hydrodynamic fields in different cells. The solution has
additional fluxes driven by these gradients for subsequent exchange of mass,
energy, and momentum to equilibrate these fields to their uniform values (or
to steady values if the system is driven). The first stage, approach to a
universal form for the velocity distribution, occurs after a few collisions.
This establishes the normal form of the solution where the hydrodynamic
fields and their gradients characterize the state. Deviations from the Gibbs
density are due to fluxes of mass, momentum, and energy across the cells.
These fluxes are proportional to the differences in values of the fields
(i.e., to their spatial gradients). The second stage is the evolution of the
distribution through the changing values of the fields, according to the
hydrodynamic equations.

This two-stage evolution can be expected for granular fluids as well. The
initial velocity relaxation will not approach the local Gibbs density, but
some other corresponding local normal state determined from the inelastic
Liouville equation (see below). Subsequently, the deviations from this local
normal state characterizing spatial inhomogeneities will again be via the
macroscopic balance equations for the granular fluid. This is the space and
time scale for a hydrodynamic description.

\section{Navier-Stokes Approximation}

\label{sec6}The self-consistent solution to (\ref{5.9}) and determination of 
$N_{\alpha }(\mathbf{r},t\mid \left\{ y_{\alpha }\right\} )$ is a formidable
problem in general. Specific cases of interest may provide simplifications
that allow further progress. Consider the example of uniform shear flow,
where the system is driven by Lees - Edwards boundary conditions (simple
periodic boundary conditions in the local Lagrangian frame \cite{Lees}). At
the macroscopic level this state is characterized by a uniform density and
temperature, and a constant $y$ derivative of $U_{x}$ - the shear rate. This
is an example for which all spatial gradients vanish, except the first order
derivative of $U_{x}$. The latter can be small, so that all properties
depend nonlinearly on the shear rate, but clearly the problem is
considerably simplified.

The class of states to be considered here are those for which all spatial
gradients of first order can occur, but which are small and all higher order
derivatives are negligible. These are weakly inhomogeneous states. It is
expected under these conditions that the normal solution to the Liouville
equation can be represented by an expansion to first order in the gradients%
\begin{equation}
\rho _{n}\left( \Gamma \mid \left\{ y_{\alpha }\right\} \right) =\rho
_{h\ell }\left( \Gamma \mid \left\{ y_{\alpha }\right\} \right) +\mathbf{G}%
_{\alpha }\left( \Gamma ,\mathbf{r}\mid \left\{ y_{\alpha }\right\} \right)
\cdot \boldsymbol{\nabla }y_{\alpha }\left( \mathbf{r},t\right) +\cdot \cdot
\label{6.1}
\end{equation}%
The dots denote contributions from $\boldsymbol{\nabla }y_{\alpha }%
\boldsymbol{\nabla }y_{\alpha }$, $\boldsymbol{\nabla \nabla }y_{\alpha }$
and higher order gradients. The reference state $\rho _{h\ell }\left( \Gamma
\mid \left\{ y_{\alpha }\right\} \right) $ is the \textit{local homogeneous
cooling state}, analogous to the local equilibrium state for a molecular
fluid. As a normal state, it is a functional of the exact hydrodynamic
fields $\left\{ y_{\alpha }\right\} $ and must yield the exact averages for
the associated microscopic phase functions%
\begin{equation}
\int d\Gamma \left( \rho _{n}-\rho _{h\ell }\right) a_{\alpha }=0,\hspace{%
0.3in}a_{\alpha }=\left( \widehat{n}\left( \mathbf{r}\right) ,\widehat{e}%
\left( \mathbf{r}\right) ,\widehat{\mathbf{g}}\left( \mathbf{r}\right)
\right) .  \label{6.2}
\end{equation}%
For spatially constant fields it must reduce to the HCS of (\ref{3.21})%
\begin{equation}
\rho _{h}\left( \Gamma ;\left\{ y_{0\beta }\right\} \right) =\rho _{h\ell
}\left( \Gamma \mid \left\{ y_{0\beta }+\delta y_{\beta }\right\} \right)
\mid _{\delta y=0},\hspace{0.3in}\frac{\partial \rho _{h}}{\partial
y_{0\alpha }}=\int d\mathbf{r}\frac{\delta \rho _{h\ell }\left( \Gamma \mid
\left\{ y_{0\beta }+\delta y_{\beta }\right\} \right) }{\delta y_{\alpha
}\left( \mathbf{r}\right) }\mid _{\delta y=0},\cdot \cdot  \label{6.3}
\end{equation}%
where $y_{0\alpha }$ denotes arbitrary homogeneous values. Thus, the choice
for the local reference state is not arbitrary and other perturbations of
the HCS consistent with (\ref{6.2}) are not consistent with the second term
of (\ref{6.1}) being of first order in the gradients. Further
characterization of the local HCS distribution is given in Appendix \ref%
{appC}. The determination of $\mathbf{G}_{\alpha }\left( \Gamma ,\mathbf{r}%
\mid \left\{ y_{\alpha }\right\} \right) $ is fixed by the requirement that (%
\ref{6.1}) be a solution to the Liouville equation (\ref{5.9}) to the same
order in the gradients. This is discussed in the next section.

Small gradients means that the relative change in the hydrodynamic fields
over the largest microscopic length scale $\ell _{0}$ is small: $\ell
_{0}\partial _{r}\ln y_{\alpha }<<1.$ There are two characteristic length
scales, the mean free path and the grain diameter. For a dilute gas the mean
free path is largest, while for a dense fluid the grain size is largest. In
many cases, the condition for small gradients can be determined by control
over the initial or boundary conditions and it turns out to be the usual
experimental condition for simple normal fluids. As noted in the
introduction, some special states for granular fluids entail a balance of
the intrinsic internal cooling and the boundary or initial conditions such
that control over the gradients is lost (two examples are noted in the
discussion). Thus a careful analysis of each case is required before making
the assumption of small gradients. In the following it is assumed this has
been assured.

The constitutive equations follow from substitution of (\ref{6.1}) into (\ref%
{5.6})-(\ref{5.8}) 
\begin{equation}
\zeta (\mathbf{r},t\mid \left\{ y_{\alpha }\right\} )\rightarrow \zeta
_{\ell }(\mathbf{r},t\mid \left\{ y_{\alpha }\right\} )+\frac{2}{3n(\mathbf{r%
},t)T(\mathbf{r},t)}\int d\Gamma w(\mathbf{r})\mathbf{G}_{\alpha }\left(
\Gamma \mid \left\{ y_{\alpha }\right\} \right) \cdot \boldsymbol{\nabla }%
y_{\alpha }\left( \mathbf{r},t\right)  \label{6.4}
\end{equation}%
\begin{equation}
\mathbf{q}(\mathbf{r},t\mid \left\{ y_{\alpha }\right\} )\rightarrow \mathbf{%
q}_{\ell }(\mathbf{r},t\mid \left\{ y_{\alpha }\right\} )+\int d\Gamma 
\mathbf{s}^{\prime }(\mathbf{r})\mathbf{G}_{\alpha }\left( \Gamma \mid
\left\{ y_{\alpha }\right\} \right) \cdot \boldsymbol{\nabla }y_{\alpha
}\left( \mathbf{r},t\right)  \label{6.5}
\end{equation}%
\begin{equation}
P_{\alpha \beta }(\mathbf{r},t\mid \left\{ y_{\alpha }\right\} )\rightarrow
P_{\ell \alpha \beta }(\mathbf{r},t\mid \left\{ y_{\alpha }\right\} )+\int
d\Gamma h_{\alpha \beta }^{\prime }(\mathbf{r})\mathbf{G}_{\nu }\left(
\Gamma \mid \left\{ y_{\alpha }\right\} \right) \cdot \boldsymbol{\nabla }%
y_{\nu }\left( \mathbf{r},t\right)  \label{6.6}
\end{equation}%
The subscript $\ell $ on the first terms denote their local HCS averages,
i.e. (\ref{5.6})-(\ref{5.8}) evaluated with $\rho _{n}\left( \Gamma \mid
\left\{ y_{\alpha }\right\} \right) \rightarrow \rho _{h\ell }\left( \Gamma
\mid \left\{ y_{\alpha }\right\} \right) $. By construction, these
constitutive relations are correct to first order in the gradients. However,
the fields at different space points of the functional also differ by terms
of first order in the gradients. Therefore, when an average at some chosen
space point is calculated a further simplification is possible by retaining
only terms of linear order in gradients at that point. To illustrate,
consider some local property represented by $a(\Gamma ,\mathbf{r})$. Its
local HCS average can be evaluated to first order using%
\begin{eqnarray}
\rho _{h\ell } &=&\rho _{h}\left( \left\{ y_{\alpha }(\mathbf{r},t\mathbf{)}%
\right\} \right) +\int d\mathbf{r}^{\prime }\left( \frac{\delta \rho _{h\ell
}}{\delta y_{\beta }\left( \mathbf{r}^{\prime },t\right) }\right) _{\delta
y=0}\left( y_{\beta }\left( \mathbf{r}^{\prime },t\right) -y_{\beta }\left( 
\mathbf{r},t\right) \right) +\cdot \cdot  \notag \\
&=&\rho _{h}\left( \left\{ y_{\alpha }(\mathbf{r},t\mathbf{)}\right\}
\right) +\left( \mathbf{M}_{\nu }\left( \left\{ y_{\alpha }(\mathbf{r},t%
\mathbf{)}\right\} \right) -\mathbf{r}\frac{\partial \rho _{h}\left( \left\{
y_{\alpha }(\mathbf{r},t\mathbf{)}\right\} \right) }{\partial y_{\alpha
}\left( \mathbf{r},t\right) }\right) \cdot \nabla y_{\beta }\left( \mathbf{r}%
,t\right) +\cdot \cdot  \label{6.6c}
\end{eqnarray}%
where the dots denote terms of higher order in the gradients, and $\mathbf{M}%
_{\nu }$ is defined by%
\begin{equation}
\mathbf{M}_{\nu }=\int d\mathbf{r}^{\prime }\left( \frac{\delta \left( \rho
_{\ell n}\right) }{\delta y_{\nu }\left( \mathbf{r}^{\prime },t\right) }%
\right) _{\delta y=0}\mathbf{r}^{\prime }  \label{6.6b}
\end{equation}%
Then the average of $a(\Gamma ,\mathbf{r})$ becomes 
\begin{eqnarray}
\int d\Gamma a(\Gamma ,\mathbf{r})\rho _{h\ell }\left( \Gamma \mid \left\{
y_{\alpha }(\mathbf{r},t\mathbf{)+}\delta y_{\alpha }\right\} \right)
&\rightarrow &\frac{1}{V}\int d\Gamma A(\Gamma )\rho _{h}\left( \Gamma
,\left\{ y_{\alpha }(\mathbf{r},t\mathbf{)}\right\} \right)  \notag \\
&&+\int d\Gamma a(\Gamma ,\mathbf{0})\mathbf{M}_{\beta }\cdot \nabla
y_{\beta }\left( \mathbf{r},t\right)  \label{6.6d}
\end{eqnarray}%
The first term is the local HCS functional evaluated at the "constant" value
of $\left\{ y_{\alpha }(\mathbf{r},t\mathbf{)}\right\} $, in which case it
becomes the HCS cooling function of these values, according to (\ref{6.3}), $%
\rho _{h\ell }\left( \Gamma \mid \left\{ y_{\alpha }(\mathbf{r},t\mathbf{)}%
\right\} \right) =\rho _{h}\left( \Gamma ,\left\{ y_{\alpha }(\mathbf{r},t%
\mathbf{)}\right\} \right) .$ The second term is the contribution from $%
\delta y_{\alpha }\left( \mathbf{r}^{\prime },t\right) $ which is of first
order in the gradient $\boldsymbol{\nabla }y_{\nu }\left( \mathbf{r}%
,t\right) $. \ 

The first terms on the right sides of (\ref{6.4})-(\ref{6.6}) therefore have
two terms at this order, one evaluated at the HCS and the second of first
order in the gradient. These latter combine with the second terms of (\ref%
{6.4})-(\ref{6.6}). These expressions then have their final forms to first
order in the gradients 
\begin{equation}
\zeta (\mathbf{r},t\mid \left\{ y_{\alpha }\right\} )\rightarrow \zeta
_{h}(\left\{ y_{\alpha }(\mathbf{r},t\mathbf{)}\right\} )+\zeta _{U}(\left\{
y_{\alpha }(\mathbf{r},t\mathbf{)}\right\} )\boldsymbol{\nabla }\cdot 
\mathbf{U}(\mathbf{r},t\mathbf{)}  \label{6.8}
\end{equation}%
\begin{equation}
\mathbf{q}(\mathbf{r},t\mid \left\{ y_{\alpha }\right\} )\rightarrow \mathbf{%
-}\lambda \left( \left\{ y_{\alpha }(\mathbf{r},t\mathbf{)}\right\} \right) 
\boldsymbol{\nabla }T\left( \mathbf{r},t\right) \mathbf{-}\mu \left( \left\{
y_{\alpha }(\mathbf{r},t\mathbf{)}\right\} \right) \boldsymbol{\nabla }%
n\left( \mathbf{r},t\right)  \label{6.9}
\end{equation}%
\begin{eqnarray}
P_{\alpha \beta }(\mathbf{r},t &\mid &\left\{ y_{\alpha }\right\}
)\rightarrow p_{h}(\left\{ y_{\alpha }(\mathbf{r},t\mathbf{)}\right\}
)\delta _{\alpha \beta }-\kappa (\left\{ y_{\alpha }(\mathbf{r},t\mathbf{)}%
\right\} )\boldsymbol{\nabla }\cdot \mathbf{U}(\mathbf{r},t\mathbf{)}\delta
_{\alpha \beta }  \notag \\
&&-\eta (\left\{ y_{\alpha }(\mathbf{r},t\mathbf{)}\right\} )\left( \partial
_{\alpha }U_{\beta }(\mathbf{r},t\mathbf{)+}\partial _{\beta }U_{\alpha }(%
\mathbf{r},t\mathbf{)-}\frac{2}{3}\boldsymbol{\nabla }\cdot \mathbf{U}(%
\mathbf{r},t\mathbf{)}\delta _{\alpha \beta }\right)  \label{6.10}
\end{eqnarray}%
Fluid symmetry (rotational invariance of $\rho _{h}$) has been used to
recognize $\mathbf{q}_{h}(\mathbf{r},t\mid \left\{ y_{\alpha }\right\} )=0$,
and that contributions to the scalar $\zeta $ from gradients of the density
and temperature must vanish. Similarly, contributions to the vector $\mathbf{%
q}$ from velocity gradients must vanish, and no second order tensor
contributions to $P_{\alpha \beta }$ can be constructed from density and
temperature gradients. Equation (\ref{6.10}) has the familiar form of
Newton's viscosity law, while (\ref{6.9}) is Fourier's law with an
additional contribution from the density gradient.

These results, together with the macroscopic balance equations (\ref{4.12})-(%
\ref{4.14}) provide the closed set of hydrodynamic equations with
constitutive relations calculated determined to first order in the
gradients. "Determined" means that formally exact expressions are now
available for calculation all parameters in these equations. For example, $%
p_{h}$ and $\zeta _{h}$ are properties of the HCS%
\begin{equation}
p_{h}(\left\{ y_{\alpha }\right\} )=\frac{1}{3V}\int d\Gamma \rho _{h}\left(
\Gamma ;\left\{ y_{\alpha }\right\} \right) H_{\alpha \alpha }^{\prime },%
\hspace{0.3in}H_{\alpha \alpha }^{\prime }=\int d\mathbf{r}h_{\alpha \alpha
}^{\prime }(\mathbf{r}).  \label{6.11}
\end{equation}%
\begin{equation}
\zeta _{h}(\left\{ y_{\alpha }\right\} )=\frac{2}{3nTV}\int d\Gamma \rho
_{h}\left( \Gamma ;\left\{ y_{\alpha }\right\} \right) W_{\zeta },\hspace{%
0.3in}W_{\zeta }=\int d\mathbf{r}w^{\prime }(\mathbf{r})  \label{6.12}
\end{equation}%
Recall that a prime on the phase function denotes the local rest frame, $%
\mathbf{v}_{s}\rightarrow \mathbf{V}_{s}=\mathbf{v}_{s}-\mathbf{U}_{s}(%
\mathbf{r},t\mathbf{)}$. Equation (\ref{6.11}) now defines the hydrostatic
pressure for a granular fluid as a function of the local density and
temperature. The scaling property of $\rho _{h}$ displayed in (\ref{3.21})
confirm that $p_{h}(\left\{ y_{\alpha }\right\} )\propto T$ and $\zeta
_{h}\propto T$ $^{1/2}$. Similarly, expressions for the transport
coefficients $\zeta _{U}$, $\lambda $, $\mu $, $\kappa $, and $\eta $ as
phase space averages follow from this analysis. Their explicit expressions
are deferred to the next section.

It is appropriate to register at this point the explicit form for the full
nonlinear granular Navier - Stokes equations \cite{Navier-Stokes}%
\begin{equation}
D_{t}n+n\nabla _{\mathbf{r}}\cdot \mathbf{U}=0  \label{6.13}
\end{equation}%
\begin{equation*}
\left( D_{t}+\zeta _{h}\right) T+\frac{2}{3n}\left( p+\frac{3nT}{2}\zeta
_{U}+\left( \frac{2}{3}\eta -\kappa \right) \nabla \cdot \mathbf{U}\right)
\nabla \cdot \mathbf{U}
\end{equation*}%
\begin{equation}
-\frac{2}{3n}(\eta \left( \partial _{\alpha }U_{\beta }+\partial _{\beta
}U_{\alpha }\right) \partial _{\alpha }U_{\beta }+\boldsymbol{\nabla }\cdot
\left( \lambda \mathbf{\nabla }T+\mu \boldsymbol{\nabla }n\right) )=0,
\label{6.14}
\end{equation}%
\begin{equation}
D_{t}U_{\alpha }+(mn)^{-1}\partial _{\alpha }\left( p-\left( \frac{2}{3}\eta
+\kappa \right) \nabla \cdot \mathbf{U}\right) -(mn)^{-1}\partial _{\beta
}\eta \left( \partial _{\alpha }U_{\beta }+\partial _{\beta }U_{\alpha
}\right) =0,  \label{6.15}
\end{equation}%
These are almost the same as the Navier-Stokes equations for a molecular
fluid, except for the presence of the cooling rate $\zeta _{h}$ and two new
transport coefficients, $\zeta _{U}$ and $\mu $ in the temperature equation.
Perhaps the most significant of these is the cooling rate, which leads to
new instabilities and new stationary states. To illustrate, consider small
homogeneous perturbations $\left\{ \delta y_{\alpha }\right\} $ of the HCS
solution. The linear equations for the perturbations are%
\begin{equation}
\partial _{t}n=0=\partial _{t}\delta U_{\alpha },\hspace{0.3in}\left(
\partial _{t}+\zeta _{0h}+T_{h}\frac{\partial \zeta _{0h}}{\partial T_{h}}%
\right) \delta T+T_{h}\frac{\partial \zeta _{0h}}{\partial n_{h}}\delta n=0
\label{6.16}
\end{equation}%
Using the scaling $\zeta _{0h}\propto \sqrt{T_{h}}$, and introducing the
dimensionless variables $\delta T^{\ast }/\delta T/T_{h}$, $\delta U_{\alpha
}^{\ast }=\delta U_{\alpha }/v_{h}(t)$, these become linear equations with
time independent coefficients that are easily solved. It is found that there
is one decaying mode, one constant mode, and three growing modes. A similar
result is found for finite spatial gradients, where the same modes are
unstable at sufficiently long wavelengths. A second interesting effect of
the cooling rate is the existence of new steady states, that are possible
when external work done on the system or energy input is balanced by the
inherent cooling from collisions. For example, the temperature equation for
steady, simple shear flow becomes%
\begin{equation}
\zeta _{h}=\frac{2}{3nT}\eta \partial _{y}U_{x}.  \label{6.17}
\end{equation}%
It is noted in the discussion that for many of these steady states, it is
not possible to control the size of the spatial gradients and higher order
hydrodynamic effects beyond Navier-Stokes order are required.

To summarize, a hydrodynamic description for normal and granular fluids has
been given from the exact macroscopic balance equations and the assumption
of a normal solution to the Liouville equation. For the class of fluid
states with small spatial gradients in the hydrodynamic fields, this normal
solutions is constructed to leading order in the form (\ref{6.1}) which
provides constitutive relations in terms of the fields and their gradients.
In this way, the macroscopic balance equations become a closed set of
hydrodynamic equations. The parameters of this description (e.g., pressure,
transport coefficients) are given in terms of averages over the solution (%
\ref{6.1}). It remains to make that solution more explicit and this is the
topic of the next section.

\section{Construction of the Normal Solution and Transport Coefficients}

\label{sec7}

The objective now is to construct a solution of the form (\ref{6.1}) that is
exact up through first order in the gradients. This extends recent results
for a corresponding solution to the Liouville equation resulting from small
initial perturbations of the HCS \cite{DBB06,DBB07}. In that case the
linearized Navier-Stokes equations are obtained for states close to strictly
homogeneous systems. Here that assumption holds only locally, so the full
nonlinear Navier-Stokes equations result. The transport coefficients
obtained by both methods are the same, except for the values of the fields
on which they depend.

As described above, the ultimate use of this solution is to calculate local
properties of the form%
\begin{equation}
A(\mathbf{r}\mid \left\{ y_{\alpha }\left( t\right) \right\} )=\int d\Gamma
a(\Gamma ,\mathbf{r})\rho _{n}\left( \Gamma \mid \left\{ y_{\alpha }(\mathbf{%
r},t\mathbf{)+}\delta y_{\alpha }\left( t\right) \right\} \right) .
\label{7.0}
\end{equation}%
Therefore, in the following analysis the gradient expansion are referred to
the field point $\mathbf{r}$ of interest, $y_{\alpha }=y_{\alpha }(\mathbf{%
r)+}\delta y_{\alpha }$. Of course the results will be general and
applicable to any choice for $\mathbf{r}$. Consider first a general solution
of the form%
\begin{equation}
\rho \left( \Gamma ,t\mid \left\{ y_{\alpha }\left( t\right) \right\}
\right) =\rho _{h\ell }\left( \Gamma \mid \left\{ y_{\alpha }\left( t\right)
\right\} \right) +\Delta \left( \Gamma ,t\mid \left\{ y_{\alpha }\left(
t\right) \right\} \right) .  \label{7.1}
\end{equation}%
The notation makes explicit the fact that there is both explicit time
dependence and that which occurs through $\left\{ y_{\alpha }\left( \mathbf{r%
},t\right) \right\} $ in both $\rho $ and $\Delta $, while $\rho _{h\ell }$
depends on $t$ only through $\left\{ y_{\alpha }\left( \mathbf{r},t\right)
\right\} $ by construction. The Liouville equation gives%
\begin{equation}
\partial _{t}\Delta +\int d\mathbf{r}\frac{\delta \Delta }{\delta y_{\alpha
}\left( \mathbf{r},t\right) }N_{\alpha }(\mathbf{r},t\mid \left\{ y_{\alpha
}\left( t\right) \right\} )+\overline{L}\Delta =-\int d\mathbf{r}\frac{%
\delta \rho _{h\ell }}{\delta y_{\alpha }\left( \mathbf{r},t\right) }%
N_{\alpha }(\mathbf{r},t\mid \left\{ y_{\alpha }\left( t\right) \right\} )-%
\overline{L}\rho _{h\ell }.  \label{7.2}
\end{equation}%
It is understood here that the time derivative is taken at constant $\left\{
y_{\alpha }\left( \mathbf{r},t\right) \right\} $. It is expected that on
some time scale this time dependence goes to zero and (\ref{7.2}) becomes
the same as (\ref{5.9}) for a normal solution.

An approximate solution is sought by expanding in the gradients. The right
side of (\ref{7.2}) is evaluated in Appendix \ref{appE} with the result%
\begin{equation}
\partial _{t}\Delta +\int d\mathbf{r}\frac{\delta \Delta }{\delta y_{\alpha
}\left( \mathbf{r},t\right) }N_{\alpha }(\mathbf{r},t\mid \left\{ y_{\alpha
}\left( t\right) \right\} )+\left( 1-\mathcal{P}\right) \overline{L}\Delta
=-\left( 1-\mathcal{P}\right) \mathbf{\Upsilon }_{\nu }\left( \Gamma
,\left\{ y_{\alpha }\left( \mathbf{r},t\right) \right\} \right) \cdot 
\boldsymbol{\nabla }y_{\nu }\left( \mathbf{r},t\right)  \label{7.3}
\end{equation}%
with "fluxes" $\mathbf{\Upsilon }_{\nu }$ given by 
\begin{equation*}
\mathbf{\Upsilon }_{\nu }\left( \Gamma ,\left\{ y_{\alpha }\left( \mathbf{r}%
,t\right) \right\} \right) \mathbf{\equiv }\left( \overline{\mathcal{L}}%
_{T}\delta _{\nu \mu }-K_{\nu \mu }^{T}\left( \left\{ y_{\alpha }\left( 
\mathbf{r},t\right) \right\} \right) \mathbf{M}_{\mu }\left( \Gamma ,\left\{
y_{\alpha }\left( \mathbf{r},t\right) \right\} \right) \right)
\end{equation*}%
\begin{equation}
\mathbf{M}_{\nu }\left( \Gamma ,\left\{ y_{\alpha }\left( \mathbf{r}%
,t\right) \right\} \right) =\int d\mathbf{r}\left( \frac{\delta \rho _{h\ell
}\left( \Gamma \mid \left\{ y_{\alpha }\left( t\right) \right\} \right) }{%
\delta y_{\alpha }\left( \mathbf{r},t\right) }\right) _{\delta y=0}\mathbf{r}%
.  \label{7.4}
\end{equation}%
Also $\overline{\mathcal{L}}_{T}$ is the operator of (\ref{3.25}) with $%
T_{h} $ replaced by $T\left( \mathbf{r},t\right) $ and $K^{T}$ is the
transpose of the matrix%
\begin{equation}
K=\left( 
\begin{array}{ccc}
0 & 0 & 0 \\ 
\frac{\partial \left( \zeta _{h}T\right) }{\partial n} & \frac{\partial
\left( \zeta _{h}T\right) }{\partial T} & 0 \\ 
0 & 0 & 0%
\end{array}%
\right) .  \label{7.5}
\end{equation}%
The matrix $K$ generates the solution for homogeneous perturbations of the
hydrodynamic equations%
\begin{eqnarray}
\partial _{t}\delta y_{\nu }\left( t\right) &=&N_{\nu }(\mathbf{r},t\mid
\left\{ y_{\alpha }(\mathbf{r},t\mathbf{)+}\delta y_{\alpha }\left( t\right)
\right\} )-N_{\nu }(\mathbf{r},t\mid y_{\alpha }(\mathbf{r},t\mathbf{)}) 
\notag \\
&\rightarrow &-K_{\nu \mu }\left( \left\{ y_{\alpha }\left( \mathbf{r}%
,t\right) \right\} \right) \delta y_{\mu }\left( t\right) .  \label{7.6}
\end{eqnarray}%
Further interpretation of its occurrence here is given below. There are two
important observations that are a consequence of the special choice of $\rho
_{h\ell }$ as the reference state. First, the right side of (\ref{7.3}) is
proportional to the gradients. This admits solutions of the form (\ref{6.1}) 
\begin{equation}
\Delta \left( \Gamma ,t\mid \left\{ y_{\alpha }\left( t\right) \right\}
\right) \rightarrow \mathbf{G}_{\alpha }\left( \Gamma ,t,\left\{ y_{\alpha }(%
\mathbf{r},t\mathbf{)}\right\} \right) \cdot \boldsymbol{\nabla }y_{\alpha
}\left( \mathbf{r},t\right) .  \label{7.7}
\end{equation}%
The second observation is the occurrence of the orthogonal projection $%
\left( 1-\mathcal{P}\right) $, defined below, which excludes the invariants
of the generator for dynamics in (\ref{7.3}). This will be shown to be
essential for the existence of a normal solution.

Substitution of the form (\ref{7.7}) into (\ref{7.3}) leads to an equation
for $\mathbf{G}_{\alpha }$ (see Appendix \ref{appE} for details)%
\begin{equation}
\partial _{t}\mathbf{G}\left( t\right) +\left( 1-\mathcal{P}\right) \left( I%
\overline{\mathcal{L}}_{T}+K^{T}\right) \mathbf{G}\left( t\right) =-\left( 1-%
\mathcal{P}\right) \mathbf{\Upsilon },  \label{7.8}
\end{equation}%
where a compact matrix notation has been introduced. The time derivative is
again taken at constant $\left\{ y_{\alpha }\left( \mathbf{r},t\right)
\right\} $ and the dependence on these fields has been suppressed since they
are simply parameters of the equation. Integrating this equation with the
choice $\mathbf{G}\left( 0\right) =0$ (initial local HCS) gives the formal
solution%
\begin{equation}
\mathbf{G}\left( t\right) =-\left( 1-\mathcal{P}\right)
\int_{0}^{t}dt^{\prime }e^{-\left( I\overline{\mathcal{L}}_{T}+K^{T}\right)
t^{\prime }}\left( 1-\mathcal{P}\right) \mathbf{\Upsilon .}  \label{7.9}
\end{equation}%
Use has been made of the property%
\begin{equation}
\left( 1-\mathcal{P}\right) \left( I\overline{\mathcal{L}}_{T}+K^{T}\right) 
\mathcal{P}=0,  \label{7.9a}
\end{equation}%
which is also proved in the Appendix.

To interpret the role of $\left( 1-\mathcal{P}\right) $ in these
expressions, defined in (\ref{7.14}) below, consider the average of some
property represented by $X\left( \Gamma \right) $ 
\begin{equation}
\left\langle X;t\right\rangle =\left\langle X\right\rangle _{h\ell }+\left(
\int_{0}^{t}dt^{\prime }\mathbf{C}_{\nu }^{X}\left( t^{\prime }\right)
\right) \cdot \nabla y_{\nu }  \label{7.10}
\end{equation}%
where the first term is the local HCS average and the correlation function $%
\mathbf{C}_{\nu }\left( t\right) $ is 
\begin{equation}
\mathbf{C}_{\nu }^{X}\left( t\right) =-\int d\Gamma X\left( \Gamma \right)
\left( 1-\mathcal{P}\right) \left( e^{-\left( I\overline{\mathcal{L}}%
_{T}+K^{T}\right) t}\left( 1-\mathcal{P}\right) \mathbf{\Upsilon }\left(
\Gamma \right) \right) _{\nu }.  \label{7.11}
\end{equation}%
The only explicit time dependence on the right side of (\ref{7.10}) occurs
in the upper limit of the time integral. Suppose that $\mathbf{C}_{\nu
}^{X}\left( t\right) \rightarrow 0$ for $t>>t_{m}$. Then for $t>>t_{m}$ the
upper limit of the time integral can be extended to $\infty $ and the
average of $X$ becomes normal - all of its time dependence is through the
parameters $\left\{ y_{\alpha }\left( \mathbf{r},t\right) \right\} $%
\begin{equation}
\left\langle X;t\right\rangle \rightarrow \left\langle X\right\rangle
_{n}=\left\langle X\right\rangle _{h\ell }+\left( \int_{0}^{\infty
}dt^{\prime }\mathbf{C}_{\nu }^{X}\left( t^{\prime }\right) \right) \cdot
\nabla y_{\nu }  \label{7.12}
\end{equation}%
Thus $t_{m}$ sets the time scale for the normal solution, and hence for the
onset of hydrodynamics. A necessary condition for the existence of a $t_{m}$
is the convergence of the time integral in (\ref{7.12}). This means there
should be no invariants of the generator for time dependence, $I\overline{%
\mathcal{L}}_{T}+K^{T}$ in its domain of action. It is shown in Appendix \ref%
{appE} that such invariants exist 
\begin{equation}
\left( I\overline{\mathcal{L}}_{T}+K^{T}\right) \Psi =0,\hspace{0.3in}\Psi
_{\nu }=\frac{\partial \rho _{h}\left( \left\{ y_{\alpha }\right\} \right) }{%
\partial y_{\nu }}.  \label{7.13}
\end{equation}%
The existence of these invariants of (\ref{7.13}) provide a microscopic
representation of these long wavelength hydrodynamic excitations at long
wavelength in the spectrum of $\overline{\mathcal{L}}_{T}$ \cite{Kandrup}.
The effective generator for the dynamics, $\overline{\mathcal{L}}_{T}+K^{T},$
gives a dynamics with these homogeneous perturbations subtracted out.
However, the projection operator $\left( 1-\mathcal{P}\right) $ projects out
such contributions from the definition (\ref{e.24})%
\begin{equation}
\left( 1-\mathcal{P}\right) X=X-\Psi _{\alpha }\frac{1}{V}\int d\Gamma 
\widetilde{A}_{\alpha }X,\hspace{0.3in}\frac{1}{V}\int d\Gamma \widetilde{A}%
_{\alpha }\Psi _{\beta }=\delta _{\alpha \beta }.  \label{7.14}
\end{equation}%
The biorthogonal set $\widetilde{A}_{\alpha }$ are linear combinations of
the total particle number, energy, and momentum defined in (\ref{e.19}).
According to (\ref{7.6}) $K$ is the generator for homogeneous hydrodynamic
perturbations.

A technical complication is the occurrence of period time dependence, the
Poincare recurrence time. This can be removed by considering the
thermodynamic limit of $V\rightarrow \infty ,N\rightarrow \infty $ at
constant $N/V$. Therefore, the normal solution to the Liouville equation to
first order in the gradients is written%
\begin{equation}
\rho _{n}\left( \Gamma \mid \left\{ y_{\alpha }\left( t\right) \right\}
\right) =\rho _{h\ell }\left( \Gamma \mid \left\{ y_{\alpha }\left( t\right)
\right\} \right) -\left( \lim \int_{0}^{t}dt^{\prime }e^{-\left( I\overline{%
\mathcal{L}}_{T}+K^{T}\right) t^{\prime }}\left( 1-\mathcal{P}\right) 
\mathbf{\Upsilon }\right) _{\nu }\cdot \nabla y_{\nu }.  \label{7.15}
\end{equation}%
It is understood that the limit is taken in the context of an average like (%
\ref{7.12}) with the thermodynamic limit followed by the long time limit,
all at constant $\left\{ y_{\alpha }\right\} =\left\{ y_{\alpha }(\mathbf{r}%
,t\mathbf{)}\right\} $.

The constitutive equations now follow directly from (\ref{6.4}) - (\ref{6.6}%
) and fluid symmetry to get the final forms (\ref{6.8}) - (\ref{6.10}) with
the transport coefficients identified as%
\begin{equation}
\zeta _{U}(\left\{ y_{\alpha }\right\} )=\frac{2}{3nTV}\int d\Gamma W^{-}%
\mathbf{M}_{3xx}+\lim \int_{0}^{t}dt^{\prime }C_{\nu }^{\zeta }\left(
t^{\prime }\right)  \label{7.16}
\end{equation}%
\begin{equation}
\lambda (\left\{ y_{\alpha }\right\} )=\frac{1}{3V}\int d\Gamma \mathbf{S}%
^{-}\cdot \mathbf{M}_{2}+\lim \int_{0}^{t}dt^{\prime }C_{\nu }^{\lambda
}\left( t^{\prime }\right)  \label{7.17}
\end{equation}%
\begin{equation}
\mu (\left\{ y_{\alpha }\right\} )=\frac{1}{3V}\int d\Gamma \mathbf{S}^{-}%
\mathbf{\cdot M}_{1}+\lim \int_{0}^{t}dt^{\prime }C^{\mu }\left( t^{\prime
}\right)  \label{7.18}
\end{equation}%
\begin{equation}
\eta (\left\{ y_{\alpha }\right\} )=\frac{1}{V}\int d\Gamma
H_{xy}^{-}M_{3xy}+\lim \int_{0}^{t}dt^{\prime }C^{\eta }\left( t^{\prime
}\right)  \label{7.19}
\end{equation}%
\begin{equation}
\kappa (\left\{ y_{\alpha }\right\} )=\frac{1}{3V}\int d\Gamma
H_{xx}^{-}M_{3xx}+\lim \int_{0}^{t}dt^{\prime }C^{\kappa }\left( t^{\prime
}\right)  \label{7.20}
\end{equation}%
with the correlation functions%
\begin{equation}
C^{\zeta }\left( t^{\prime }\right) =-\frac{2}{3nTV}\int d\Gamma \widetilde{W%
}e^{-\left( \overline{\mathcal{L}}-\lambda _{3}\right) t^{\prime }}\left(
e^{-\left( I\overline{\mathcal{L}}_{T}+K^{T}\right) t^{\prime }}\left( 1-%
\mathcal{P}\right) \Upsilon \right) _{3xx}  \label{7.21}
\end{equation}%
\begin{equation}
C^{\lambda }\left( t^{\prime }\right) =-\frac{1}{3V}\int d\Gamma \widetilde{%
\mathbf{S}}\cdot \left( e^{-\left( I\overline{\mathcal{L}}_{T}+K^{T}\right)
t^{\prime }}\left( 1-\mathcal{P}\right) \mathbf{\Upsilon }\right) _{2}
\label{7.22}
\end{equation}%
\begin{equation}
C^{\mu }\left( t^{\prime }\right) =-\frac{1}{3V}\int d\Gamma \widetilde{%
\mathbf{S}}\cdot \left( e^{-\left( I\overline{\mathcal{L}}_{T}+K^{T}\right)
t^{\prime }}\left( 1-\mathcal{P}\right) \mathbf{\Upsilon }\right) _{1} 
\notag
\end{equation}%
\begin{equation}
C^{\eta }\left( t^{\prime }\right) =-\frac{1}{V}\int d\Gamma \widetilde{H}%
_{xy}\left( e^{-\left( I\overline{\mathcal{L}}_{T}+K^{T}\right) t^{\prime
}}\left( 1-\mathcal{P}\right) \Upsilon \right) _{3xy}  \label{7.24}
\end{equation}%
\begin{equation}
C^{\kappa }\left( t^{\prime }\right) =-\frac{1}{3V}\int d\Gamma
H_{xx}^{-}\left( e^{-\left( I\overline{\mathcal{L}}_{T}+K^{T}\right)
t^{\prime }}\left( 1-\mathcal{P}\right) \Upsilon \right) _{3xx}  \label{7.25}
\end{equation}%
and%
\begin{equation}
\widetilde{W}=\int d\mathbf{r}\left( 1-\mathcal{P}^{\dagger }\right) w(%
\mathbf{r}),\hspace{0.3in}\widetilde{\mathbf{S}}=\int d\mathbf{r}\left( 1-%
\mathcal{P}^{\dagger }\right) \mathbf{s}^{\prime }(\mathbf{r}),\hspace{0.3in}%
\widetilde{H}_{ij}=\int d\mathbf{r}\left( 1-\mathcal{P}^{\dagger }\right)
h_{ij}^{\prime }(\mathbf{r})  \label{7.26}
\end{equation}%
The adjoint projection operator $\mathcal{P}^{\dagger }$ is defined by $\int
d\Gamma X\mathcal{P}Y\equiv \int d\Gamma \left( \mathcal{P}^{\dagger
}X\right) Y$. The first terms of (\ref{7.16}) - (\ref{7.20}) come from the
expansion of $\rho _{h\ell }$ to first order in the gradients, (\ref{6.6c}).
They vanish for non-singular, conservative forces but are non-zero for
granular fluids and for the elastic hard sphere fluid. Also, they vanish in
the low density limit but can be dominant at high densities.

The appearance of $\overline{\mathcal{L}}_{T}$ in (\ref{7.8}), (\ref{7.15})
and the correlation functions for the transport coefficients is awkward as
it implies using the temperature as a dynamical variable in addition to the
phase space variables $\Gamma $. This difficulty can be avoided by
transforming to the dimensionless representation described at the end of
Section \ref{sec3}. The transformation is described in Appendix \ref{appE}
with the results%
\begin{equation}
\left( \partial _{s}+\left( 1-\mathcal{P}^{\ast }\right) \left( I\overline{%
\mathcal{L}}^{\ast }-\Lambda ^{\ast }\right) \right) \mathbf{G}^{\ast
}=-\left( 1-\mathcal{P}^{\ast }\right) \left( I\overline{\mathcal{L}}^{\ast
}-\Lambda ^{\ast }\right) \mathbf{M}^{\ast }\left( \left\{ y_{\alpha
}\right\} \right) ,  \label{7.15a}
\end{equation}%
where an asterisk denotes the function, operator, or matrix expresses in
terms of the dimensionless variables. The matrix $\Lambda ^{\ast }$ is 
\begin{equation}
\Lambda ^{\ast }=\left( 
\begin{array}{ccc}
0 & \frac{\partial \left( \zeta _{h}T\right) }{Tv_{h}\ell ^{2}\partial n} & 0
\\ 
0 & \frac{1}{2}\zeta _{h}^{\ast } & 0 \\ 
0 & 0 & -\frac{1}{2}\zeta _{h}^{\ast }%
\end{array}%
\right) .  \label{7.15b}
\end{equation}%
Note that now $\overline{\mathcal{L}}_{T}$ has been replaced by the phase
space operator $\overline{\mathcal{L}}$ of (\ref{3.25}) and the temperature
no longer appears in this dimensionless form. The variable $s$ is the same
as that of (\ref{3.34}), with $T_{h}\rightarrow T$, and represents the
average number of collisions per particle. The corresponding solution to the
dimensionless Liouville equation is 
\begin{equation}
\rho _{n}^{\ast }\left( s,\Gamma ^{\ast }\mid \left\{ y_{\alpha }^{\ast
}\left( t\right) \right\} \right) =\rho _{h\ell }^{\ast }\left( \Gamma
^{\ast }\mid \left\{ y_{\alpha }^{\ast }\left( s\right) \right\} \right)
-\left( \lim \int_{0}^{s}ds^{\prime }\left( e^{-\left( I\overline{\mathcal{L}%
}^{\ast }-\Lambda ^{\ast }\right) s^{\prime }}\left( 1-\mathcal{P}^{\ast
}\right) \mathbf{\Upsilon }^{\ast }\right) _{\nu }\right) \cdot \nabla
^{\ast }y_{\nu }^{\ast }.  \label{7.15c}
\end{equation}%
The eigenvalues of $\Lambda $ are $0,\frac{1}{2}\zeta _{h}^{\ast },-\frac{1}{%
2}\zeta _{h}^{\ast }.$ These are excitations for small homogeneous
perturbations of the hydrodynamic equations in this representation. The
invariants of (\ref{7.13}) become 
\begin{equation}
\left( I\overline{\mathcal{L}}^{\ast }-\Lambda ^{\ast }\right) \Psi ^{\ast
}=0.  \label{7.15d}
\end{equation}%
The interpretation of $\left( 1-\mathcal{P}^{\ast }\right) $ excluding these
invariants of the dynamics is the same as discussed above. For practical
purposes, both in theoretical and simulation applications, it is usually
most convenient to use the dimensionless forms.

This completes the formal derivation of the nonlinear Navier-Stokes
equations, including expressions for the pressure tensor, cooling rate, and
energy flux including contributions up through first order in the gradients
of the hydrodynamic fields. These expressions are functions of the
hydrodynamic fields to be determined by their detailed calculation. The
scaling property of hard spheres allows the temperature dependence to be
obtained directly from dimensional analysis, but the density dependence
requires confrontation of the full many-body problem for calculation.

\subsection{Example: Shear Viscosity}

To illustrate the results for the transport coefficients, the shear
viscosity is considered in more detail. The dimensionless form is used, but
to simplify the notation the asterisk is suppressed. The shear viscosity is
found to be%
\begin{equation}
\eta (\left\{ y_{\alpha }\right\} )=\frac{1}{V}\int d\Gamma H_{xy}M_{\eta
}-\lim \frac{1}{V}\int_{0}^{s}ds^{\prime }\int d\Gamma H_{xy}e^{-\left( 
\overline{\mathcal{L}}-\lambda _{3}\right) s^{\prime }}\Upsilon _{\eta }.
\label{7.27}
\end{equation}%
with $\lambda _{3}=-\frac{1}{2}\zeta _{h}$ and 
\begin{equation}
M_{\eta }=-\frac{1}{2}\sum_{s=1}^{N}q_{xs}\partial _{v_{ys}}\rho _{h},%
\hspace{0.3in}\Upsilon _{\eta }=\left( \overline{\mathcal{L}}-\lambda
_{3}\right) M_{\eta }.  \label{7.28}
\end{equation}%
The projection operators vanish in this case from fluid symmetry. The volume
integrated momentum flux $H_{xy}$ is given from (\ref{b.15}) as%
\begin{equation}
H_{xy}=\sum_{r=1}^{N}v_{rx}v_{ry}-\frac{1}{2}%
\sum_{r,s}^{N}q_{rsx}T(r,s)v_{sy}.  \label{7.29}
\end{equation}

In the elastic limit 
\begin{equation}
\left( \overline{\mathcal{L}}-\lambda _{3}\right) \rightarrow \overline{L},%
\hspace{0.3in}\rho _{h}\rightarrow \rho _{e},\hspace{0.3in}\Upsilon _{\eta
}\rightarrow -H_{xy}^{-}\rho _{e},\hspace{0.3in}M_{\eta }\rightarrow -\frac{1%
}{2}\sum_{s=1}^{N}q_{xs}v_{ys},  \label{7.30}
\end{equation}%
This result for the flux $\Upsilon _{\eta }$ follows from (\ref{a.21}) 
\begin{equation}
\Upsilon _{\eta }\rightarrow \overline{L}\left( M_{\eta }\rho _{e}\right)
=\left( M_{\eta }\overline{L}\rho _{e}\right) +\left( L_{-}M_{\eta }\right)
\rho _{e}=-H_{xy}^{-}\rho _{e}.  \label{7.31}
\end{equation}%
Interestingly, this introduces the generator for time reversed dynamics and
the associated momentum flux $H_{xy}^{-}$. The expression for the shear
viscosity becomes%
\begin{equation}
\eta (\left\{ y_{\alpha }\right\} )\rightarrow \frac{1}{V}\int d\Gamma
H_{xy}M_{3xy}+\lim \frac{1}{V}\int_{0}^{s}ds^{\prime }\int d\Gamma H_{xy}e^{-%
\overline{L}s^{\prime }}H_{xy}^{-}.  \label{7.32}
\end{equation}%
This is the Green-Kubo shear viscosity for a normal fluid. It differs from
that for a fluid whose particles interact via non-singular forces by the
presence of the first term on the right side, and by the difference between
the two fluxes (forward and reversed dynamics) in the flux-flux time
correlation function. As noted above, this is a peculiarity of
non-conservative and/or singular forces.

Comparison of (\ref{7.27}) and (\ref{7.32}) shows the most significant new
affects of dissipative dynamics: 1) one of the fluxes, $\mathbf{\Upsilon }%
_{\eta }$ is generated from the reference homogeneous state. This
representation holds as well for normal fluids, but for granular fluids the
homogeneous state is no longer given by an equilibrium distribution. Hence
the form of the flux is quite different from that for a normal fluid; 2) the
generator of the dynamics, $\left( \overline{\mathcal{L}}-\lambda
_{3}\right) $, is modified in two important ways. The operator $\overline{%
\mathcal{L}}$ generates the usual hard sphere trajectories, but modified by
an additional scaling operator to account for the dominant effects of
collisional cooling. In addition, the dynamics of homogeneous perturbations
of the reference state are compensated by the subtraction of an appropriate
eigenvalue for that dynamics, $\lambda _{3}$. This effectively shifts the
spectrum of $\left( \overline{\mathcal{L}}-\lambda _{3}\right) $ to exclude
the homogeneous dynamics of the reference state - cooling and its
homogeneous perturbations.

\subsection{Impurity Diffusion}

Perhaps the simplest example of hydrodynamics is the diffusion of an
impurity particle in a host fluid. To illustrate briefly, the host fluid is
taken to be in its HCS and unperturbed by a single impurity particle. The
impurity is taken to be a hard sphere, although its mass, size and
restitution coefficient for collisions with particles of the host fluid may
be different from those for fluid particle pairs. The macroscopic balance
equation is that for conservation of the probability density $P(\mathbf{r}%
,t) $ for the location of the impurity at time $t$ 
\begin{equation}
\partial _{t}P(\mathbf{r},t)+\nabla \cdot \mathbf{J}(\mathbf{r},t)=0.
\label{7.33}
\end{equation}%
This follows from averaging the corresponding microscopic conservation law,
where $P(\mathbf{r},t)$ and the flux $\mathbf{J}(\mathbf{r},t)$ are 
\begin{equation}
P(\mathbf{r},t)=\left\langle \delta \left( \mathbf{r}-\mathbf{q}_{0}\right)
;t\right\rangle ,\hspace{0.3in}\mathbf{J}(\mathbf{r},t)=\left\langle \mathbf{%
v}_{0}\delta \left( \mathbf{r}-\mathbf{q}_{0}\right) ;t\right\rangle .
\label{7.34}
\end{equation}%
The subscript zero distinguishes the position and velocity of the impurity
particle. The average flux $\mathbf{J}(\mathbf{r},t)$ is given by Fick's law
in the hydrodynamic limit and for small gradients%
\begin{equation}
\mathbf{J}(\mathbf{r},t)\rightarrow -D(\left\{ y_{\alpha }\right\} )\nabla P(%
\mathbf{r},t).  \label{7.35}
\end{equation}

The above analysis can be applied directly to this case as well with the
resulting Green-Kubo expression for the dimensionless diffusion coefficient 
\cite{Dufty02} (asterisks suppressed)%
\begin{equation}
D=\lim \frac{mT_{0}}{2m_{0}T}\int_{0}^{s}ds^{\prime }C_{vv}(s^{\prime }).
\label{7.36}
\end{equation}%
The normalized velocity autocorrelation function is 
\begin{equation}
C_{vv}(s)\equiv \frac{\langle \mathbf{v}_{0}\left( s\right) \cdot \mathbf{v}%
_{0}\rangle }{\langle v_{0}^{2}\rangle }=\frac{2m_{0}T}{3mT_{0}}\int d\Gamma
\left( e^{s\mathcal{L}}\mathbf{v}_{0}\right) \cdot \mathbf{v}_{0}\rho _{hcs}.
\label{7.37}
\end{equation}%
This expression for the diffusion coefficient is very similar to the
corresponding result for normal fluids except for the changes noted above
for the shear viscosity - a different generator for the dynamics and a
different reference fluid state. In addition, the appearance of two
different temperatures here is due to the different mechanical properties of
the host and impurity particles (e.g., mass, size, restitution coefficient).
This is a reflection of the failure of equipartition for granular mixtures,
as is expected for any non equilibrium state \cite{GD99}.

A simple estimate for the diffusion coefficient can be obtained from
truncation of the cumulant expansion for the correlation function \cite%
{Dufty02,Dufty01}

\begin{equation}
C_{vv}(s)=\exp \left[ \sum_{p=1}^{\infty }\frac{1}{p!}\omega _{p}\left(
-s\right) ^{p}\right] \rightarrow e^{-\omega _{1}s}  \label{7.38}
\end{equation}%
where the first cumulant is%
\begin{equation}
\omega _{1}=-\frac{<\left( \mathcal{L}\mathbf{v}_{0}\right) \cdot \mathbf{v}%
_{0}>}{\langle v_{0}^{2}\rangle }\rightarrow -\frac{1}{2}\zeta _{h}+\nu
\left( 1+\frac{mT_{0}}{m_{0}T}\right) ^{1/2}.  \label{7.39}
\end{equation}%
The arrow indicates an evaluation of the average using the first term in an
expansion of the reduced pair distribution function for the impurity and
host particle in Sonine polynomials, and a neglect of velocity correlations.
The dimensionless collision frequency $\nu $ and cooling rate $\zeta _{h}$
evaluated in the same approximation are 
\begin{equation}
\nu =\frac{2\pi \left( 1+\alpha _{0}\right) }{3\Gamma (3/2)}\frac{m}{m+m_{0}}%
\left( \frac{\overline{\sigma }}{\sigma }\right) ^{2}\,\chi _{0},\hspace{%
0.3in}\zeta _{h}=\frac{2^{1/2}\pi }{3\Gamma (3/2)}\,\chi (1-\alpha ^{2}),
\label{7.40}
\end{equation}%
and $\chi _{0}$ is the pair correlation function at contact. This gives the
diffusion coefficient as%
\begin{equation}
D\rightarrow \lim \frac{mT_{0}}{m_{0}T}\left( 2\left( 1+\frac{mT_{0}}{m_{0}T}%
\right) ^{1/2}\nu -\zeta _{h}\right) ^{-1}(1-e^{-\omega _{1}s}).
\label{7.41}
\end{equation}%
It is seen that $t_{m}=\omega _{1}^{-1}$ sets the time scale for cross over
to a normal solution and hydrodynamics.

This result has been compared with molecular dynamics simulations for the
special case of self-diffusion (mechanically equivalent particles) \cite%
{Dufty02}. It is found to give excellent agreement at low densities and weak
inelasticity, while differences grow at both higher densities and stronger
inelasticity. This result also agrees with the Enskog kinetic theory for
both normal and granular gases.

\section{Kinetic Theory and Boltzmann Limit}

\label{sec8}The above analysis has provided an exact formal derivation of
the hydrodynamic equations and associated expressions for the transport
coefficients in terms of HCS correlation functions. The difficult many-body
problem is postponed to the final stage of evaluating these expressions.
Approximations are introduced only at this final stage. An alternative
approach is that of kinetic theory where the many-body problem is confronted
at the outset, and approximations introduced at an early stage. \ Of course,
the two should yield equivalent results to the extent that consistent
approximations are used in each. In this section, the ideas behind a kinetic
theory derivation of the hydrodynamic equations are reviewed and an
approximation expected to be valid for low density gases is introduced.
Although attention is focused here on low density, kinetic theory methods
with different approximations can be applied to higher density fluids as
well; for references to granular gases see \cite{Poschel1}. The macroscopic
balance equations are obtained from the resulting kinetic equation, and the
associated constitutive equations are expressed in terms of the solution to
that equation. The construction of a normal solution is described in analogy
to that given above for the Liouville equation, and expressions for the
transport coefficients of a low density gas are obtained.

The motivation for a kinetic theory representation is based on the fact that
hydrodynamic fields are averages of sums of single particle functions of the
form%
\begin{equation}
A=\sum_{r=1}^{N}a(x_{r}),  \label{8.1}
\end{equation}%
where $x_{r}\longleftrightarrow (\mathbf{q}_{r},\mathbf{v}_{r})$ is used to
denote both the position and velocity of particle $r$. The $N$ particle
average of such functions can be reduced to an average over the single
particle phase space%
\begin{equation}
\left\langle A;t\right\rangle =\int dx_{1}a(x_{1})N\int dx_{2}..dx_{N}\rho
(\Gamma ,t)=n\int dx_{1}a(x_{1})f^{(1)}(x_{1},t),  \label{8.2}
\end{equation}%
where $f^{(1)}(x_{1},t)$ is the first member of a family of reduced
distribution functions obtained by integrating out some degrees of freedom 
\begin{equation}
n^{m}f^{(m)}(x_{1},..,x_{m},s)\equiv \frac{N!}{(N-m)!}\int
dx_{m+1}..dx_{N}\rho (\Gamma ,s).  \label{8.3}
\end{equation}%
Clearly, $f^{(1)}(x_{1},t)$ is proportional to the exact probability density
to find a particle with $(\mathbf{q}_{1},\mathbf{v}_{1})$ at time $t$,
regardless of the phase of all other particles. The representation (\ref{8.2}%
) expresses the fact the much less information is required for such averages
than is contained in the full $N$ particle state of the system $\rho (\Gamma
,t)$, and that it is sufficient to know the dynamics in the single particle
phase space.

Equations for the dynamics in the reduced distribution functions follow from
the Liouville equation by integrating it over some degrees of freedom%
\begin{eqnarray}
&&\left( \partial _{t}+\sum_{i=1}^{m}\mathbf{v}_{i}\cdot \nabla
_{i}-\sum_{i<j}^{m}\overline{T}(i,j)\right) f^{(m)}(x_{1},..,x_{m},t)  \notag
\\
&=&n\sum_{i=1}^{m}\int dx_{m+1}\,\overline{T}%
(i,m+1)f^{(m+1)}(x_{1},..,x_{m+1},t).  \label{8.4}
\end{eqnarray}%
The left side of this equation describes the dynamics of $m$ particles, just
as the Liouville equation for an isolated system of $m$ particles. The right
side represents the effects due to interactions with the remaining $N-m$
particles. It describes this interaction for each of the $m$ particles with
another with $x_{m+1}$ times the probability density that there is such a
particle with this phase. The latter is the joint probability for $m+1$
particles. It is seen that the equation for $f^{(m)}$ is coupled in this way
to that for $f^{(m+1)}$. This family of equations is known as the Born,
Bogoliubov, Green, Kirkwood, Yvon (BBGKY) hierarchy \cite{McL}. The
simplicity of the representation (\ref{8.2}) is somewhat misleading, since
the many-body problem is simply transferred to the problem of solving this
hierarchy.

There have been many attempts to obtain approximate solutions for $%
f^{(1)}(x_{1},t)$. Most of these entail a "closure" approximation so that
the first hierarchy equation becomes a closed equation for $f^{(1)}$ alone.
This is accomplished by some form of functional assumption in which it is
assumed that the two particle distribution function can be expressed as a
functional of the one particle distribution%
\begin{equation}
f^{(2)}(x_{1},x_{1},t)\rightarrow F(x_{1},x_{2},t\mid f^{(1)}(t))
\label{8.5}
\end{equation}%
If the functional form $F(x_{1},x_{2},t\mid \cdot )$ can be discovered then
the first BBGKY hierarchy becomes a \textit{kinetic equation}%
\begin{equation}
\left( \partial _{t}+\mathbf{v}_{1}\cdot \nabla _{1}\right)
f^{(1)}(x_{1},t)=J(x_{1},t\mid f^{(1)}(t)),  \label{8.6}
\end{equation}%
where the collision operator $J(x_{1},t\mid f^{(1)}(t))$ is identified as 
\begin{equation}
J(x_{1},t\mid f^{(1)}(t))\equiv n\int dx_{2}\,\overline{T}%
(1,2)F(x_{1},x_{2},t\mid f^{(1)}(t)).  \label{8.7}
\end{equation}

In fact, (\ref{8.5}) is so general that such a construction is always
possible in principle. A meaningful physical approximation requires further
guidance. Bogoliubov argued \cite{Bogoliubov} that in many cases there is an
initial "synchronization" time after which this functional exists as a time
independent functional%
\begin{equation}
f^{(2)}(x_{1},x_{1},t)\rightarrow F(x_{1},x_{2}\mid f^{(1)}(t)).  \label{8.8}
\end{equation}%
In this case the kinetic equation is Markovian. Bogoliubov then went on to
construct this functional formally as an expansion in the reduced density as
a small parameter. This procedure was formalized by others via cluster
expansions, which allowed a more penetrating analysis of terms in the
expansion. It was found that the expansion is not uniform in time, due to an
unexpected class of recollisions of particles ("rings") leading to secular
contributions. A similar detailed analysis for granular systems has not yet
been carried out.

Here, a formal solution to the hierarchy will be constructed for low density
gases using again the reduced density as a small parameter and for the
special case of inelastic hard spheres. This expansion is initiated by
returning to the dimensionless form of the Liouville equation (\ref{3.39}).
The corresponding dimensionless form of the hierarchy is found to be%
\begin{equation}
\left( \partial _{s}+\overline{\mathcal{L}}_{m}^{\ast }\left( \epsilon
\right) \right) f^{\ast (m)}(x_{1}^{\ast },..,x_{m}^{\ast
},s)=\sum_{i=1}^{m}\int dx_{m+1}\,\overline{T}^{\ast }(i,m+1)f^{\ast
(m+1)}(x_{1}^{\ast },..,x_{s+1}^{\ast },s).  \label{8.9}
\end{equation}%
with the definitions%
\begin{equation}
f^{\ast (m)}(x_{1}^{\ast },..,x_{m}^{\ast },s)\equiv \left( n\ell
^{3}v_{h}^{3}(t)\right) ^{m}f^{(m)}(x_{1},..,x_{m},t)  \label{8.10}
\end{equation}%
\begin{equation}
\overline{\mathcal{L}}_{m}^{\ast }\left( \epsilon \right) X\equiv
\sum_{i=1}^{m}\left( \mathbf{v}_{i}^{\ast }\cdot \boldsymbol{\nabla }_{%
\mathbf{q}_{i}^{\ast }}X+\frac{\zeta _{h}^{\ast }}{2}\boldsymbol{\nabla }_{%
\mathbf{v}_{i}^{\ast }}\cdot \left[ \mathbf{v}_{i}^{\ast }X\right] \right)
-\epsilon ^{2}\sum_{i<j}^{m}\overline{T}^{\ast }(i,j)X  \label{8.11}
\end{equation}%
\begin{eqnarray}
\overline{T}^{\ast }(i,j) &\equiv &\frac{\ell }{v_{h}(t)}\left( \frac{\ell }{%
\sigma }\right) ^{2}\overline{T}(i,j)  \notag \\
&=&\int \ d\hat{\mathbf{\sigma }}\ \Theta (\mathbf{g}_{ij}^{\ast }\cdot \hat{%
\mathbf{\sigma }})(\mathbf{g}_{ij}^{\ast }\cdot \hat{\mathbf{\sigma }})\left[
\alpha ^{-2}\delta (\mathbf{q}_{ij}^{\ast }-\epsilon \hat{\mathbf{\sigma }}%
)b_{ij}^{-1}-\delta (\mathbf{q}_{ij}^{\ast }+\epsilon \hat{\mathbf{\sigma }})%
\right]  \label{8.12}
\end{eqnarray}%
\begin{equation}
\epsilon \equiv \frac{\sigma }{\ell }=n\sigma ^{3}  \label{8.13}
\end{equation}%
The length scale has been chosen to be the mean free path $\ell \equiv
1/(n\sigma ^{2})$, where $n$ is the density. Low density is now defined by $%
\epsilon <<1$, i.e. the grain size is small compared to the mean free path.
This suggests looking for a solution to the hierarchy as an expansion in
this small parameter%
\begin{equation}
f^{\ast (m)}=f_{0}^{\ast (m)}+\epsilon f_{1}^{\ast (m)}+..  \label{8.14}
\end{equation}

It is readily shown \cite{Dufty(KT)01} that the hierarchy is solved exactly
to zeroth order in $\epsilon $ by 
\begin{equation}
f_{0}^{(m)}(x_{1},\cdots ,x_{m},s)=\prod_{i=1}^{s}f_{0}^{\ast (1)}(x_{i},s),
\label{8.15}
\end{equation}%
where $f_{0}^{\ast (1)}(x_{i},s)$ is the solution to%
\begin{equation}
\left( \partial _{s}+\mathbf{v}_{1}^{\ast }\cdot \nabla _{\mathbf{q}%
_{1}^{\ast }}\right) f_{0}^{\ast (1)}(x_{1}^{\ast },s)+\frac{\zeta
_{0h}^{\ast }}{2}\boldsymbol{\nabla }_{\mathbf{v}_{i}^{\ast }}\cdot \left[ 
\mathbf{v}_{i}^{\ast }f_{0}^{\ast (1)}(x_{1}^{\ast },s)\right] =J_{0}^{\ast
}(\mathbf{v}_{1}^{\ast }\mid f_{0}^{\ast (1)}\left( s\right) ).  \label{8.16}
\end{equation}%
The operator $J^{\ast }$ is the Boltzmann collision operator for inelastic
hard spheres \cite{McL}%
\begin{equation}
J^{\ast }(\mathbf{v}_{1}^{\ast }\mid f_{0}^{\ast (1)}\left( s\right) )=\int
dx_{2}^{\ast }\,\overline{T}_{0}^{\ast }(1,2)f_{0}^{\ast (1)}(x_{1}^{\ast
},s)f_{0}^{\ast (1)}(x_{2}^{\ast },s),  \label{8.17}
\end{equation}%
where $\overline{T}_{0}^{\ast }(1,2)$ is $\overline{T}^{\ast }(i,j)$ at $%
\epsilon =0$%
\begin{equation}
\overline{T}_{0}^{\ast }(i,j)=\int \ d\hat{\mathbf{\sigma }}\ \Theta (%
\mathbf{g}_{ij}\cdot \hat{\mathbf{\sigma }})(\mathbf{g}_{ij}\cdot \hat{%
\mathbf{\sigma }})\delta (\mathbf{q}_{ij}^{\ast })\left[ \alpha
^{-2}b_{ij}^{-1}-1\right] .  \label{8.18}
\end{equation}%
The effect of different locations of the colliding particles (contained in
the Boltzmann-Bogoliubov collision operator \cite{Bogoliubov}) does not
contribute at this order. Finally, $\zeta _{0h}^{\ast }$ is given by (\ref%
{3.30}) at $\epsilon =0$, which reduces to (using the symmetry of the HCS) 
\begin{equation}
\zeta _{h}^{\ast }=\left( 1-\alpha ^{2}\right) \frac{\pi }{12}\int
dv_{1}^{\ast }dv_{2}^{\ast }f_{0h}^{\ast (1)}(v_{1}^{\ast })f_{0h}^{\ast
(1)}(v_{2}^{\ast })g_{12}^{\ast 3},  \label{8.19}
\end{equation}%
where $f_{0h}^{\ast (1)}$ is the stationary HCS solution to the Boltzmann
equation (\ref{8.16})%
\begin{equation}
\frac{\zeta _{0h}^{\ast }}{2}\boldsymbol{\nabla }_{\mathbf{v}_{i}^{\ast
}}\cdot \left[ \mathbf{v}_{i}^{\ast }f_{0h}^{\ast (1)}(v_{1}^{\ast })\right]
=J^{\ast }(v_{1}^{\ast }\mid f_{0h}^{\ast (1)}\left( s\right) ).
\label{8.20}
\end{equation}%
This provides a complete description for the determination of $f_{0}^{\ast
(1)}(x_{1}^{\ast },s)$ and the calculation of averages such as (\ref{8.2})
at low density. It is an exceptional non-trivial result, and somewhat
special to the hard sphere interactions. A similar analysis for non-singular
forces leads to a collisionless mean field theory.

It is sufficient for the purposes here to stop at this point and inquire
about solutions to this kinetic equation leading to the hydrodynamics
appropriate for a low density gas. However, it is instructive to digress
briefly for a discussion of higher order terms. These consist of two types,
those that correct the single particle distribution in the product form (\ref%
{8.15}), and those that generate correlations. One class of terms of the
first type are those coming from the $\epsilon $ dependence of $\overline{T}%
^{\ast }(i,j)$ which give the "collision transfer" corrections to the
Boltzmann equation. This suggests a "renormalization" of the expansion to
one in powers of $\epsilon ^{2}$ at constant $\overline{T}^{\ast }(i,j)$.
Then, to lowest order the above results are obtained again but with the
replacement $\overline{T}_{0}^{\ast }(1,2)\rightarrow \overline{T}^{\ast
}(i,j)$. This is the Boltzmann-Bogoliubov low density kinetic theory that
retains the different locations of the colliding particles. The first
corrections to this theory have the form 
\begin{equation}
f_{1}^{\ast (m)}(x_{1}^{\ast },..,x_{m}^{\ast
},s)=\sum_{j=1}^{m}\prod_{i\neq j}^{m}f_{0}^{\ast (1)}(x_{i}^{\ast
},s)f_{1}^{\ast (1)}(x_{j}^{\ast },s)+\sum_{i<j}^{m}\prod_{k\neq
i,j}^{m}f_{0}^{\ast (1)}(x_{k}^{\ast },s)G^{\ast }(x_{i}^{\ast },x_{j}^{\ast
},s),  \label{8.21}
\end{equation}%
where the expression for $f_{1}^{\ast (m)}$ holds for $m\geq 2$. Thus, the
reduced distribution functions for any number of particles are determined as
a sum of products of the single particle functions $f_{0}^{\ast
(1)}(x_{1}^{\ast },s)$ and $f_{1}^{\ast (1)}(x_{1}^{\ast },s)$, and pair
function $G^{\ast }(x_{1}^{\ast },x_{2}^{\ast },s)$. These are determined
from the set of three fundamental kinetic equations \cite{Dufty(KT)01}%
\begin{eqnarray}
&&\left( \partial _{s}+\mathbf{v}_{1}^{\ast }\cdot \nabla _{\mathbf{q}%
_{1}^{\ast }}\right) f_{0}^{\ast (1)}(x_{1}^{\ast },s)+\frac{\zeta
_{0h}^{\ast }}{2}\boldsymbol{\nabla }_{\mathbf{v}_{i}^{\ast }}\cdot \left[ 
\mathbf{v}_{i}^{\ast }f_{0}^{\ast (1)}(x_{1}^{\ast },s)\right]  \notag \\
&=&\int dx_{2}^{\ast }\,\overline{T}^{\ast }(1,2)f_{0}^{\ast
(1)}(x_{1}^{\ast },s)f_{0}^{\ast (1)}(x_{2}^{\ast },s).  \label{8.22}
\end{eqnarray}%
\begin{equation}
\left( \partial _{s}+\mathcal{L}_{1}^{\ast }\right) f_{1}^{\ast
(1)}(x_{1}^{\ast },s)=\int dx_{2}^{\ast }\,\overline{T}^{\ast }(1,2)G^{\ast
}(x_{1}^{\ast },x_{2}^{\ast },s).  \label{8.23}
\end{equation}%
\begin{equation}
\left( \partial _{s}+\mathcal{L}_{1}^{\ast }+\mathcal{L}_{2}^{\ast }\right)
G^{\ast }(x_{1}^{\ast },x_{2}^{\ast },s)=\overline{T}^{\ast
}(1,2)f_{0}^{\ast (1)}(x_{1}^{\ast },s)f_{0}^{\ast (1)}(x_{2}^{\ast },s)
\label{8.24}
\end{equation}%
The operator $\mathcal{L}_{1}^{\ast }$ is defined over functions of $%
x_{1}^{\ast }$ by(\ref{8.22})%
\begin{eqnarray}
\mathcal{L}_{1}^{\ast }h^{\ast }(x_{1}^{\ast }) &\equiv &\mathbf{v}%
_{1}^{\ast }\cdot \nabla _{\mathbf{q}_{1}^{\ast }}+\frac{\zeta _{0h}^{\ast }%
}{2}\boldsymbol{\nabla }_{\mathbf{v}_{i}^{\ast }}\cdot \left[ \mathbf{v}%
_{i}^{\ast }h^{\ast }(x_{1}^{\ast })\right]  \notag \\
&&-\int dx_{2}^{\ast }\,\overline{T}^{\ast }(1,2)\left( f_{0}^{\ast
(1)}(x_{1}^{\ast },s)h^{\ast }(x_{2}^{\ast })+h^{\ast }(x_{1}^{\ast
})f_{0}^{\ast (1)}(x_{2}^{\ast },s)\right) .  \label{8.25}
\end{eqnarray}

These low density results (\ref{8.22})-(\ref{8.24}) are remarkably rich. The
first equation for the single particle distribution function is given by the
Boltzmann equation, as expected. The second equation provides corrections to
the Boltzmann results in terms of the two particle correlations. This
provides a means to understand the limitations of the Boltzmann equation.
The third equation gives the dynamics of these pair correlation. They are
driven by a source term which is the result of the binary collision operator
acting on uncorrelated single particle distributions that are solutions to
the Boltzmann equation. In more detail, it is possible to show that the
solution to this pair correlation equation, together with the equation for $%
f_{1}^{\ast (1)}$ give corrections to the Boltzmann equation due to
correlated recollisions ("ring" collisions) among many particles. These
effects dominate at long times and therefore show that the Boltzmann theory
is not accurate at asymptotically long times, no matter how low the density.
The third equation also gives the means to study the dynamics of pair
correlations, which also are characterized by correlation recollisions.
Finally, not shown is the equation for pair correlations at two different
times, which also follows from this analysis \cite{Dufty(KT)01}.

\subsection{Hydrodynamics from Kinetic Theory}

Consider now the Boltzmann kinetic equation for the one particle
distribution function (\ref{8.22}). The dimensionless hydrodynamic fields,
scaled relative to their values in the corresponding HCS, (\ref{8.20}), are%
\begin{equation}
\left( 
\begin{array}{c}
n^{\ast }(\mathbf{r}^{\ast },s) \\ 
n^{\ast }(\mathbf{r}^{\ast },s)T^{\ast }(\mathbf{r}^{\ast },s) \\ 
n^{\ast }(\mathbf{r}^{\ast },s)\mathbf{U}^{\ast }(\mathbf{r}^{\ast },s)%
\end{array}%
\right) =\int d\mathbf{v}^{\ast }\left( 
\begin{array}{c}
1 \\ 
\frac{1}{3}v^{\ast 2} \\ 
\mathbf{v}^{\ast }%
\end{array}%
\right) f^{\ast }(\mathbf{r}^{\ast },\mathbf{v}^{\ast },s)  \label{8.27}
\end{equation}%
Multiplying by (\ref{8.22}) by $1,v^{\ast 2},$and $\mathbf{v}^{\ast }$ and
integrating, successively, gives the macroscopic balance equations. They are
the same as those of (\ref{4.12})-(\ref{4.14}), in dimensionless form,
except for expressions defining the cooling rate, heat flux, and pressure
tensor. Here they are integrals over $f^{\ast }(x_{1}^{\ast },s)$ 
\begin{equation}
\zeta ^{\ast }=\left( 1-\alpha ^{2}\right) \frac{\pi }{12T^{\ast }}\int
dv_{1}^{\ast }dv_{2}^{\ast }f^{\ast }(\mathbf{r}^{\ast },\mathbf{v}^{\ast
},s)f^{\ast }(\mathbf{r}^{\ast },\mathbf{v}^{\ast },s)g_{12}^{\ast 3}
\label{8.28}
\end{equation}%
\begin{equation}
P_{ij}^{\ast }=2\int d\mathbf{v}\,^{\ast }V_{i}^{\ast }V_{j}^{\ast
}\,f^{\ast }(\mathbf{r}^{\ast },\mathbf{v}^{\ast },s),\hspace{0.5in}\mathbf{q%
}^{\ast }=\int d\mathbf{v}\,^{\ast }V^{\ast 2}\mathbf{V}^{\ast }f^{\ast }(%
\mathbf{r}^{\ast },\mathbf{v}^{\ast },s),  \label{8.29}
\end{equation}%
and $\mathbf{V}^{\ast }=\mathbf{v}^{\ast }-\mathbf{U}^{\ast }(\mathbf{r}%
^{\ast },s).$

The conceptual discussion in the sections above showing that hydrodynamics
follows from the existence of a normal state applies here as well \cite%
{Origins}. A normal solution to the Boltzmann equation for first order in
the gradients (Navier-Stokes hydrodynamics) is obtained in a way analogous
to that for the Liouville equation in Section \ref{sec7} \cite{DuftyGubbins}%
. The solution is written as%
\begin{equation}
f^{\ast }(\mathbf{r}^{\ast },\mathbf{v}^{\ast },s)=f_{\ell h}^{\ast
}(V^{\ast },\left\{ y_{\alpha }^{\ast }(\mathbf{r}^{\ast },s)\right\}
)+\Delta (\mathbf{r}^{\ast },\mathbf{v}^{\ast }\mid \left\{ y_{\alpha
}^{\ast }(s)\right\} ).  \label{8.30}
\end{equation}%
The reference state $f_{\ell h}^{\ast }$ is the local HCS distribution given
by%
\begin{equation}
f_{\ell h}^{\ast }(V^{\ast },\left\{ y_{\alpha }^{\ast }(\mathbf{r}^{\ast
},s)\right\} )=n^{\ast }(\mathbf{r}^{\ast },s)f_{h}^{\ast }(\frac{V^{\ast }}{%
\sqrt{T^{\ast }(\mathbf{r}^{\ast },s)}}),  \label{8.31}
\end{equation}%
where $f_{h}^{\ast }(v^{\ast })$ is the HCS solution given by (\ref{8.20}).
Because of this choice for the reference state it is found that $\Delta (%
\mathbf{r}^{\ast },\mathbf{v}^{\ast }\mid \left\{ y_{\alpha }^{\ast
}(s)\right\} )$ is of first order in the gradients%
\begin{eqnarray}
\left( \partial _{s}+\mathcal{L}_{01}^{\ast }\right) \Delta &=&-\left(
\partial _{s}+\mathbf{v}^{\ast }\cdot \nabla _{\mathbf{r}^{\ast }}\right)
f_{\ell h}^{\ast }  \notag \\
&=&-\frac{\partial f_{\ell h}^{\ast }}{\partial y_{\nu }^{\ast }}\left(
N_{\nu }^{\ast }+\delta _{2\nu }\zeta ^{\ast }T^{\ast }+\mathbf{v}^{\ast
}\cdot \nabla _{\mathbf{r}^{\ast }}y_{\nu }^{\ast }\right)  \label{8.32}
\end{eqnarray}%
The subscript $0$ on $\mathcal{L}_{01}^{\ast }$ means that the spatial
gradient contribution to $\mathcal{L}_{1}^{\ast }$ in (\ref{8.25}) is
excluded ($\mathcal{L}_{01}^{\ast }=\mathcal{L}_{1}^{\ast }-\mathbf{v}^{\ast
}\cdot \nabla _{\mathbf{r}^{\ast }}$). In the first equality, use has been
made of the fact $f_{\ell h}^{\ast }$ is related to the HCS solution by (\ref%
{8.31}). The second term contains the hydrodynamic fluxes $N_{\nu }^{\ast
}+\delta _{2\nu }\zeta ^{\ast }T^{\ast }$ (recall equation (\ref{5.2}))
which are first order in the gradients. As in Section \ref{sec7}, it is
essential that the reference state is an exact solution to zeroth order in
the gradients to assure that $\Delta $ is of first order%
\begin{equation}
\Delta (\mathbf{r}^{\ast },\mathbf{v}^{\ast }\mid \left\{ y_{\alpha }^{\ast
}(s)\right\} )\rightarrow \mathbf{G}_{\alpha }^{\ast }(s,\mathbf{V}^{\ast
},\left\{ y_{\alpha }^{\ast }(\mathbf{r}^{\ast },s)\right\} )\cdot 
\boldsymbol{\nabla }^{\ast }y_{\alpha }^{\ast }\left( \mathbf{r}^{\ast
},s\right) .  \label{8.33}
\end{equation}%
Substituting this form into (\ref{8.31}), the analysis proceeds in a similar
manner to that described for the Liouville equation in Appendix \ref{appE}.
The details will not be given and only the result is quoted for the normal
solution%
\begin{equation}
f^{\ast }(\mathbf{V}^{\ast },\left\{ y_{\alpha }^{\ast }\right\} )=f_{\ell
h}^{\ast }(V^{\ast },\left\{ y_{\alpha }^{\ast }\right\} )-\left( \lim
\int_{0}^{s}ds^{\prime }\left( e^{-\left( I\overline{\mathcal{L}}_{01}^{\ast
}-\Lambda ^{\ast }\right) s^{\prime }}\left( 1-\mathcal{P}_{1}^{\ast
}\right) \mathbf{\Upsilon }^{\ast }\right) _{\nu }\right) \cdot \nabla
^{\ast }y_{\nu }^{\ast }.  \label{8.34}
\end{equation}

The corresponding normal average for some property $X\left( \mathbf{v}%
\right) $ 
\begin{equation}
\left\langle X;\left\{ y_{\alpha }^{\ast }\right\} \right\rangle _{n}^{\ast
}=\left\langle X;\left\{ y_{\alpha }^{\ast }\right\} \right\rangle _{h\ell
}^{\ast }+\left( \int_{0}^{\infty }ds\mathbf{C}_{\nu }^{\ast X}\left(
s;\left\{ y_{\alpha }^{\ast }\right\} \right) \right) \cdot \nabla y_{\nu
}^{\ast },  \label{8.35}
\end{equation}%
\begin{equation}
\mathbf{C}_{\nu }^{\ast X}\left( s;\left\{ y_{\alpha }^{\ast }\right\}
\right) =-\int d\mathbf{v}X\left( \mathbf{v}\right) \left( 1-\mathcal{P}%
_{1}^{\ast }\right) \left( e^{-\left( I\overline{\mathcal{L}}_{01}^{\ast
}-\Lambda ^{\ast }\right) s}\left( 1-\mathcal{P}_{1}^{\ast }\right) \mathbf{%
\Upsilon }^{\ast }\left( \mathbf{v}\right) \right) _{\nu }.  \label{8.36}
\end{equation}%
These results should be compared with those of Section \ref{sec7} to note
that they have the same form as those for the Liouville equation. The single
particle projection operator now projects orthogonal to the invariants of $%
\left( I\overline{\mathcal{L}}_{01}^{\ast }-\Lambda ^{\ast }\right) $, where 
$\Lambda ^{\ast }$ is the same matrix as in (\ref{7.15b}), 
\begin{equation}
\left( I\overline{\mathcal{L}}_{01}^{\ast }-\Lambda ^{\ast }\right) \Psi
_{\alpha }=0,  \label{8.39}
\end{equation}%
\begin{equation}
\mathcal{P}^{\left( 1\right) }X\left( \mathbf{v}\right) =\Psi _{\alpha
}\left( \mathbf{V,}\left\{ y_{\alpha }\right\} \right) \int d\mathbf{v}_{1}%
\widetilde{A}_{\beta }\left( \mathbf{V}_{1}\mathbf{,}\left\{ y_{\alpha
}\right\} \right) X\left( \mathbf{v}_{1}\right) ,  \label{8.37}
\end{equation}%
\begin{equation}
\Psi _{\alpha }\left( \mathbf{V,}\left\{ y_{\alpha }\right\} \right) =\frac{%
\partial f_{\ell h}\left( V\mathbf{,}\left\{ y_{\alpha }\right\} \right) }{%
\partial y_{\alpha }},\hspace{0.3in}\widetilde{A}_{\beta }\left( \mathbf{V,}%
\left\{ y_{\alpha }\right\} \right) =\left( 
\begin{array}{c}
0 \\ 
\frac{2}{3n^{\ast }}\left( V^{\ast 2}-\frac{3}{2}T^{\ast }\right) \\ 
\frac{1}{n^{\ast }}\mathbf{V}^{\ast }%
\end{array}%
\right) .  \label{8.38}
\end{equation}%
Also, the flux $\mathbf{\Upsilon }^{\ast }\left( \mathbf{v}\right) $ is
generated from the HCS in much the same way%
\begin{eqnarray}
\left( 1-\mathcal{P}_{1}^{\ast }\right) \mathbf{\Upsilon }^{\ast }\left( 
\mathbf{v}\right) &=&-\left( 1-\mathcal{P}_{1}^{\ast }\right) \left( I%
\overline{\mathcal{L}}_{1}^{\ast }-\Lambda ^{\ast }\right) \mathbf{q}\Psi
_{\alpha }  \notag \\
&=&-\left( 1-\mathcal{P}_{1}^{\ast }\right) \mathbf{V}\Psi _{\alpha }
\label{8.40}
\end{eqnarray}

The constitutive equations for Navier-Stokes hydrodynamics also are the same
as those of Section \ref{sec7}, and follow from insertion of (\ref{8.34})
into (\ref{8.28}) and (\ref{8.29}). The transport coefficients are of course
different, due to the limitations imposed by the low density approximations.
For example, the shear viscosity is now%
\begin{equation}
\eta ^{\ast }(\left\{ y_{\alpha }\right\} )=-\lim \int_{0}^{s}ds^{\prime
}\int d\mathbf{v}H_{xy}e^{-\left( \overline{\mathcal{L}}_{01}^{\ast
}-\lambda _{3}\right) s^{\prime }}\Upsilon _{\eta },  \label{8.41}
\end{equation}%
\begin{equation}
H_{xy}=v_{x}v_{y},\hspace{0.3in}\Upsilon =v_{x}\partial _{v_{y}}f_{h}^{\ast
}\left( V^{\ast }\right)  \label{8.42}
\end{equation}%
This is quite similar to the general result (\ref{7.27}). The first term on
the right side of that result vanishes in the low density limit and does not
appear here. It has been shown that the general result reduces to this
obtained from kinetic theory when the correlation function there is
evaluated with the same low density assumptions as applied here \cite%
{BDB(KT)07}. The details for other transport coefficients are described in 
\cite{DB(GK)01,Bari} and will not be repeated here. It is found that the
results agree in detail with those obtained from the alternative
Chapman-Enskog method to construct a normal solution (properly adapted to
the granular Boltzmann equation \cite{BDKyS98}). The Chapman-Enskog
representation requires solution to an integral equation. This has been done
approximately, using a truncated polynomial expansion for the solution (or
equivalently, using the cumulant expansion approximation described for the
diffusion coefficient above). The Green-Kubo representation given here
provides another approach to evaluation using direct Monte Carlo simulation
of the correlation function \cite{Brey(GK)}.

\section{Discussion}

\label{sec9}

The presentation given here has addressed the origins of hydrodynamics for a
granular fluid. There have been three components to the discussion: 1) an
exact set of balance equations, based on the starting microscopic
description - either Liouville dynamics or kinetic theory; 2) the concept of
a normal state which implies constitutive equations in terms of the
hydrodynamic fields, and hence a hydrodynamic description; and 3) an
approximate construction of the normal solution for small local gradients,
leading to the Navier-Stokes hydrodynamic equations. Only the last component
entails specific limitations on the state conditions considered, with
explicit neglect of higher order gradients in the fields. It is presumed
that these state conditions can be controlled to assure small gradients, and
then Navier-Stokes hydrodynamics would apply. Below, it will be noted that
some new kinds of steady states for granular fluids preclude the control
over gradients and that conditions for Navier-Stokes hydrodynamics cannot be
attained in these cases. But there are many other cases, verified in both
simulations and experiments, for which a Navier-Stokes order hydrodynamics
is found to give an accurate description \cite{BRMC99,Brey,Caldera,Swinney}.
It is worth emphasizing again that the failure of a Navier-Stokes
description does not mean the absence of a hydrodynamic description, only
that the constitutive equations are more complex.

Even when the gradients are small, it is expected that the hydrodynamic
description will dominate all other "microscopic" excitations only on some
sufficiently long time scale. Thus the characteristic hydrodynamic times
should be long compared to the decay times for all other excitations. For
normal fluids this is assured by the fact that the hydrodynamic fields are
associated with global conserved quantities (number, energy, and momentum),
whose lifetimes become infinite in the long wavelength limit, while all
other characteristic times remain finite. This is not the case for granular
fluids, since the energy is no longer conserved. Instead, in the long
wavelength limit there is a new characteristic time determined by
homogeneous cooling, $\zeta _{h}^{-1}$. Some have argued that for this
reason the energy or temperature should not be considered among the
hydrodynamic fields for a granular gas. However, there are many good reasons
why the temperature should remain one of the fields in the long time, long
wavelength description of granular fluids. For example, consider the
simplest case of a general homogeneous state. On the time scale of a few
collisions it approaches the HCS for which the temperature obeys Haff's law (%
\ref{3.33}). The latter is the exact hydrodynamic description in this case
which is seen to dominate even though it has a time scale $\zeta _{h}^{-1}$.
For spatially inhomogeneous states, it has been shown here that the dominant
background cooling can be suppressed by the dimensionless representation.
This means that the relevant approach to hydrodynamics is relative to this
background state. The picture described above of each cell in the fluid
rapidly approaching a local HCS due to velocity relaxation, followed by
approach to spatial uniformity by fluxes between the homogeneously cooling
cells is consistent with this.

A more precise study of this problem is possible within the simplifying
features of the Boltzmann kinetic theory. There the response to an initial
small spatial perturbation of the HCS can be studied to see if there is a
dominant set of collective modes at long times and long wavelengths. For a
normal fluid this is established by showing that the spectrum of the
generator for dynamics in the linearized Boltzmann equation has five
isolated points that are smaller in magnitude than all other spectral
points, and therefore dominate at long times. These are shown to exist and
to be the same as the eigenvalues of the linearized hydrodynamic equations.
A similar analysis of the linearized granular Boltzmann operator has shown
the existence of hydrodynamic eigenvalues in its spectrum in the long
wavelength limit \cite{Modes,Bari}. Further progress can be made using
kinetic models for this Boltzmann operator, showing that the hydrodynamic
modes of the spectrum remain isolated and smallest, even for strong
dissipation \cite{Modes,Lipari}. In summary, there is very good evidence for
the dominance of a hydrodynamic description for granular gases in the long
time, long wavelength limit.

As with normal fluids, there are states for which the simple Navier-Stokes
form for hydrodynamics does not apply because the gradients are no longer
small. In the case of normal fluids steady states can generally be studied
in the Navier-Stokes domain by external control of boundary conditions. For
granular fluids, there are new types of steady states associated with an
autonomous balance of external constraints and the internal cooling. For
example, a fluid under shear has viscous heating due to the work done on it
at the boundaries. Normally this is compensated by a temperature gradient
that induces a compensating heat flux to produce the steady state. However,
a granular fluid can compensate for the external work done via its
collisional cooling. The latter scales as $T^{3/2}$ so any amount of work
done can be accommodated by the system choosing an appropriate steady state
temperature. This balance is given by the energy balance equation (\ref{4.13}%
) whose steady state form becomes, for uniform temperature and density 
\begin{equation}
\zeta (n,T_{s})T_{s}=-\frac{2}{3n}P_{\alpha \beta }\partial _{\alpha
}U_{\beta }.  \label{9.1}
\end{equation}%
This imposes a relationship of the given boundary shear, the coefficient of
restitution $\alpha $, and the steady state temperature $T_{s}$. Any attempt
to produce small gradients by decreasing the boundary shear also decreases $%
T_{s}$ at constant $\alpha $. Since the dimensionless measure of small shear
scales as $T^{-1/2}$ it is found that this dimensionless shear can never be
brought within the accuracy of the Navier-Stokes hydrodynamics \cite%
{Inherent}. Another example of this is a granular fluid pinned between two
walls at constant temperature. The internal cooling introduces a temperature
gradient toward lower temperatures between the walls. The gradients
established are controlled by the cooling and cannot be made small by
changing the boundary temperatures. It can be shown that the steady state
temperature profile is never that of the Navier-Stokes hydrodynamic
prediction \cite{Inherent}. Such discrepancies have been seen in recent
molecular dynamics simulations \cite{Hrenya}.

These last observations justify characterizing granular fluids as "complex".
Under some conditions they are well-described by the local partial
differential equations of Navier-Stokes form; under others they have a
rheology or complex dissipation that is not observed in normal fluids, or
only for those with structural complexity. Studies to date suggest that a
hydrodynamic description in its most general sense - closed equations for
the hydrodynamic fields - is a reasonable expectation, although the
associated constitutive equations may be complicated and state dependent.
Discovery of generic constitutive equations (e.g., for shear flow) is both
the challenge and opportunity posed by granular fluids.

\section{Acknowledgments}

The results presented and the author's understanding of them have been
obtained in collaboration and discussion with many others over the past
decade. Particular acknowledgement is given to Professors Javier Brey and
Maria Jose Ruiz Montero of the Universidad de Sevilla, Professors Andres
Santos and Vicente Garzo of the Universidad de Extremadura, Dr. James Lutsko
of the Universite Libre de Bruxelle, and Dr. Aparna Baskaran of Syracuse
University.

\appendix

\section{Hard Sphere Dynamics}

\label{appA}

\subsection{Generator of Trajectories}

The phase function $A\left( \Gamma \right) $ in (\ref{3.4}) is evaluated at
a phase point $\Gamma _{t}=\left\{ \mathbf{q}_{1}(t),\ldots ,\mathbf{q}%
_{N}(t),\mathbf{v}_{1}(t),\dots ,\mathbf{v}_{N}(t)\right\} $ which has
evolved from its initial point at $t=0$, $\Gamma =\left\{ \mathbf{q}%
_{1},\ldots ,\mathbf{q}_{N},\mathbf{v}_{1},\dots ,\mathbf{v}_{N}\right\} $.
Therefore $A\left( \Gamma _{t}\right) $ can be considered as a function of
the initial point and time, $A\left( \Gamma _{t}\right) =A\left( \Gamma
,t\right) $. The simplest example to illustrate this is a gas of non
interacting particles for which $\Gamma _{t}=\left\{ \mathbf{q}_{1}+\mathbf{v%
}_{1}t,\ldots ,\mathbf{q}_{N}+\mathbf{v}_{N}t,\mathbf{v}_{1},\dots ,\mathbf{v%
}_{N}\right\} $. This corresponds to free streaming of all particles at
constant velocities. A phase function at time $t$ can therefore be expressed
in terms of a generator for the dynamics in the following way%
\begin{eqnarray}
A\left( \Gamma _{t}\right) &=&A\left( \left\{ \mathbf{q}_{1}+\mathbf{v}%
_{1}t,\ldots ,\mathbf{q}_{N}+\mathbf{v}_{N}t,\mathbf{v}_{1},\dots ,\mathbf{v}%
_{N}\right\} \right)  \notag \\
&=&e^{L_{0}t}A\left( \left\{ \mathbf{q}_{1},\ldots ,\mathbf{q}_{N},\mathbf{v}%
_{1},\dots ,\mathbf{v}_{N}\right\} \right) ,  \label{a.1}
\end{eqnarray}%
where $L$ is the sum of generators for translations for each particle
coordinate%
\begin{equation}
L_{0}=\sum_{r}\mathbf{v}_{r}\cdot \nabla _{\mathbf{q}_{r}}.  \label{a.2}
\end{equation}%
In this form the dependence on $\Gamma $ and $t$ is made explicit.

In the presence of interactions among the particles the time dependence of
the positions and velocities is more complex, but still given by a
deterministic rule for evolution of the initial point. If the particle
interaction is pairwise additive and due to conservative, non-singular
forces $\mathbf{F}(q_{ij}),$ then (\ref{a.1}) is replaced by 
\begin{equation}
A\left( \Gamma _{t}\right) =e^{Lt}A\left( \Gamma \right) ,  \label{a.3}
\end{equation}%
where the generator $L$%
\begin{equation}
L=L_{0}+\frac{1}{2}\sum_{r,s}^{N}m^{-1}\mathbf{F}(q_{ij})\cdot \left( 
\boldsymbol{\nabla }_{\mathbf{v}_{i}}-\boldsymbol{\nabla }_{\mathbf{v}%
_{j}}\right) .  \label{a.4}
\end{equation}%
It is easily verified that this generator changes the positions and
velocities according to Hamilton's equations ($L$ is the linear operator
representing the Poisson bracket of its operand with the Hamiltonian for the
system). Generalization to include non-singular, non-conservative forces is
straightforward.

Hard sphere dynamics is also represented by a rule for evolution of the
initial phase point: particles freely stream until one pair is at contact.
At that time the velocities of that pair are changed according to a
specified rule (e.g., that of (\ref{3.2})) constrained to conserve total
momentum for the pair. Subsequently, all particles freely stream again until
another pair is at contact, when the corresponding velocity transformation
of that pair is introduced. The generator for this process cannot be
represented simply as in (\ref{a.4}) using the corresponding singular force.
To discover the correct form \cite{lutsko1}, assume the first collision is
between particles $1$ and $2$, and let $\tau _{1}(\mathbf{q}_{12},\mathbf{g}%
_{12})$ denote the time at which it occurs as a function of the initial
separation and relative velocity. Therefore, the time evolution of any phase
function $A\left( \Gamma \right) $ can be given compactly by 
\begin{equation}
A\left( \Gamma _{t}\right) =\left( 1-\Theta \left( t-\tau _{1}\right)
\right) A\left( \Gamma _{t}\right) +\Theta \left( t-\tau _{1}\right) A\left(
\Gamma _{t}\right) +\cdot \cdot  \label{a.5}
\end{equation}%
where $\Theta $ is the Heaviside step function,%
\begin{equation*}
\Theta \left( x\right) =\left( 
\begin{array}{c}
1,\hspace{0.3in}x\geq 1 \\ 
0,\hspace{0.3in}x<1%
\end{array}%
\right) .
\end{equation*}%
The dots in (\ref{a.5}) denote contributions that arise only on the time
scale of the third and later collisions. The first term on the right side of
the first line contributes only for times before the first collision, so
that $\mathbf{q}_{1,2}(t)=\mathbf{q}_{1,2}+\mathbf{v}_{1,2}t,$ $\mathbf{v}%
_{1,2}(t)=\mathbf{v}_{1,2}$. The second term contributes only after that
collision, so $\mathbf{q}_{1,2}(t)=\mathbf{q}_{1,2}+\mathbf{v}_{1,2}\tau
_{1}+\mathbf{v}_{1,2}^{\prime }\left( t-\tau _{1}\right) ,$ $\mathbf{v}%
_{1,2}(t)=\mathbf{v}_{1,2}^{\prime }$. The prime indicates that the
velocities of the colliding pair have been changed for $t>\tau _{1}$
according to the \ collision rule (\ref{3.2})%
\begin{equation}
\mathbf{v}_{1}^{\prime }=\mathbf{v}_{1}-\frac{1}{2}\left( 1+\alpha \right)
\left( \widehat{\mathbf{q}}_{12}\cdot \mathbf{g}_{12}\right) \widehat{%
\mathbf{q}}_{12},\hspace{0.3in}\mathbf{v}_{2}^{\prime }=\mathbf{v}_{2}+\frac{%
1}{2}\left( 1+\alpha \right) \left( \widehat{\mathbf{q}}_{12}\cdot \mathbf{g}%
_{12}\right) \widehat{\mathbf{q}}_{12}.  \label{a.6}
\end{equation}%
This distinction between $\Gamma =\left\{ \mathbf{q}_{1}(t),\ldots ,\mathbf{q%
}_{N}(t),\mathbf{v}_{1}(t),\dots ,\mathbf{v}_{N}(t)\right\} $ for $t<\tau
_{1}$ and $t\geq \tau _{1}$ reflects the discontinuity in the velocity at $%
t=\tau _{1}$.

Differentiation of (\ref{a.6}) with respect to time gives 
\begin{equation}
\partial _{t}A\left( \Gamma _{t}\right) =\left( \mathbf{v}_{1}(t)\cdot
\nabla _{\mathbf{q}_{1}(t)}+\mathbf{v}_{2}(t)\cdot \nabla _{\mathbf{q}%
_{2}(t)}\right) A\left( \Gamma _{t}\right) +\delta \left( t-\tau _{1}\right)
\left( \Delta A\left( \Gamma _{\tau _{1}}\right) -A\left( \Gamma _{\tau
_{1}-\epsilon }\right) \right) +\cdot \cdot  \label{a.7}
\end{equation}%
The first term is defined for times before and after the collision, with
appropriate changes in $\mathbf{q}_{1,2}(t)$ and $\mathbf{v}_{1,2}(t)$ as
given above. The second term occurs only at the time of the collision and is
proportional to the discontinuity in $\Delta A\left( \Gamma _{\tau
_{1}}\right) =\lim_{\epsilon \rightarrow 0}\left( A\left( \Gamma _{\tau
_{1}}\right) -A\left( \Gamma _{\tau _{1}-\epsilon }\right) \right) $ due to
the instantaneous change in the velocities of the colliding pair. This can
be represented by a substitution operator $b_{12}\left( \mathbf{g}_{12}(\tau
_{1})\right) $ as in (\ref{3.9}), but here defined for functions of the
position and velocity at time $\tau _{1}$%
\begin{equation}
\Delta A\left( \Gamma _{\tau _{1}}\right) =\left( b_{12}\left( \mathbf{g}%
_{12}(\tau _{1})\right) -1\right) A\left( \Gamma _{\tau _{1}-\epsilon
}\right) .  \label{a.8}
\end{equation}%
Note that $b_{12}\left( \mathbf{g}_{12}(\tau )\right) $ changes $\mathbf{v}%
_{1,2}(t)$ at constant $\mathbf{q}_{1,2}(t)$. The time dependence of the
delta function can be expressed through $\mathbf{q}_{12}(t)-\mathbf{q}%
_{12}(\tau _{1})=\mathbf{g}_{12}(t-\tau _{1})$, where $\mathbf{g}_{12}$ is
the relative velocity before the collision. By definition of the hard sphere
collision $\mathbf{q}_{12}(\tau _{1})=$ $\mathbf{\sigma }$ is a vector of
length $\sigma $ implying contact between the two particles. This can occur
only if the particles are directed toward each other, i.e. for $\widehat{%
\mathbf{\sigma }}\mathbf{\cdot g}_{12}<0$. Thus the delta function in time
can be written in terms of a delta function for position 
\begin{eqnarray}
\delta \left( t-\tau _{1}\right) &=&\Theta \left( -\widehat{\boldsymbol{%
\sigma }}\mathbf{\cdot g}_{12}\right) \delta \left( \frac{\widehat{%
\boldsymbol{\sigma }}\cdot \left( \mathbf{q}_{12}(t)-\mathbf{q}_{12}(\tau
_{1})\right) }{\left\vert \widehat{\boldsymbol{\sigma }}\cdot \mathbf{g}%
_{12}\right\vert }\right)  \notag \\
&=&\Theta \left( -\widehat{\boldsymbol{\sigma }}\mathbf{\cdot g}_{12}\right)
\left\vert \widehat{\boldsymbol{\sigma }}\cdot \mathbf{g}_{12}\right\vert
\delta \left( q_{12}(t)-\sigma \right) .  \label{a.9}
\end{eqnarray}%
Hence (\ref{a.11}) takes the form 
\begin{eqnarray}
\partial _{t}A\left( \Gamma _{t}\right) &=&\left( \mathbf{v}_{1}(t)\cdot
\nabla _{\mathbf{q}_{1}(t)}+\mathbf{v}_{2}(t)\cdot \nabla _{\mathbf{q}%
_{2}(t)}\right) A\left( \Gamma _{t}\right)  \notag \\
&&+\delta \left( q_{12}(t)-\sigma \right) \Theta \left( -\widehat{%
\boldsymbol{\sigma }}\mathbf{\cdot g}_{12}\left( \tau _{1}\right) \right)
\left\vert \widehat{\boldsymbol{\sigma }}\cdot \mathbf{g}_{12}\left( \tau
_{1}\right) \right\vert  \notag \\
&&\times \left( b_{12}\left( \mathbf{g}_{12}(\tau _{1})\right) -1\right)
A\left( \Gamma \left( \tau _{1}\right) \right) +\cdot \cdot  \label{a.10}
\end{eqnarray}%
Since the delta function enforces $t=\tau _{1}$ the latter can be expressed
equivalently as $t$ in the second term. With the representation (\ref{a.3})
the generator of trajectories $L$ is identified as 
\begin{eqnarray}
L &=&\mathbf{v}_{1}\cdot \nabla _{\mathbf{q}_{1}}+\mathbf{v}_{2}\cdot \nabla
_{\mathbf{q}_{2}}+T\left( 1,2\right) +\cdot \cdot  \notag \\
&=&L_{0}+\frac{1}{2}\sum_{r,s}^{N}T\left( r,s\right)  \label{a.11}
\end{eqnarray}%
with the binary collision operator $T(r,s)$ given by 
\begin{equation}
T(r,s)=\delta \left( q_{rs}-\sigma \right) \Theta \left( -\widehat{%
\boldsymbol{\sigma }}_{rs}\cdot \mathbf{g}_{rs}\right) \left\vert \widehat{%
\boldsymbol{\sigma }}_{rs}\cdot \mathbf{g}_{rs}\right\vert \left(
b_{rs}-1\right) .  \label{a.12}
\end{equation}%
The second line of (\ref{a.11}) extends the analysis to longer times where
there is sufficient time for many collisions.

\subsection{Two adjoint generators}

Next, consider the generator $\overline{L}$ defined in (\ref{3.10}). This
can be identified from 
\begin{eqnarray}
\int d\Gamma A(\Gamma )\overline{L}\rho \left( \Gamma \right) &\equiv &-\int
d\Gamma \left( LA(\Gamma )\right) \rho \Gamma )  \notag \\
&=&-\sum_{r=1}^{N}\int d\Gamma \rho \left( \Gamma \right) \left( \mathbf{v}%
_{r}\cdot \nabla _{\mathbf{q}_{r}}A\left( \Gamma \right) \right) -\frac{1}{2}%
\sum_{r,s}^{N}\int d\Gamma \rho \left( \Gamma \right) \delta \left(
q_{rs}-\sigma \right)  \notag \\
&&\times \Theta \left( -\widehat{\boldsymbol{\sigma }}_{rs}\cdot \mathbf{g}%
_{rs}\right) \left\vert \widehat{\boldsymbol{\sigma }}_{rs}\cdot \mathbf{g}%
_{rs}\right\vert \left( b_{rs}-1\right) A\left( \Gamma \right) .
\label{a.13}
\end{eqnarray}%
Define the inverse of $b_{rs}$ by $b_{rs}^{-1}b_{rs}=b_{rs}b_{rs}^{-1}=1$, 
\begin{equation}
b_{rs}^{-1}\mathbf{g}_{rs}\equiv \mathbf{g}_{rs}^{\prime \prime }=\mathbf{g}%
_{rs}-\left( 1+\alpha \right) \alpha ^{-1}\left( \widehat{\boldsymbol{\sigma 
}}_{rs}\cdot \mathbf{g}_{rs}\right) \widehat{\boldsymbol{\sigma }}_{rs}.
\label{a.14}
\end{equation}%
A useful identity is given by 
\begin{equation}
\int d\Gamma X\left( \Gamma \right) b_{rs}Y\left( \Gamma \right) =\int
d\Gamma \alpha ^{-1}X\left( b_{rs}^{-1}\Gamma \right) Y\left( \Gamma \right)
.  \label{a.15}
\end{equation}%
This follows by changing integration variables from $\left( \mathbf{v}_{r},%
\mathbf{v}_{s}\right) $ to $\left( b_{rs}\mathbf{v}_{r},b_{rs}\mathbf{v}%
_{s}\right) $. The factor $\alpha ^{-1}$ is the Jacobian for this change of
variables. Also 
\begin{equation}
b_{rs}^{-1}\left( \widehat{\boldsymbol{\sigma }}_{rs}\cdot \mathbf{g}%
_{rs}\right) =-\alpha ^{-1}\widehat{\boldsymbol{\sigma }}_{rs}\cdot \mathbf{g%
}_{rs}.  \label{a.16}
\end{equation}%
With these results (\ref{a.13}) becomes 
\begin{eqnarray}
\int d\Gamma A(\Gamma )\overline{L}\rho \left( \Gamma \right)
&=&\sum_{i=1}^{N}\left( \int d\Gamma \rho \left( \Gamma \right) \left( 
\mathbf{v}_{r}\cdot \nabla _{\mathbf{q}_{r}}A\left( \Gamma \right) \right)
+\int_{S}d\widehat{\mathbf{q}}_{r}\cdot \int d\mathbf{v}_{r}d\Gamma _{s\neq
r}\left( \mathbf{v}_{r}\rho \left( \Gamma \right) A\left( \Gamma \right)
\right) \right)  \notag \\
&&+\frac{1}{2}\sum_{r,s}\int d\Gamma A\left( \Gamma \right) \delta \left(
q_{rs}-\sigma \right) \left\vert \widehat{\boldsymbol{\sigma }}_{rs}\cdot 
\mathbf{g}_{rs}\right\vert  \notag \\
&&\times \left( \Theta \left( \widehat{\boldsymbol{\sigma }}_{rs}\cdot 
\mathbf{g}_{rs}\right) \alpha ^{-2}b_{rs}^{-1}-\Theta \left( -\widehat{%
\boldsymbol{\sigma }}_{rs}\cdot \mathbf{g}_{rs}\right) \right) \rho \left(
\Gamma \right) .  \label{a.17}
\end{eqnarray}%
An integration by parts has been performed and the formal adjoint Liouville
operator is identified as 
\begin{equation}
\overline{L}=L_{0}-\frac{1}{2}\sum_{r,s}^{N}\overline{T}(r,s),  \label{a.18}
\end{equation}%
where, the new binary collision operator is 
\begin{equation}
\overline{T}(r,s)=\delta (q_{rs}-\sigma )|\mathbf{g}_{rs}\cdot \mathbf{\hat{q%
}}_{rs}|\left( \Theta \left( \mathbf{q}_{rs}\cdot \mathbf{g}_{rs}\right)
\alpha ^{-2}b_{rs}^{-1}-\Theta \left( -\mathbf{q}_{rs}\cdot \mathbf{g}%
_{rs}\right) \right) .  \label{a.19}
\end{equation}

A second adjoint generator can be interpreted as that for reversed dynamics 
\cite{DBB07}. It is defined by (\ref{3.15})%
\begin{equation}
e^{-t\overline{L}}\left( \rho (\Gamma )B(\Gamma )\right) =\left( e^{-t%
\overline{L}}\rho (\Gamma )\right) \left( e^{-tL_{-}}B(\Gamma )\right) ,
\label{a.20}
\end{equation}%
or equivalently%
\begin{equation}
\overline{L}\left( \rho (\Gamma )B(\Gamma )\right) =\left( \overline{L}\rho
(\Gamma )\right) B(\Gamma )+\rho (\Gamma )\left( L_{-}B(\Gamma )\right) ,
\label{a.21}
\end{equation}%
The distributive property of (\ref{a.22}) is clearly satisfied by $L_{0}$,
so $L_{-}$ can be written in the form 
\begin{equation}
L_{-}\equiv L_{0}-\frac{1}{2}\sum_{r,s=1}^{N}T_{-}(r,s).  \label{a.22}
\end{equation}%
The new binary operator $T_{-}(i,j)$ is identified from from the
corresponding condition%
\begin{eqnarray*}
\rho (\Gamma )T_{-}(r,s)B(\Gamma ) &=&\overline{T}(r,s)\left( \rho (\Gamma
)B(\Gamma )\right) -\left( \overline{T}(r,s)\rho (\Gamma )\right) B(\Gamma )
\\
&=&\delta (q_{r,s}-\sigma )\alpha ^{-1}b_{rs}^{-1}|\mathbf{g}_{rs}\cdot 
\mathbf{\hat{q}}_{rs}|\Theta (-\mathbf{\hat{g}}_{rs}\cdot \mathbf{\hat{q}}%
_{rs})\left( \rho (\Gamma )B(\Gamma )\right) \\
&&-B(\Gamma )\delta (q_{rs}-\sigma )\alpha ^{-1}b_{rs}^{-1}|\mathbf{g}%
_{rs}\cdot \mathbf{\hat{q}}_{rs}|\Theta (-\mathbf{\hat{g}}_{rs}\cdot \mathbf{%
\hat{q}}_{rs})\rho (\Gamma ).
\end{eqnarray*}%
\begin{eqnarray}
&=&\left[ \delta (q_{rs}-\sigma )\alpha ^{-1}b_{rs}^{-1}|\mathbf{g}%
_{rs}\cdot \mathbf{\hat{q}}_{rs}|\Theta (-\mathbf{\hat{g}}_{rs}\cdot \mathbf{%
\hat{q}}_{rs})\rho (\Gamma )\right] \left( b_{rs}^{-1}-1\right) B(\Gamma ) 
\notag \\
&=&\rho (\Gamma )\delta (q_{rs}-\sigma )|\mathbf{g}_{rs}\cdot \mathbf{\hat{q}%
}_{rs}|\Theta (\mathbf{\hat{g}}_{rs}\cdot \mathbf{\hat{q}}_{rs})\left(
b_{rs}^{-1}-1\right) B(\Gamma ).  \label{a.23}
\end{eqnarray}%
The last equality follows from a boundary condition for hard particles at
contact 
\begin{equation}
\delta (q_{rs}-\sigma )\alpha ^{-1}b_{rs}^{-1}|\mathbf{g}_{rs}\cdot \mathbf{%
\hat{q}}_{rs}|\Theta (-\mathbf{\hat{g}}_{rs}\cdot \mathbf{\hat{q}}_{rs})\rho
(\Gamma )=\delta (q_{rs}-\sigma )|\mathbf{g}_{rs}\cdot \mathbf{\hat{q}}%
_{rs}|\Theta (\mathbf{\hat{g}}_{rs}\cdot \mathbf{\hat{q}}_{rs})\rho (\Gamma
).  \label{a.24}
\end{equation}%
This can be understood from the fact that the flux of particles moving
towards each other at contact must equal the flux of particles separating at
contact. Since the velocities of the former and latter are, respectively, $%
\mathbf{g}_{ij}^{\prime }=b_{ij}^{-1}\mathbf{g}_{ij}$ and $\mathbf{g}_{ij}$,
this gives \cite{lutsko01} 
\begin{equation}
\delta (q_{rs}-\sigma )|\mathbf{g}_{rs}^{\prime }\cdot \mathbf{\hat{q}}%
_{rs}|\Theta (-\mathbf{g}_{rs}^{\prime }\cdot \mathbf{\hat{q}}_{rs})\rho
(\Gamma ^{\prime })d\Gamma ^{\prime }=\delta (q_{rs}-\sigma )|\mathbf{g}%
_{rs}\cdot \mathbf{\hat{q}}_{rs}|\Theta (\mathbf{\hat{g}}_{rs}\cdot \mathbf{%
\hat{q}}_{rs})\rho (\Gamma )d\Gamma .  \label{a.25}
\end{equation}%
This result implies (\ref{a.24}). The form for $T_{-}(i,j)$ is then
identified from (\ref{a.23}) as

\begin{equation}
T_{-}(r,s)=\delta (q_{rs}-\sigma )|\mathbf{g}_{rs}\cdot \mathbf{\hat{q}}%
_{rs}|\Theta (\mathbf{\hat{g}}_{rs}\cdot \mathbf{\hat{q}}_{rs})\left(
b_{rs}^{-1}-1\right)  \label{a.26}
\end{equation}

\section{Balance Equations and Fluxes}

\label{appB}

In this appendix, the forms of the fluxes and sources associated with the
local number density $\hat{n}$, energy density $\hat{e}$, and momentum
density $\mathbf{\hat{g}}$ are identified from the microscopic balance
equations associated with these densities. Consider first the case for phase
functions of the form

\begin{equation}
A(\mathbf{r},t)=e^{Lt}A(\mathbf{r}),\hspace{0.3in}A(\mathbf{r}%
)=\sum_{s=1}^{N}a(\mathbf{v}_{s})\delta (\mathbf{r}-\mathbf{q}_{s}).
\label{b.1}
\end{equation}%
Its forward time evolution is%
\begin{equation}
\left( \partial _{t}-L\right) A(\mathbf{r},t)=0  \label{b.2}
\end{equation}%
The action of the Liouville operator is%
\begin{eqnarray}
LA(\mathbf{r}) &=&-\nabla _{\mathbf{r}}\cdot \sum_{s=1}^{N}\mathbf{v}_{s}a(%
\mathbf{v}_{s})\delta (\mathbf{r}-\mathbf{q}_{s})  \notag \\
&&+\frac{1}{2}\sum_{r,s}^{N}T(r,s)\left( \delta (\mathbf{r}-\mathbf{q}_{r})a(%
\mathbf{v}_{r})+\delta (\mathbf{r}-\mathbf{q}_{s})a(\mathbf{v}_{s})\right)
\label{b.3}
\end{eqnarray}%
Next, use the identity%
\begin{eqnarray}
\delta (\mathbf{r}-\mathbf{q}_{r}) &=&\delta (\mathbf{r}-\mathbf{q}%
_{s})+\int_{0}^{1}d\gamma \frac{\partial }{\partial \gamma }\delta \left( 
\mathbf{r}-\mathbf{q}_{s}+\gamma \mathbf{q}_{sr}\right)  \notag \\
&=&\delta (\mathbf{r}-\mathbf{q}_{s})+\nabla _{\mathbf{r}_{\beta }}\cdot
\int_{0}^{1}d\gamma \delta \left( \mathbf{r}-\mathbf{q}_{s}+\gamma \mathbf{q}%
_{sr}\right) \mathbf{q}_{sr},  \label{b.4}
\end{eqnarray}%
to obtain the microscopic balance equation for $A(\mathbf{r},t)$ 
\begin{equation}
\partial _{t}A(\mathbf{r},t)+\nabla _{\mathbf{r}}\cdot \mathbf{B}(\mathbf{r}%
,t)=S(\mathbf{r}).  \label{b.5}
\end{equation}%
The flux $\mathbf{B}(\mathbf{r},t)$ is identified as 
\begin{equation}
\mathbf{B}(\mathbf{r})=\sum_{k=1}^{N}\mathbf{v}_{k}a(\mathbf{v}_{k})\delta (%
\mathbf{r}-\mathbf{q}_{k})-\frac{1}{2}\int_{0}^{1}d\gamma
\sum_{r,s}^{N}\delta \left( \mathbf{r}-\mathbf{q}_{r}+\gamma \mathbf{q}%
_{rs}\right) \mathbf{q}_{rs}T(r,s)a(\mathbf{v}_{s}).  \label{b.7}
\end{equation}%
The first term on the right side is the \textquotedblright
kinetic\textquotedblright\ contribution due to the kinetic motion of the
particles, while the second term is the \textquotedblright collisional
transfer\textquotedblright\ part of the flux that requires no motion. The
source term in the balance equation is 
\begin{equation}
S(\mathbf{r})=\frac{1}{2}\sum_{r=1}^{N}\sum_{j\neq s}^{N}\delta (\mathbf{r}-%
\mathbf{q}_{r})T(r,s)\left( a(\mathbf{v}_{r})+a(\mathbf{v}_{s})\right)
\label{b.8}
\end{equation}%
It vanishes for any two particle \textquotedblright collisional
invariant\textquotedblright\ $a(\mathbf{v}_{s})+a(\mathbf{v}_{r})$ which is
the same before and after a binary collision. For elastic collisions or
conservative forces these are given by $a(\mathbf{v}_{s})=1,\mathbf{v},v^{2}$%
.

Consider now the local microscopic number, energy, and momentum densities
defined by 
\begin{equation}
\widehat{n}=\sum_{s=1}^{N}\delta (\mathbf{r}-\mathbf{q}_{s}),\hspace{0.3in}%
\widehat{e}=\sum_{s=1}^{N}\frac{1}{2}m\mathbf{v}_{s}^{2}\delta (\mathbf{r}-%
\mathbf{q}_{s}),\hspace{0.3in}\text{\ }\widehat{\mathbf{g}}=\sum_{s=1}^{N}m%
\mathbf{v}_{s}\delta (\mathbf{r}-\mathbf{q}_{s}).  \label{b.9}
\end{equation}%
The microscopic balance equations become 
\begin{equation}
\partial _{t}\hat{n}(\mathbf{r},t)+m^{-1}\nabla _{\mathbf{r}}\cdot \mathbf{%
\hat{g}}(\mathbf{r,}t)=0  \label{b.10}
\end{equation}%
\begin{equation}
\partial _{t}\hat{e}(\mathbf{r},t)+\nabla _{\mathbf{r}}\cdot \mathbf{s}(%
\mathbf{r,}t)=w(\mathbf{r,}t)  \label{b.11}
\end{equation}%
\begin{equation}
\partial _{t}\hat{g}_{\alpha }(\mathbf{r,}t)+\nabla _{\mathbf{r}_{\beta
}}h_{\alpha \beta }(\mathbf{r,}t)=0  \label{b.12}
\end{equation}%
The absence of sources on the right sides of (\ref{b.10}) and (\ref{b.12})
occurs since $1$ and $\mathbf{v}$ are still summational invariants for
inelastic collisions. The loss function in the energy equation is%
\begin{eqnarray}
w(\mathbf{r}) &=&\frac{1}{2}\sum_{r,s}^{N}\delta (\mathbf{r}-\mathbf{q}%
_{i})T(r,s)\left( \frac{1}{2}mv_{r}^{2}+\frac{1}{2}mv_{s}^{2}\right)  \notag
\\
&=&\frac{1}{2}\sum_{r,s}^{N}\delta (\mathbf{r}-\mathbf{q}_{i})T(r,s)\frac{1}{%
4}mg_{rs}^{2}.  \label{b.13}
\end{eqnarray}%
The second equality follows from the fact that the center of mass energy is
conserved, by momentum conservation. The energy and momentum fluxes are%
\begin{equation}
\mathbf{s}(\mathbf{r})=\sum_{s=1}^{N}\mathbf{v}_{s}\frac{1}{2}m\mathbf{v}%
_{s}^{2}\delta (\mathbf{r}-\mathbf{q}_{s})-\frac{1}{2}\int_{0}^{1}d\gamma
\sum_{r,s}^{N}\delta \left( \mathbf{r}-\mathbf{q}_{r}+\gamma \mathbf{q}%
_{rs}\right) \mathbf{q}_{rs}T(r,s)\frac{1}{2}m\mathbf{v}_{s}^{2}.
\label{b.14}
\end{equation}%
\begin{equation}
h_{\alpha \beta }(\mathbf{r})=\sum_{s=1}^{N}mv_{s\alpha }v_{s\beta }\delta (%
\mathbf{r}-\mathbf{q}_{r})-\frac{1}{2}\int_{0}^{1}d\gamma
\sum_{r,s}^{N}\delta \left( \mathbf{r}-\mathbf{q}_{r}+\gamma \mathbf{q}%
_{rs}\right) q_{rs\alpha }T(r,s)mv_{s\beta }.  \label{b.15}
\end{equation}

\subsection{Time reversed dynamics}

Now consider the same phase function as in (\ref{b.1}) except for the
reversed trajectory (still with $t\geq 0$) 
\begin{equation}
\left( -\partial _{t}-L_{-}\right) A(\mathbf{r},-t)=0,\hspace{0.3in}A(%
\mathbf{r},-t)=e^{-L_{-}t}A(\mathbf{r}).  \label{b.21}
\end{equation}%
\begin{equation}
L_{-}=\sum_{s=1}^{N}\mathbf{v}_{s}\cdot \mathbf{\nabla }_{\mathbf{q}_{s}}-%
\frac{1}{2}\sum_{r,s}^{N}T_{-}(r,s)  \label{b.22}
\end{equation}%
The derivation of balance equations for the backward dynamics is similar to
that above with the result%
\begin{equation}
-\partial _{t}A(\mathbf{r},-t)+\nabla _{\mathbf{r}}\cdot \mathbf{B}_{-}(%
\mathbf{r},-t)=S_{-}(\mathbf{r}).  \label{b.23}
\end{equation}%
The flux $\mathbf{B}_{-}(\mathbf{r})$ is 
\begin{equation}
\mathbf{B}(\mathbf{r})=\sum_{s=1}^{N}\mathbf{v}_{s}a(\mathbf{v}_{s})\delta (%
\mathbf{r}-\mathbf{q}_{s})-\frac{1}{2}\int_{0}^{1}d\gamma
\sum_{r,s}^{N}\delta \left( \mathbf{r}-\mathbf{q}_{r}+\gamma \mathbf{q}%
_{rs}\right) \mathbf{q}_{rs}T_{-}(r,s)a(\mathbf{v}_{s}),  \label{b.24}
\end{equation}%
and the source $S_{-}(\mathbf{r})$ is 
\begin{equation}
S_{-}(\mathbf{r})=-\frac{1}{2}\sum_{s=1}^{N}\delta (\mathbf{r}-\mathbf{q}%
_{s})\sum_{r,s}^{N}T_{-}(r,s)\left( a(\mathbf{v}_{r})+a(\mathbf{v}%
_{s})\right) .  \label{b.25}
\end{equation}%
Interestingly, the fluxes and sources for the forward and backward balance
equations are not the same due to the differences between $T(r,s)$ and $%
T_{-}(r,s)$. This is a peculiarity of hard spheres for both normal and
granular fluids. It occurs for non-singular, non-conservative dynamics as
well.

The corresponding microscopic balance equations for reversed dynamics are 
\begin{equation}
-\frac{\partial \hat{n}(\mathbf{r},-t)}{\partial t}+m^{-1}\nabla _{\mathbf{r}%
}\cdot \mathbf{\hat{g}}(\mathbf{r,-}t)=0  \label{b.26}
\end{equation}%
\begin{equation}
-\frac{\partial \hat{e}(\mathbf{r},-t)}{\partial t}+\nabla _{\mathbf{r}%
}\cdot \mathbf{s}_{-}(\mathbf{r,-}t)=w_{-}(\mathbf{r,-}t)  \label{b.27}
\end{equation}%
\begin{equation}
-\frac{\partial \hat{g}_{\alpha }(\mathbf{r,-}t)}{\partial t}+\nabla _{%
\mathbf{r}_{\beta }}h_{-\alpha \beta }(\mathbf{r,-}t)=0  \label{b.28}
\end{equation}%
\begin{equation}
w_{-}(\mathbf{r})=-\frac{1}{2}\sum_{r,s}^{N}\delta (\mathbf{r}-\mathbf{q}%
_{r})T_{-}(r,s)\left( \frac{1}{2}m\mathbf{v}_{r}^{2}+\frac{1}{2}m\mathbf{v}%
_{s}^{2}\right) .  \label{b.29}
\end{equation}%
The energy and momentum fluxes are%
\begin{equation}
\mathbf{s}_{-}(\mathbf{r})=\sum_{s}^{N}\mathbf{v}_{s}\frac{1}{2}m\mathbf{v}%
_{s}^{2}\delta (\mathbf{r}-\mathbf{q}_{s})+\frac{1}{2}\int_{0}^{1}d\gamma
\sum_{r,s}^{N}\delta \left( \mathbf{r}-\mathbf{q}_{r}+\gamma \mathbf{q}%
_{rs}\right) \mathbf{q}_{rs}T_{-}(r,s)\frac{1}{2}m\mathbf{v}_{s}^{2}.
\label{b.30}
\end{equation}%
\begin{equation}
h_{-\alpha \beta }(\mathbf{r})=\sum_{s=1}^{N}mv_{s\alpha }v_{s\beta }\delta (%
\mathbf{r}-\mathbf{q}_{s})+\frac{1}{2}\int_{0}^{1}d\gamma
\sum_{r,s}^{N}\delta \left( \mathbf{r}-\mathbf{q}_{r}+\gamma \mathbf{q}%
_{rs}\right) q_{rs\alpha }T_{-}(r,s)mv_{s\beta }.  \label{b.31}
\end{equation}%
Note that the fluxes are not the same as those for the forward time
conservation laws, (\ref{b.14}) and (\ref{b.15}), because of the differences
between the binary collision operators $T(r,s)$ and $T_{-}(r,s)$. This
observation is not purely of academic interest, as both sets of fluxes occur
in the Green-Kubo expressions for transport coefficients for a normal fluid
(see for example (\ref{7.32})).

\section{Local Rest Frame Transformation}

\label{appC}The microscopic densities and fluxes can be represented in terms
of contributions from local convective flow plus their values in the local
rest frame. This is accomplished by a partial Galilean transformation on all
velocities, $\mathbf{v}_{s}^{\prime }=\mathbf{v}_{s}-\mathbf{U}(\mathbf{r}%
,t) $, where $\mathbf{U}(\mathbf{r},t)$ is the macroscopic flow field
defined by Eq. (\ref{4.6}) and the point $\mathbf{r}$ is the same as that of
the local property being considered. A straightforward calculation for the
densities gives the results $\widehat{n}(\mathbf{r})=\widehat{n}^{\prime }(%
\mathbf{r}), $ $l(\mathbf{r})=l^{\prime }(\mathbf{r})$ and 
\begin{equation}
\widehat{e}(\mathbf{r})=\widehat{e}^{\prime }(\mathbf{r})+\widehat{\mathbf{g}%
}^{\prime }(\mathbf{r})\cdot \mathbf{U(\mathbf{r}},t\mathbf{)+}\frac{1}{2}m%
\widehat{n}(\mathbf{r})U^{2}(\mathbf{r},t),\hspace{0.3in}\widehat{\mathbf{g}}%
(\mathbf{r})=\widehat{\mathbf{g}}^{\prime }(\mathbf{r})+m\widehat{n}(\mathbf{%
r})\mathbf{U}(\mathbf{r},t).  \label{c.1}
\end{equation}%
A prime on a phase function indicates that same function evaluated at $%
\mathbf{v}_{s}\rightarrow \mathbf{v}_{s}^{\prime }.$ The energy and momentum
fluxes transform according to%
\begin{eqnarray}
\mathbf{s}(\mathbf{r}) &=&\mathbf{s}^{\prime }(\mathbf{r})+\mathbf{U}(%
\mathbf{r},t)\left( \widehat{e}^{\prime }(\mathbf{r})+\widehat{\mathbf{g}}%
^{\prime }(\mathbf{r})\cdot \mathbf{U(\mathbf{r}},t\mathbf{)+}\frac{1}{2}m%
\widehat{n}^{\prime }(\mathbf{r})U^{2}(\mathbf{r},t)\right)  \notag \\
&&+\frac{1}{2}\widehat{\mathbf{g}}^{\prime }(\mathbf{r})U^{2}(\mathbf{r}%
,t)+h_{\alpha \beta }^{\prime }(\mathbf{r})  \label{c.2}
\end{eqnarray}%
\begin{equation}
h_{\alpha \beta }(\mathbf{r})=h_{\alpha \beta }^{\prime }(\mathbf{r})+%
\widehat{g}_{\alpha }^{\prime }(\mathbf{r})U_{\beta }(\mathbf{r},t)+\widehat{%
g}_{\beta }^{\prime }(\mathbf{r})U_{\alpha }(\mathbf{r},t)+m\widehat{n}%
^{\prime }(\mathbf{r})U_{\alpha }(\mathbf{r},t)U_{\beta }(\mathbf{r},t)
\label{c.3}
\end{equation}%
The averages of these are then $n(\mathbf{r},t)=n^{\prime }(\mathbf{r},t),$ $%
\left\langle w(\mathbf{r});t\right\rangle =\left\langle w^{\prime }(\mathbf{r%
});t\right\rangle $ and 
\begin{equation}
e(\mathbf{r},t)=e^{\prime }(\mathbf{r},t)\mathbf{+}\frac{1}{2}mn(\mathbf{r}%
,t)U^{2}(\mathbf{r},t),\hspace{0.3in}\mathbf{g}(\mathbf{r},t)=mn(\mathbf{r}%
,t)\mathbf{U}(\mathbf{r},t).  \label{c.4}
\end{equation}%
\begin{equation}
\left\langle \mathbf{s}(\mathbf{r});t\right\rangle =\left\langle \mathbf{s}%
^{\prime }(\mathbf{r});t\right\rangle +\mathbf{U}(\mathbf{r},t)\left(
e^{\prime }(\mathbf{r},t)\mathbf{+}\frac{1}{2}mn(\mathbf{r},t)U^{2}(\mathbf{r%
},t)\right) +\left\langle h_{\alpha \beta }^{\prime }(\mathbf{r}%
);t\right\rangle  \label{c.5}
\end{equation}

\section{Local Homogeneous Cooling State}

\label{appD}

The HCS solution to the Liouville equation corresponds to spatially constant
hydrodynamic fields. The \textit{local} HCS is a reference state
distribution that is not a solution to the Liouville equation but
approximates the true nonequilibrium normal solution. Physically, it
represents a partitioning of the system into cells such that each cell is in
a HCS at its own values for the fields. These values are chosen to be same
as the exact values for the nonequilibrium state. This condition is
expressed in (\ref{6.2}). Recall that the HCS distribution has the scaling
property expressed by (\ref{3.21}) 
\begin{equation}
\rho _{h}\left( \Gamma ;t\right) =(lv_{h}\left( t\right) )^{-Nd}\rho
_{h}^{\ast }\left( \left\{ \frac{\mathbf{q}_{rs}}{l},\frac{\mathbf{v}_{r}-%
\mathbf{U}_{h}}{v_{h}(t)}\right\} ,n_{h}l^{3}\right) ,\hspace{0.3in}v_{h}(t)=%
\sqrt{2T_{h}(t)/m}  \label{d.1}
\end{equation}%
This suggests that the local HCS distribution should be defined as 
\begin{eqnarray}
\rho _{\ell h}\left( \Gamma ,|\left\{ y_{\alpha }\right\} \right) &\equiv
&\prod_{s=1}^{N}(lv(\mathbf{q}_{s},t))^{-3}\rho _{h}^{\ast }\left( \{\frac{%
\mathbf{q}_{rs}}{l},\frac{\mathbf{v}_{r}-\mathbf{U}\left( \mathbf{q}%
_{r},t\right) }{v(\mathbf{q}_{r},t)}\},n\left( \mathbf{q}_{r},t\right)
l^{3}\right) ,  \label{d.2} \\
\hspace{0.3in}v(\mathbf{q}_{r},t) &=&\sqrt{2T(\mathbf{q}_{r},t)/m}.
\label{d.2a}
\end{eqnarray}%
The dimensionless function $\rho _{h}^{\ast }$ has its arguments changed to
reflect the fact that the hydrodynamic fields at the positions of each
particle can be different. Thus translational invariance is broken only by
the spatial dependence of the hydrodynamic fields. It is verified that (\ref%
{3.21}) is satisfied, and clearly $\rho _{\ell h}\rightarrow \rho _{h}$ if
the value of the fields is the same everywhere. The stronger requirements in
(\ref{6.3}) are also satisfied. To illustrate, let the reference value for
the fields be evaluated at a chosen point $\mathbf{r}$ and write $y_{\alpha
}\left( \mathbf{q}_{r},t\right) =y_{\alpha }\left( \mathbf{r},t\right) 
\mathbf{+}\delta y_{\alpha }\left( \mathbf{q}_{r},t\right) $. Then direct
calculation gives 
\begin{equation}
\int d\mathbf{r}^{\prime }\frac{\delta \rho _{h\ell }\left( \Gamma \mid
\left\{ y_{\beta }\right\} \right) }{\delta U_{i}\left( \mathbf{r}^{\prime
},t\right) }\mid _{\delta y=0}=\frac{\partial \rho _{h}}{\partial
U_{i}\left( \mathbf{r},t\right) }.  \label{d.3}
\end{equation}%
In the same way higher functional derivatives have similar properties%
\begin{equation}
\int d\mathbf{r}^{\prime }d\mathbf{r}^{\prime \prime }\frac{\delta \rho
_{h\ell }\left( \Gamma \mid \left\{ y_{0\beta }+\delta y_{\beta }\right\}
\right) }{\delta U_{i}\left( \mathbf{r}^{\prime },t\right) \delta
U_{j}\left( \mathbf{r}^{\prime \prime },t\right) }\mid _{\delta y=0}=\frac{%
\partial ^{2}\rho _{h}}{\partial U_{i}\left( \mathbf{r},t\right) \partial
U_{j}\left( \mathbf{r},t\right) }.  \label{d.4}
\end{equation}%
These requirements exclude other choices for $\rho _{\ell h}$ that are
simply perturbations of $\rho _{h}$. The essential role of (\ref{d.3}) is
demonstrated in the next appendix.

The modified scaling of $\rho _{h\ell }\left( \Gamma ,|\left\{ y_{\alpha
}\right\} \right) $ provides alternative forms for the functional
derivatives with respect to temperature and velocity fields. For example,%
\begin{eqnarray}
\frac{\delta \rho _{h\ell }\left( \Gamma \mid \left\{ y_{\beta }\right\}
\right) }{\delta T\left( \mathbf{r}^{\prime },t\right) } &\mid &_{\delta
y=0}=-\sum_{s=1}^{N}\frac{\delta \ln v(\mathbf{q}_{s},t)}{\delta T\left( 
\mathbf{r}^{\prime },t\right) }\nabla _{\mathbf{v}_{s}}\cdot \left( \mathbf{v%
}_{s}-\mathbf{U}\left( \mathbf{q}_{s},t\right) \right) \rho _{h}\left(
\left\{ y_{\beta }\left( \mathbf{r},t\right) \right\} \right)  \notag \\
&=&-\frac{1}{2T\left( \mathbf{r}^{\prime },t\right) }\sum_{s=1}^{N}\delta (%
\mathbf{q}_{s}-\mathbf{r}^{\prime })\nabla _{\mathbf{v}_{s}}\cdot \left( 
\mathbf{v}_{s}-\mathbf{U}\left( \mathbf{r}^{\prime },t\right) \right) \rho
_{h}\left( \left\{ y_{\beta }\left( \mathbf{r},t\right) \right\} \right)
\label{d.5}
\end{eqnarray}%
Similarly, the corresponding velocity derivative can be written%
\begin{equation}
\frac{\delta \rho _{h\ell }\left( \Gamma \mid \left\{ y_{\beta }\right\}
\right) }{\delta U_{i}\left( \mathbf{r}^{\prime },t\right) }\mid _{\delta
y=0}=-\sum_{s=1}^{N}\delta (\mathbf{q}_{s}-\mathbf{r}^{\prime })\frac{%
\partial \rho _{h}\left( \left\{ y_{\beta }\left( \mathbf{r},t\right)
\right\} \right) }{\partial v_{si}}  \label{d.6}
\end{equation}%
These functional derivatives are therefore phase space densities derived
from derivatives of $\rho _{h}$ with respect to the density.

As stated above, the local HCS distribution is not a solution to the
Liouville equation. Instead, it differs from a solution by terms
proportional to gradients in the hydrodynamic fields.

\section{Solution to the Liouville Equation}

\label{appE}

In this Appendix the normal solution to the Liouville equation $\rho _{n}$
is obtained to first order in the gradients. First write it as a deviation
from the local HCS defined in Appendix \ref{appD} 
\begin{equation}
\rho =\rho _{h\ell }+\Delta .  \label{e.0}
\end{equation}%
Substitution into the Liouville equation gives (\ref{7.2})%
\begin{equation}
\partial _{t}\Delta +\int d\mathbf{r}\frac{\delta \Delta }{\delta y_{\alpha
}\left( \mathbf{r},t\right) }N_{\alpha }(\mathbf{r},t\mid \left\{ y_{\alpha
}\left( t\right) \right\} )+\overline{L}\Delta =-\int d\mathbf{r}\frac{%
\delta \rho _{h\ell }}{\delta y_{\alpha }\left( \mathbf{r},t\right) }%
N_{\alpha }(\mathbf{r},t\mid \left\{ y_{\alpha }\left( t\right) \right\} )-%
\overline{L}\rho _{h\ell }.  \label{e.1}
\end{equation}%
where the time derivative is taken at constant $\left\{ y_{\beta }\right\} $%
. The assumption is that $\rho _{h\ell }$ has been chosen as the correct
reference state in order that $\Delta $ is of first order in the gradients.
This requires that the right side of (\ref{5.9}) should be proportional to
the gradients. To show this, represent $\rho _{h\ell }$ as an expansion to
first order in the gradients about the reference values $\left\{ y_{\alpha }(%
\mathbf{r},t\mathbf{)}\right\} $ 
\begin{eqnarray}
\rho _{h\ell } &=&\rho _{h}\left( \left\{ y_{\alpha }(\mathbf{r},t\mathbf{)}%
\right\} \right) +\int d\mathbf{r}^{\prime }\left( \frac{\delta \rho _{h\ell
}}{\delta y_{\beta }\left( \mathbf{r}^{\prime },t\right) }\right) _{\delta
y=0}\left( y_{\beta }\left( \mathbf{r}^{\prime },t\right) -y_{\beta }\left( 
\mathbf{r},t\right) \right) +\cdot \cdot  \notag \\
&=&\rho _{h}\left( \left\{ y_{\alpha }(\mathbf{r},t\mathbf{)}\right\}
\right) +\mathbf{m}_{\beta }\left( \mathbf{r},\left\{ y_{\alpha }(\mathbf{r}%
,t\mathbf{)}\right\} \right) \cdot \nabla y_{\beta }\left( \mathbf{r}%
,t\right) +\cdot \cdot  \label{e.2}
\end{eqnarray}%
where the dots denote terms of higher order in the gradients, and $\mathbf{m}%
_{\nu }$ is defined by%
\begin{eqnarray}
\mathbf{m}_{\nu }\left( \mathbf{r},\left\{ y_{\alpha }(\mathbf{r},t\mathbf{)}%
\right\} \right) &\equiv &\int d\mathbf{r}^{\prime \prime }\left( \frac{%
\delta \rho _{h\ell }}{\delta y_{\nu }\left( \mathbf{r}^{\prime \prime
},t\right) }\right) _{\delta y=0}\left( \mathbf{r}^{\prime \prime }\mathbf{-r%
}\right)  \notag \\
&\equiv &\mathbf{M}_{\nu }\left( \left\{ y_{\alpha }(\mathbf{r},t\mathbf{)}%
\right\} \right) -\mathbf{r}\frac{\partial \rho _{h}\left( \left\{ y_{\alpha
}(\mathbf{r},t\mathbf{)}\right\} \right) }{\partial y_{\alpha }\left( 
\mathbf{r},t\right) }  \label{e.3}
\end{eqnarray}

With this expansion for $\rho _{h\ell }$ the functional derivative on the
right side can be evaluated%
\begin{eqnarray}
\int d\mathbf{r}^{\prime }\frac{\delta \rho _{h\ell }}{\delta y_{\alpha
}\left( \mathbf{r}^{\prime },t\right) }N_{\alpha }(\mathbf{r}^{\prime },t
&\mid &\left\{ y_{\alpha }\right\} )=\frac{\partial \rho _{h}\left( \left\{
y_{\alpha }(\mathbf{r},t\mathbf{)}\right\} \right) }{\partial y_{\alpha
}\left( \mathbf{r},t\right) }N_{\alpha }(\mathbf{r},t\mid \left\{ y_{\alpha
}\right\} )  \notag \\
+\frac{\partial \mathbf{m}_{\beta }\left( \mathbf{r},\left\{ y_{\alpha }(%
\mathbf{r},t\mathbf{)}\right\} \right) }{\partial y_{\alpha }\left( \mathbf{r%
},t\right) }N_{\alpha }(\mathbf{r},t &\mid &\left\{ y_{\alpha }\right\}
)\cdot \nabla y_{\beta }\left( \mathbf{r},t\right)  \notag \\
+\mathbf{m}_{\alpha }\left( \mathbf{r},\left\{ y_{\alpha }(\mathbf{r},t%
\mathbf{)}\right\} \right) \cdot \nabla N_{\alpha }(\mathbf{r},t &\mid
&\left\{ y_{\alpha }\right\} )  \label{e.4}
\end{eqnarray}%
The last two terms are of first order in the gradients so it is sufficient
to use the lowest order form $N_{\alpha }(\mathbf{r},t\mid \left\{ y_{\alpha
}\right\} )\rightarrow -\delta _{\alpha 2}\zeta _{h}\left( \left\{ y_{\alpha
}(\mathbf{r},t\mathbf{)}\right\} \right) T\left( \mathbf{r},t\right) $ to get%
\begin{eqnarray}
\int d\mathbf{r}^{\prime }\frac{\delta \rho _{h\ell }}{\delta y_{\alpha
}\left( \mathbf{r}^{\prime },t\right) }N_{\alpha }(\mathbf{r}^{\prime },t
&\mid &\left\{ y_{\alpha }\right\} )=\frac{\partial \rho _{h}\left( \left\{
y_{\alpha }(\mathbf{r},t\mathbf{)}\right\} \right) }{\partial y_{\alpha
}\left( \mathbf{r},t\right) }N_{\alpha }(\mathbf{r},t\mid \left\{ y_{\alpha
}\right\} )  \notag \\
&&-\zeta _{h}\left( \left\{ y_{\alpha }(\mathbf{r},t\mathbf{)}\right\}
\right) T\left( \mathbf{r},t\right) \frac{\partial \mathbf{m}_{\beta }\left( 
\mathbf{r},\left\{ y_{\alpha }(\mathbf{r},t\mathbf{)}\right\} \right) }{%
\partial T\left( \mathbf{r},t\right) }  \notag \\
&&+\mathbf{m}_{2}\left( \mathbf{r},\left\{ y_{\alpha }(\mathbf{r},t\mathbf{)}%
\right\} \right) \cdot \frac{\partial \zeta _{h}\left( \left\{ y_{\alpha }(%
\mathbf{r},t\mathbf{)}\right\} \right) T\left( \mathbf{r},t\right) }{%
\partial y_{\beta }\left( \mathbf{r},t\right) }\cdot \nabla y_{\beta }\left( 
\mathbf{r},t\right)  \label{e.5}
\end{eqnarray}%
The right side of (\ref{e.1}) is now%
\begin{eqnarray}
-\int d\mathbf{r}^{\prime }\frac{\delta \rho _{h\ell }}{\delta y_{\alpha
}\left( \mathbf{r}^{\prime },t\right) }N_{\alpha }(\mathbf{r}^{\prime },t
&\mid &\left\{ y_{\alpha }\right\} )-\overline{L}\rho _{h\ell }=-\overline{%
\mathcal{L}}_{T}\rho _{h}-\left( \overline{\mathcal{L}}_{T}\mathbf{m}_{\beta
}\left( \mathbf{r},\left\{ y_{\alpha }(\mathbf{r},t\mathbf{)}\right\}
\right) \right.  \notag \\
&&\left. +\mathbf{m}_{2}\left( \mathbf{r},\left\{ y_{\alpha }(\mathbf{r},t%
\mathbf{)}\right\} \right) \frac{\partial \zeta _{h}\left( \left\{ y_{\alpha
}(\mathbf{r},t\mathbf{)}\right\} \right) T\left( \mathbf{r},t\right) }{%
\partial y_{\beta }\left( \mathbf{r},t\right) }\right) \cdot \nabla y_{\beta
}\left( \mathbf{r},t\right)  \notag \\
&&-\frac{\partial \rho _{h}\left( \left\{ y_{\alpha }(\mathbf{r},t\mathbf{)}%
\right\} \right) }{\partial y_{\alpha }\left( \mathbf{r},t\right) }\left(
N_{\alpha }(\mathbf{r},t\mid \left\{ y_{\alpha }\right\} )\right.  \notag \\
&&\left. +\delta _{\alpha 2}\zeta _{h}\left( \left\{ y_{\alpha }(\mathbf{r},t%
\mathbf{)}\right\} \right) T\left( \mathbf{r},t\right) \right) .  \label{e.6}
\end{eqnarray}%
The operator $\overline{\mathcal{L}}_{T}$ is the same as that introduced in
Section \ref{sec3}, except with the HCS values $\left\{ y_{h\alpha }\right\} 
$ replaced by the true values at the point of interest, $\left\{ y_{\alpha }(%
\mathbf{r},t\mathbf{)}\right\} $%
\begin{equation}
\overline{\mathcal{L}}_{T}X\left( \left\{ y_{\alpha }(\mathbf{r},t\mathbf{)}%
\right\} \right) \equiv -\zeta _{h}\left( \left\{ y_{\alpha }(\mathbf{r},t%
\mathbf{)}\right\} \right) T\left( \mathbf{r},t\right) \frac{\partial
X\left( \left\{ y_{\alpha }(\mathbf{r},t\mathbf{)}\right\} \right) }{%
\partial T\left( \mathbf{r},t\right) }+\overline{L}X.  \label{e.7}
\end{equation}%
The first term of (\ref{e.6}) vanishes, $\overline{\mathcal{L}}_{T}\rho
_{h}=0$. This is the first important consequence of the choice $\rho _{h\ell
}$ for the reference state; it is a solution to the Liouville equation to
first order in the gradients.

Next consider the last term of (\ref{e.6}) and recall that $N_{\alpha }(%
\mathbf{r},t\mid \left\{ y_{\alpha }\right\} )$ arises from averaging the
microscopic conservation laws%
\begin{eqnarray}
N_{\alpha }(\mathbf{r},t &\mid &\left\{ y_{\alpha }\right\} )=\int d\Gamma
\left( L\widetilde{a}_{\alpha }\left( \mathbf{r,}\left\{ y_{\alpha }(\mathbf{%
r},t\mathbf{)}\right\} \right) \right) \rho _{n}=-\int d\Gamma \widetilde{a}%
_{\alpha }\left( \mathbf{r,}\left\{ y_{\alpha }(\mathbf{r},t\mathbf{)}%
\right\} \right) \overline{L}\rho _{n}  \notag \\
&=&-\int d\Gamma \widetilde{a}_{\alpha }\left( \mathbf{r,}\left\{ y_{\alpha
}(\mathbf{r},t\mathbf{)}\right\} \right) \left( \overline{\mathcal{L}}%
_{T}\rho _{n}+\zeta _{h}\left( \left\{ y_{\alpha }\right\} \right) T\frac{%
\partial \rho _{n}}{\partial T}\right)  \label{e.8}
\end{eqnarray}%
where $\widetilde{a}_{\alpha }\left( \mathbf{r,}\left\{ y_{\alpha }(\mathbf{r%
},t\mathbf{)}\right\} \right) $ are linear combinations of the local
densities of number, energy, and momentum%
\begin{equation}
\widetilde{a}_{\alpha }\left( \mathbf{r,}\left\{ y_{\alpha }(\mathbf{r},t%
\mathbf{)}\right\} \right) \mathbf{=}\left( 
\begin{array}{c}
\widehat{n}\left( \mathbf{r}\right) -n\left( \mathbf{r},t\right) \\ 
\frac{2}{3n\left( \mathbf{r},t\right) }\left( \widehat{e}^{\prime }\left( 
\mathbf{r}\right) -\frac{3}{2}T\left( \mathbf{r},t\right) \widehat{n}\left( 
\mathbf{r}\right) \right) \\ 
\frac{1}{n\left( \mathbf{r},t\right) m}\widehat{\mathbf{g}}^{\prime }\left( 
\mathbf{r}\right)%
\end{array}%
\right) .  \label{e.9}
\end{equation}%
Note that 
\begin{eqnarray}
\int d\Gamma \widetilde{a}_{\alpha }\left( \mathbf{r,}\left\{ y_{\alpha }(%
\mathbf{r},t\mathbf{)}\right\} \right) \zeta _{h}\left( \left\{ y_{\alpha
}\right\} \right) T\frac{\partial \rho _{n}}{\partial T} &=&\zeta _{h}\left(
\left\{ y_{\alpha }\right\} \right) T\frac{\partial }{\partial T}\int
d\Gamma \widetilde{a}_{\alpha }\left( \mathbf{r,}\left\{ y_{\alpha }(\mathbf{%
r},t\mathbf{)}\right\} \right) \rho _{n}  \notag \\
&&-\zeta _{h}\left( \left\{ y_{\alpha }\right\} \right) T\int d\Gamma \frac{%
\partial \widetilde{a}_{\alpha }\left( \mathbf{r,}\left\{ y_{\alpha }(%
\mathbf{r},t\mathbf{)}\right\} \right) }{\partial T}\rho _{n}  \notag \\
&=&\delta _{\alpha 2}\zeta _{h}\left( \left\{ y_{\alpha }\right\} \right) T,
\label{e.10}
\end{eqnarray}%
so%
\begin{eqnarray}
N_{\alpha }(\mathbf{r},t &\mid &\left\{ y_{\alpha }\right\} )+\delta
_{\alpha 2}\zeta _{h}\left( \left\{ y_{\alpha }\right\} \right) T=-\int
d\Gamma \widetilde{a}_{\alpha }\left( \mathbf{r,}\left\{ y_{\alpha }(\mathbf{%
r},t\mathbf{)}\right\} \right) \overline{\mathcal{L}}_{T}\left( \rho _{h\ell
}+\Delta \right)  \notag \\
&=&-\int d\Gamma \widetilde{a}_{\alpha }\left( \mathbf{r,}\left\{ y_{\alpha
}(\mathbf{r},t\mathbf{)}\right\} \right) \overline{\mathcal{L}}_{T}\mathbf{m}%
_{\beta }\left( \mathbf{r},\left\{ y_{\alpha }(\mathbf{r},t\mathbf{)}%
\right\} \right) \cdot \nabla y_{\beta }\left( \mathbf{r},t\right)  \notag \\
&&-\int d\Gamma \widetilde{a}_{\alpha }\left( \mathbf{r,}\left\{ y_{\alpha }(%
\mathbf{r},t\mathbf{)}\right\} \right) \overline{\mathcal{L}}_{T}\Delta
\label{e.11}
\end{eqnarray}

With these results the Liouville equation (\ref{e.1}) becomes%
\begin{equation}
\partial _{t}\Delta +\int d\mathbf{r}\frac{\delta \Delta }{\delta y_{\alpha
}\left( \mathbf{r},t\right) }N_{\alpha }(\mathbf{r},t\mid \left\{ y_{\alpha
}\right\} )+\overline{L}\Delta -\frac{\partial \rho _{h}\left( \left\{
y_{\alpha }(\mathbf{r},t\mathbf{)}\right\} \right) }{\partial y_{\alpha
}\left( \mathbf{r},t\right) }\int d\Gamma \widetilde{a}_{\alpha }\left( 
\mathbf{r,}\left\{ y_{\alpha }(\mathbf{r},t\mathbf{)}\right\} \right) 
\overline{\mathcal{L}}_{T}\Delta  \notag
\end{equation}%
\begin{eqnarray}
&=&-\left( \overline{\mathcal{L}}_{T}\mathbf{m}_{\beta }\left( \mathbf{r}%
,\left\{ y_{\alpha }(\mathbf{r},t\mathbf{)}\right\} \right) +\mathbf{m}%
_{2}\left( \mathbf{r},\left\{ y_{\alpha }(\mathbf{r},t\mathbf{)}\right\}
\right) \frac{\partial \zeta _{h}\left( \left\{ y_{\alpha }(\mathbf{r},t%
\mathbf{)}\right\} \right) T\left( \mathbf{r},t\right) }{\partial y_{\beta
}\left( \mathbf{r},t\right) }\right) \cdot \nabla y_{\beta }\left( \mathbf{r}%
,t\right)  \notag \\
&&+\frac{\partial \rho _{h}\left( \left\{ y_{\alpha }(\mathbf{r},t\mathbf{)}%
\right\} \right) }{\partial y_{\alpha }\left( \mathbf{r},t\right) }\int
d\Gamma \widetilde{a}_{\alpha }\left( \mathbf{r,}\left\{ y_{\alpha }(\mathbf{%
r},t\mathbf{)}\right\} \right) \overline{\mathcal{L}}_{T}\mathbf{m}_{\beta
}\left( \mathbf{r},\left\{ y_{\alpha }(\mathbf{r},t\mathbf{)}\right\}
\right) \cdot \nabla y_{\beta }\left( \mathbf{r},t\right)  \label{e.12}
\end{eqnarray}%
The right side is now explicitly proportional to the gradients, so a
solution of the form (\ref{6.1}) is possible%
\begin{equation}
\Delta =\int d\mathbf{rG}_{\alpha }\left( \Gamma ,\mathbf{r},t\mid \left\{
y_{\alpha }(\mathbf{r},t\mathbf{)}\right\} \right) \cdot \boldsymbol{\nabla }%
y_{\alpha }\left( \mathbf{r},t\right) .  \label{e.13}
\end{equation}%
This is a direct consequence of the choice of $\rho _{h\ell }$ as the
reference state, as seen by the above analysis where all terms of zeroth
order in the gradients cancel. Substitution of (\ref{e.13}) into (\ref{e.12}%
), evaluating the functional derivatives on the left side, and equating
coefficients of the gradients gives the desired equation for $\mathbf{G}%
_{\alpha }\left( \Gamma ,\mathbf{r},t\mid \left\{ y_{\alpha }\right\}
\right) $ (here and below it is understood that all fields are evaluated at $%
y_{\alpha }=y_{\alpha }(\mathbf{r},t\mathbf{)}$) 
\begin{eqnarray}
&&\partial _{t}\Delta +\left( \overline{\mathcal{L}}_{T}\mathbf{G}_{\beta
}\left( t,\left\{ y_{\alpha }\right\} \right) +\mathbf{G}_{2}\left(
t,\left\{ y_{\alpha }\right\} \right) \frac{\partial \zeta _{h}\left(
\left\{ y_{\alpha }\right\} \right) T}{\partial y_{\beta }}\right)  \notag \\
&&-\frac{\partial \rho _{h}\left( \left\{ y_{\alpha }\right\} \right) }{%
\partial y_{\alpha }}\int d\Gamma \widetilde{a}_{\alpha }\left( \mathbf{r,}%
\left\{ y_{\alpha }\right\} \right) \overline{\mathcal{L}}_{T}\mathbf{G}%
_{\beta }\left( t,\left\{ y_{\alpha }\right\} \right)  \notag \\
&=&-\left( \overline{\mathcal{L}}_{T}\mathbf{m}_{\beta }\left( \mathbf{r}%
,\left\{ y_{\alpha }\right\} \right) +\mathbf{m}_{2}\left( \mathbf{r}%
,\left\{ y_{\alpha }\right\} \right) \frac{\partial \zeta _{h}\left( \left\{
y_{\alpha }\right\} \right) T}{\partial y_{\beta }}\right)  \notag \\
&&+\frac{\partial \rho _{h}\left( \left\{ y_{\alpha }\right\} \right) }{%
\partial y_{\alpha }}\int d\Gamma \widetilde{a}_{\alpha }\left( \mathbf{r,}%
\left\{ y_{\alpha }\right\} \right) \overline{\mathcal{L}}_{T}\mathbf{m}%
_{\beta }\left( \mathbf{r},\left\{ y_{\alpha }\right\} \right)  \label{e.15}
\end{eqnarray}

The structure of this equation can be exposed further by introducing a set
of functions $\left\{ \psi _{\alpha }\right\} $ that are biorthogonal to the
set $\left\{ \widetilde{a}_{\alpha }\right\} $%
\begin{equation}
\psi _{\alpha }(\mathbf{r}^{\prime }\mathbf{,}\left\{ y_{\alpha }\right\} 
\mathbf{)=}\left( \frac{\delta \rho _{h\ell }\left( \left\{ y_{\alpha
}\right\} \right) }{\delta y_{\alpha }\left( \mathbf{r}^{\prime },t\right) }%
\right) _{\delta y=0}.  \label{e.16}
\end{equation}%
The biorthogonality property is%
\begin{eqnarray}
\int d\Gamma \widetilde{a}_{\alpha }\left( \mathbf{r,}\left\{ y_{\alpha
}\right\} \right) \psi _{\beta }(\mathbf{r}^{\prime }\mathbf{,}\left\{
y_{\alpha }\right\} \mathbf{)} &\mathbf{=}&\left( \frac{\delta }{\delta
y_{\beta }\left( \mathbf{r}^{\prime },t\right) }\int d\Gamma \widetilde{a}%
_{\alpha }\left( \mathbf{r,}\left\{ y_{\alpha }\right\} \right) \rho _{h\ell
}\left( \left\{ y_{\alpha }\right\} \right) \right) _{\delta y=0}  \notag \\
&&-\int d\Gamma \frac{\delta \widetilde{a}_{\alpha }\left( \mathbf{r,}%
\left\{ y_{\alpha }\right\} \right) }{\delta y_{\beta }\left( \mathbf{r}%
^{\prime },t\right) }\rho _{h}\left( \left\{ y_{\alpha }\right\} \right) 
\notag \\
&=&\delta \left( \mathbf{r}^{\prime }-\mathbf{r}\right) \delta _{\alpha
\beta }  \label{e.17}
\end{eqnarray}%
Integrating over $\mathbf{r}^{\prime }$ gives the related orthogonality
condition%
\begin{equation}
\frac{1}{V}\int d\Gamma \widetilde{A}_{\beta }\Psi _{\alpha }=\delta
_{\alpha \beta },  \label{e.18}
\end{equation}%
where 
\begin{equation}
\Psi _{\alpha }\left( \left\{ y_{\alpha }\right\} \right) =\frac{\partial
\rho _{h}\left( \left\{ y_{\alpha }\right\} \right) }{\partial y_{\alpha }},%
\hspace{0.3in}\widetilde{A}_{\beta }\left( \left\{ y_{\alpha }\right\}
\right) =\int d\mathbf{r}\left( 
\begin{array}{c}
\widehat{N}-N \\ 
\frac{2}{3n}\left( \widehat{H}^{\prime }-\frac{3}{2}T\widehat{N}\right) \\ 
\frac{1}{nm}\widehat{\mathbf{P}}^{\prime }%
\end{array}%
\right) .  \label{e.19}
\end{equation}%
Several useful identities identities follow from the condition that the
averages of fields $\widetilde{a}_{\alpha }\left( \mathbf{r,}\left\{
y_{\alpha }\right\} \right) $ vanish for $\rho _{h\ell }$ and therefore,
according to (\ref{6.2}) 
\begin{equation}
\int d\Gamma \widetilde{a}_{\alpha }\left( \mathbf{r,}\left\{ y_{\alpha }(%
\mathbf{r},t\mathbf{)}\right\} \right) \rho _{h\ell }\left( \left\{
y_{\alpha }(\mathbf{r},t\mathbf{)}\right\} \right) =0=\int d\Gamma 
\widetilde{A}_{\alpha }\left( \left\{ y_{\alpha }(\mathbf{r},t\mathbf{)}%
\right\} \right) \rho _{h\ell }\left( \left\{ y_{\alpha }(\mathbf{r},t%
\mathbf{)}\right\} \right) .  \label{e.20}
\end{equation}%
Direct calculation gives the same result for $\rho _{h}\left( \left\{
y_{\alpha }\right\} \right) $ 
\begin{equation}
\int d\Gamma \widetilde{a}_{\alpha }\left( \mathbf{r,}\left\{ y_{\alpha }(%
\mathbf{r},t\mathbf{)}\right\} \right) \rho _{h}\left( \left\{ y_{\alpha
}\right\} \right) =0=\int d\Gamma \widetilde{A}_{\alpha }\left( \left\{
y_{\alpha }(\mathbf{r},t\mathbf{)}\right\} \right) \rho _{h}\left( \left\{
y_{\alpha }(\mathbf{r},t\mathbf{)}\right\} \right) .  \label{e.21}
\end{equation}%
This implies%
\begin{equation}
\int d\Gamma \widetilde{a}_{\alpha }\left( \mathbf{r,}\left\{ y_{\alpha }(%
\mathbf{r},t\mathbf{)}\right\} \right) \mathbf{m}_{\beta }\left( \mathbf{r}%
,\left\{ y_{\alpha }(\mathbf{r},t\mathbf{)}\right\} \right) =0=\int d\Gamma 
\widetilde{A}_{\alpha }\left( \left\{ y_{\alpha }(\mathbf{r},t\mathbf{)}%
\right\} \right) \mathbf{m}_{\beta }\left( \mathbf{r},\left\{ y_{\alpha }(%
\mathbf{r},t\mathbf{)}\right\} \right)  \label{e.22}
\end{equation}%
\begin{equation}
\int d\Gamma \widetilde{a}_{\alpha }\left( \mathbf{r,}\left\{ y_{\alpha }(%
\mathbf{r},t\mathbf{)}\right\} \right) \mathbf{G}_{\beta }\left( t,\left\{
y_{\alpha }(\mathbf{r},t\mathbf{)}\right\} \right) =0=\int d\Gamma 
\widetilde{A}_{\alpha }\left( \left\{ y_{\alpha }(\mathbf{r},t\mathbf{)}%
\right\} \right) \mathbf{G}_{\beta }\left( t,\left\{ y_{\alpha }(\mathbf{r},t%
\mathbf{)}\right\} \right)  \label{e.23}
\end{equation}%
Finally, define the projection operator $\mathcal{P}$ by%
\begin{equation}
\mathcal{P}X=\Psi _{\alpha }\frac{1}{V}\int d\Gamma \widetilde{A}_{\alpha }X.
\label{e.24}
\end{equation}%
Then (\ref{e.15}) takes the final form 
\begin{equation}
\left( \partial _{t}+\left( 1-\mathcal{P}\right) \left( I\overline{\mathcal{L%
}}_{T}+K^{T}\right) \right) \mathbf{G}\left( t,\left\{ y_{\alpha }\right\}
\right) =-\left( 1-\mathcal{P}\right) \left( I\overline{\mathcal{L}}%
_{T}+K^{T}\right) \mathbf{M}\left( \left\{ y_{\alpha }\right\} \right)
\label{e.25}
\end{equation}%
Here $I$ is the unit matrix and $K^{T}$ is the transpose of the matrix%
\begin{equation}
K=\left( 
\begin{array}{ccc}
0 & 0 & 0 \\ 
\frac{\partial \left( \zeta _{h}T\right) }{\partial n} & \frac{\partial
\left( \zeta _{h}T\right) }{\partial T} & 0 \\ 
0 & 0 & 0%
\end{array}%
\right) .  \label{e.26}
\end{equation}%
The matrix $K$ generates the solution to the hydrodynamic equations in the
absence of gradients, Eqs. (\ref{6.16}).

The explicit dependence on $\mathbf{r}$ has cancelled in (\ref{e.25}) (as it
must, for a normal solution) as a consequence of 
\begin{equation}
\left( I\overline{\mathcal{L}}_{T}+K^{T}\right) \mathbf{r}\Psi =\mathbf{r}%
\left( I\overline{\mathcal{L}}_{T}+K^{T}\right) \Psi =0.  \label{e.27}
\end{equation}%
The second equality follows from the fact that $\Psi $ form the null space
for $\left( I\overline{\mathcal{L}}_{T}+K^{T}\right) $ 
\begin{equation}
\left( I\overline{\mathcal{L}}_{T}+K^{T}\right) \Psi =0.  \label{e.29}
\end{equation}%
This can be demonstrated by direct calculation%
\begin{eqnarray}
\left( I\overline{\mathcal{L}}_{T}+K^{T}\right) _{\alpha \beta }\Psi _{\beta
} &=&\left( \overline{L}\Psi _{\alpha }-\zeta _{h}T\partial _{T}+K_{\alpha
2}^{T}\Psi _{2}\right)  \notag \\
&=&\frac{\partial \overline{L}\rho _{h}}{\partial y_{\alpha }}-\zeta _{h}T%
\frac{\partial \Psi _{2}}{\partial y_{\alpha }}+K_{\alpha 2}^{T}\Psi _{2} 
\notag \\
&=&\frac{\partial \zeta _{h}T\Psi _{2}}{\partial y_{\alpha }}-\zeta _{h}T%
\frac{\partial \Psi _{2}}{\partial y_{\alpha }}+K_{\alpha 2}^{T}\Psi _{2}=0
\label{e.30}
\end{eqnarray}

Equation (\ref{e.25}) is the primary result of this Appendix, giving the
exact solution to the Liouville equation up to first order in the gradients.
As a final simplification, a transformation from the operator $\overline{%
\mathcal{L}}_{T}$ to the phase space operator $\overline{\mathcal{L}}$ of (%
\ref{3.25}) can be made by introducing dimensionless variables 
\begin{eqnarray}
\mathbf{G}_{1}\left( \Gamma ,t,\left\{ y_{\alpha }\right\} \right) &=&\frac{%
\ell }{n}\left( \ell v(T)\right) ^{-3N}\mathbf{G}_{1}^{\ast }\left( \Gamma
^{\ast },s,n\ell ^{3}\right) ,  \notag \\
\mathbf{G}_{2}\left( \Gamma ,t,\left\{ y_{\alpha }\right\} \right) &=&\frac{%
\ell }{T}\left( \ell v(T)\right) ^{-3N}\mathbf{G}_{2}^{\ast }\left( \Gamma
^{\ast },s,n\ell ^{3}\right) ,  \notag \\
\mathbf{G}_{3}\left( \Gamma ,t,\left\{ y_{\alpha }\right\} \right) &=&\frac{%
\ell }{T^{1/2}}\left( \ell v(T)\right) ^{-3N}\mathbf{G}_{2}^{\ast }\left(
\Gamma ^{\ast },s,n\ell ^{3}\right)  \label{e.31}
\end{eqnarray}%
The first factor in each case gives the dimensions to compensate for the
associate gradient multiplying $\mathbf{G}_{\alpha }$. The second time
arises in each case because the solution to the Liouville equation is a
density in phase space. The dimensionless phase point $\Gamma ^{\ast }\equiv
\left\{ \mathbf{q}_{1}^{\ast },..,\mathbf{q}_{N}^{\ast },\mathbf{V}%
_{1}^{\ast },..,\mathbf{V}_{N}^{\ast }\right\} $ is defined by%
\begin{equation}
\mathbf{q}_{r}^{\ast }=\mathbf{q}_{r}/\ell ,\hspace{0.3in}\mathbf{V}%
_{r}^{\ast }=\left( \mathbf{v}_{r}-\mathbf{U}(\mathbf{r},t)\right) /v_{h}(t)
\label{e.32}
\end{equation}%
where $v_{h}(t)$ is the thermal velocity defined in (\ref{3.22}), with $%
T_{h}\rightarrow T(\mathbf{r},t)$. With these definitions%
\begin{eqnarray}
\left( \partial _{t}\mid _{\Gamma ,\left\{ y_{\alpha }\right\} }-\zeta
_{h}T\partial _{T}\mid _{\Gamma ,t}\right) &=&\frac{v_{h}(t)}{\ell }\left( 
\frac{\ell }{v_{h}(t)}\partial _{t}\mid _{\Gamma ^{\ast },\left\{ y_{\alpha
}\right\} }-\zeta _{h}^{\ast }T\partial _{T}\mid _{\Gamma ^{\ast },t}+\frac{%
\zeta _{h}^{\ast }}{2}\sum_{r=1}^{N}\left( 3+\mathbf{V}_{r}^{\ast }\cdot 
\boldsymbol{\nabla }_{\mathbf{V}_{r}^{\ast }}\right) \right)  \notag \\
&=&\frac{v_{h}(t)}{\ell }\left( \frac{\ell }{v_{h}(t)}\partial _{t}\mid
_{\Gamma ^{\ast },\left\{ y_{\alpha }\right\} }-\zeta _{h}^{\ast }T\frac{%
\partial s}{\partial T}\mid _{t}\partial _{s}\mid _{\Gamma ^{\ast }}\right. 
\notag \\
&&\left. -\zeta _{h}^{\ast }T\partial _{T}\mid _{\Gamma ^{\ast },s}+\frac{%
\zeta _{h}^{\ast }}{2}\sum_{r=1}^{N}\left( 3+\mathbf{V}_{r}^{\ast }\cdot 
\boldsymbol{\nabla }_{\mathbf{V}_{r}^{\ast }}\right) \right)  \label{e.33}
\end{eqnarray}%
Here $s=s(t,T)$ is a dimensionless ttime scale, and $\zeta _{h}^{\ast }$ is
the constant dimensionless cooling rate of (\ref{3.30}). A judicious choice
for the dimensionless time is seen to be%
\begin{equation}
\frac{\ell }{v_{h}(t)}\partial _{t}\mid _{\Gamma ,\left\{ y_{\alpha
}\right\} }-\zeta _{h}^{\ast }T\frac{\partial s}{\partial T}\mid
_{t}\partial _{s}\mid _{\Gamma ^{\ast }}\equiv \partial _{s}\mid _{\Gamma
^{\ast }}  \label{e.34}
\end{equation}%
or%
\begin{equation}
\frac{ds}{\left( 1+\zeta _{h}^{\ast }T\frac{\partial s}{\partial T}\mid
_{t}\right) }\equiv \frac{v_{h}(t)}{\ell }dt.  \label{e.35}
\end{equation}%
It can be shown that this definition agrees with that of (\ref{3.34}) in the
sense $s(t)=s(t,T(t))$%
\begin{equation}
ds(t)=\frac{v_{h}(T(t))}{\ell }dt,\hspace{0.3in}T(t)=T(0)\left( 1+\frac{%
v_{h}(0)\zeta _{h}^{\ast }}{2\ell }t\right) ^{-2}.  \label{e.35a}
\end{equation}

Equation (\ref{e.33}) simplifies to 
\begin{equation}
\left( \partial _{t}\mid _{\Gamma ,\left\{ y_{\alpha }\right\} }-\zeta
_{h}T\partial _{T}\mid _{\Gamma ,t}\right) =\frac{v_{h}(t)}{\ell }\left(
\partial _{s}\mid _{\Gamma ^{\ast }}-\zeta _{h}^{\ast }T\partial _{T}\mid
_{\Gamma ^{\ast },s}+\frac{\zeta _{h}^{\ast }}{2}\sum_{r=1}^{N}\left( 3+%
\mathbf{V}_{r}^{\ast }\cdot \boldsymbol{\nabla }_{\mathbf{V}_{r}^{\ast
}}\right) \right) .  \label{e.36}
\end{equation}%
Now, taking into account the forms (\ref{e.31}), the equation (\ref{e.25})
for $\mathbf{G}_{\alpha }$ has the corresponding dimensionless form%
\begin{equation}
\left( \partial _{s}+\left( 1-\mathcal{P}^{\ast }\right) \left( I\overline{%
\mathcal{L}}^{\ast }-\overline{\Lambda }^{\ast }+K^{T\ast }\right) \right) 
\mathbf{G}^{\ast }=-\left( 1-\mathcal{P}^{\ast }\right) \left( I\overline{%
\mathcal{L}}^{\ast }-\overline{\Lambda }^{\ast }+K^{T\ast }\right) \mathbf{M}%
^{\ast }\left( \left\{ y_{\alpha }\right\} \right) ,  \label{e.37}
\end{equation}
where an asterisk denotes the function, operator, or matrix in terms of the
dimensionless variables. The matrix $\overline{\Lambda }^{\ast }$ arises
from the first factors of (\ref{e.31}) associated with the dimensions of the
respective gradients. 
\begin{equation}
\overline{\Lambda }^{\ast }=\left( 
\begin{array}{ccc}
0 & 0 & 0 \\ 
0 & \zeta _{h}^{\ast } & 0 \\ 
0 & 0 & \frac{1}{2}\zeta _{h}^{\ast }%
\end{array}%
\right) .  \label{e.38}
\end{equation}%
The notation is simplified by introducing the matrix $\Lambda ^{\ast }$ 
\begin{equation}
\Lambda ^{\ast }=K^{T\ast }-\overline{\Lambda }^{\ast }=\left( 
\begin{array}{ccc}
0 & \frac{\partial \left( \zeta _{h}T\right) }{Tv_{h}\ell ^{2}\partial n} & 0
\\ 
0 & \frac{1}{2}\zeta _{h}^{\ast } & 0 \\ 
0 & 0 & -\frac{1}{2}\zeta _{h}^{\ast }%
\end{array}%
\right) .  \label{e.39}
\end{equation}


\begin{thebibliography}{99}
\bibitem{Navier-Stokes} Here the Navier-Stokes approximation is defined by
calculating the cooling rate, energy flux, and momentum flux to first order
in the gradients. However, the fluxes occur under a gradient in the
macroscopic balance equations while the cooling rate does not. Hence the
equations themselves do not have all terms to second order in the gradients
(i.e., the additional terms of second order contributing to the cooling
rate).
\end{thebibliography}
\end{document}